\documentclass[11pt,reqno]{amsart}

\usepackage[T1]{fontenc}     
\usepackage{lmodern}         
\usepackage{amsmath, amsthm, amssymb,amsaddr}
\usepackage{dcolumn}
\usepackage{bm}
\usepackage{amsmath} 
\usepackage{mathrsfs}
\usepackage{mathtools}
\usepackage{algpseudocode}
\usepackage{algorithm}
\usepackage[pdftex]{graphicx}
\usepackage{subfigure}
\usepackage{epstopdf}
\usepackage{tikz}
\usepackage{xcolor}
\usepackage[export]{adjustbox}
\usepackage{caption}
\usepackage{hyperref}
\usepackage[pagewise]{lineno}
\usepackage{multirow}
\usepackage{caption}
\usepackage{tikz}
\usepackage{subcaption}
\usepackage[export]{adjustbox} 

\theoremstyle{remark}

\newtheorem{definition}{Definition}

\DeclareMathOperator*{\argmin}{arg\,min}

\newcommand{\rev}[1]{{\color{black}#1}}

\newcommand{\be}[1]{\begin{equation}\label{#1}}
\newcommand{\ee}{\end{equation}}

\newcommand{\no}[1]{#1}

\newcommand{\bx}{\boldsymbol{x}}

\newcommand{\by}{\boldsymbol{y}}

\newcommand{\bk}{\boldsymbol{k}}
\newcommand{\btk}{\mathbf{\bar{k}}}

\newcommand{\iu}{{i\mkern1mu}}
\newcommand{\fd}{\mathfrak{d}}

\renewcommand{\no}[1]{} 

\title[Hessian-free ray-Born inversion for quantitative ultrasound tomography]{Hessian-free ray-Born inversion for quantitative ultrasound tomography of weakly heterogeneous media}
\author{Ashkan Javaherian}
\address{Department of Bio-Electric, School of Electrical and Computer Engineering, University College of Engineering, University of Tehran, Tehran, Iran.
\footnote{The earlier method referenced in the Abstract is described in \cite{Javaherian2}.
The MATLAB codes supporting the findings of this study, as well as those reported in \cite{Javaherian2} and \cite{Javaherian1}, are publicly available in the following GitHub repository: \url{https://github.com/Ash1362/ray-based-quantitative-ultrasound-tomography/}
 \cite{Javaherian-toolbox}.
Results demonstrating the implementation of the image-reconstruction approach proposed in this study on in-vitro and in-vivo ultrasound datasets released by the University of Rochester Medical Center are reported in \cite{Javaherian-experimental} and are also included in the GitHub repository. The datasets themselves can be accessed via the link provided in \cite{Ali2}.}}
\email{ashkan.javaherian@ut.ac.ir; ajavaherian62@gmail.com}
\date{August 2026}

\begin{document}

\maketitle

\begin{abstract}
This study presents a frequency-domain, Hessian-free ray-Born inversion method for quantitative ultrasound tomography, extending the author's previous Hessian-based approach. Both approaches model acoustic wave propagation using a ray-based approximation of the Green's function in weakly heterogeneous media, and perform the inversion iteratively in the frequency domain, progressing from low to high frequencies. In the earlier method, each frequency subproblem was solved through iterative inversion of the Hessian operator, a process that not only increased computational cost but also made the update steps more sensitive to noise. The present work addresses these limitations by transforming a full-frequency Hessian operator into an identity operator through a specific preconditioning scheme, thereby enabling a single-step inversion for each frequency subproblem. This reformulation reduces the computational expense by approximately an order of magnitude relative to the Hessian-based approach, while yielding robust reconstructions with reduced sensitivity to noise by eliminating the ill-conditioning of the Hessian operator. Furthermore, compared with the author's previous study, the present work improves the approximation of the geometrical component of the wave amplitude by introducing a paraxial ray-tracing framework, thereby further enhancing both computational efficiency and reconstruction accuracy. The inversion approach proposed in the present study has enabled the first successful translation of acoustic inverse scattering methods relying on first-scattered waves to a clinical setting.
\end{abstract}

\section{Introduction}

Ultrasound tomography (UT) is a technique used to map the acoustic properties of an object based on ultrasonic excitations and measurements taken from outside the object \cite{Javaherian2,Duric}. This method has gained significant interest in biomedical applications, particularly for detecting malignant tumors in the breast \cite{Duric}. Ideally, ultrasound data enable the recovery of quantitative maps of sound speed, absorption, and density—collectively referred to as quantitative ultrasound tomography (QUT)—along with a qualitative image of reflectivity \cite{Synnevag, Ruiter,Gemmeke,Hopp1}. This manuscript specifically focuses on sound speed reconstruction, and thus, the term QUT will refer to this process \cite{Li1}. Approaches for sound speed imaging in UT can be categorized based on: (1) the type of data used in inversion, (2) whether the objective function to be minimized is nonlinear or linearized, (3) whether the objective function is defined in the time or frequency domain, and (4) the forward model employed.

The first class of approaches relies exclusively on direct times-of-flight data between emitters and receivers, effectively ignoring scattered waves \cite{Li3, Op}. The second class incorporates the full time-series data, including scattered waves, to provide a more comprehensive representation of the medium \cite{Sandhu, Wiskin2, Goncharsky3, Bernard, Ali}. The third class strikes a balance by leveraging direct arrivals and primarily scattered waves, offering improved accuracy compared to the first class while maintaining a reduced computational burden relative to the second class \cite{Devaney2, Devaney3, Beydoun, Borup, Mast0, Simonetti, Muller, Martiartu, Fan, Fauk, Guo}.

This study adopts a hybrid approach that leverages the strengths of both the first and third classes of methods: direct time-of-flight data are used to generate an initial estimate, while predominantly scattered waves, deaberrated using a ray-based approximation to the Green’s function, are iteratively incorporated to refine the solution to the main inverse problem. In \cite{Javaherian2}, we employed a reconstruction approach based on iterative linearization of the objective function, where the linearized subproblems were formed and solved sequentially in the frequency domain from low to high frequencies. The forward problem is modeled using a ray-approximation to heterogeneous Green’s functions. Ray theory, based on a high-frequency approximation, strikes a balance between accuracy and computational efficiency for broadband data \cite{Thierry}. While ray theory has limitations in seismic imaging, such as caustics and multivaluedness of minimal acoustic length \cite{Cerveny2001}, these challenges are less critical for imaging soft tissues, like the breast, where the refractive index typically ranges from $0.9$ to $1.1$.

The main challenges in extending high-resolution quantitative ultrasound to 3D imaging via full-wave inversion are computational cost and transducer directivity, both of which can be managed more easily with ray-based methods \cite{Javaherian2}. Previous work in \cite{Javaherian2} demonstrated numerically that ray-based methods for quantitative ultrasound tomography (QUT) can reconstruct high-resolution sound speed maps at a significantly lower computational cost than full-wave inversion methods, which solve the complete wave equation \cite{Goncharsky3, Ali}.

Common methods accounting for diffraction or singly scattered waves typically rely on the Green's function in homogeneous media, incorporating acoustic heterogeneity only within the scattering potential \cite{Devaney2, Devaney3, Beydoun, Borup, Mast0, Simonetti, Muller, Martiartu, Fan, Fauk, Guo}. However, it has been shown that combining ray theory with Born inversion can significantly improve accuracy by incorporating sound speed heterogeneity, refraction, and acoustic absorption and dispersion into the Green's functions that predict incident and scattered waves \cite{Javaherian2}. A one-step inversion method combining diffraction tomography with ray tracing was proposed in \cite{Huth}, which compensates for phase aberrations in the Green’s function due to heterogeneities by performing ray tracing on a low-resolution image, initially reconstructed using a time-of-flight-based method, and solving the inverse problem in a single step.

In the present study, for modeling the forward operator based on the ray approximation, the Green's functions are approximated along rays initialized at the emitters and intercepted by the receivers after traversing the medium, referred to as linked rays. While the phase and absorption-related components of the amplitude decay are approximated along the linked rays, evaluating the geometrical spreading contribution of the amplitude decay requires computation of the ray Jacobian, which traditionally involves tracing auxiliary rays for each linked ray.  In \cite{Javaherian2}, the auxiliary rays are traced independently of the linked rays by introducing small perturbations to the initial angles of the linked rays. In contrast, this study introduces a paraxial ray-tracing system, where a paraxial ray is traced simultaneously with the linked ray. This approach enhances both efficiency and stability in computing the rays' Jacobian.

The QUT inversion approaches proposed in \cite{Javaherian2} and extended in this study belong to the class of linearized inversion techniques known as ray-Born migration/inversion in inverse seismic theory \cite{Thierry, Jin, Lambare1, Lambare2}. In \cite{Javaherian2}, the objective function is discretized in the frequency domain, linearized, and sequentially minimized over the frequency range covered by the ultrasound transducers. The range is divided into intervals, each containing a fixed number of discretized frequencies, typically two (chosen heuristically). Minimization proceeds sequentially, starting with the low-frequency intervals. The solution from each frequency subproblem serves as the initial guess for the next frequency set. {\rev{Since reconstructing acoustic parameters from ultrasound data is a nonlinear inverse problem, a low-to-high frequency continuation strategy helps guide the optimization toward the true solution \cite{Sandhu, Wiskin2,Fauk}.}}

Each linearized subproblem in \cite{Javaherian2}, which requires approximating the action of the inverse Hessian on the gradient of the objective function, is solved using a Conjugate Gradient (CG) algorithm. This entails iterative computations of the Fréchet derivative of the forward operator and its adjoint. Each CG iteration involves two-way forward propagation of Green’s functions from the emitters to the receivers, followed by two-way back-projection from the receivers to the emitters. These steps are repeated for all emitter--receiver pairs, leading to a substantial increase in computational cost. Nevertheless, the overall cost remains significantly lower than that of full-wave inversion approaches \cite{Javaherian2}.   

{\rev{
The current manuscript aims to further reduce the computational cost by introducing a Hessian-free solution while preserving the essential information contained in the Hessian operator. To this end, the objective function is linearized once over the entire frequency range. The resulting full-frequency linearized objective function is then preconditioned such that the Hessian operator is approximated by the identity operator, while its informative content is incorporated into the corresponding gradient, thereby obtaining a single-step full-frequency update direction. This update direction is evaluated using a recursive scheme that progressively updates the wavenumber field from low to high frequencies. This framework not only improves computational efficiency by eliminating the need for the recursive computation of the Hessian operator, but also avoids the implicit inversion of the ill-conditioned Hessian operator, thereby reducing sensitivity to noise and reconstruction artifacts while preserving the fine scattering features essential for high-resolution imaging.}}

Section \ref{sec:greens} introduces a Green's function solution to the frequency-domain Helmholtz wave equation using ray theory, applicable to weakly heterogeneous and absorbing media. Section \ref{sec:minimisation_approach} details the Hessian-free ray-Born inversion approach for solving the inverse problem in QUT. Section \ref{sec:ray-tracing} discusses ray tracing and the approximation of the Green's function for heterogeneous media using ray theory. Section \ref{sec:Discretised-approximation} outlines the procedure for discretizing both the forward and inverse problems. Section \ref{sec:results} provides a numerical validation of the ray-based approximation to the heterogeneous Green’s function and presents reconstructed images illustrating the performance of the proposed Hessian-free ray-Born inversion approach, in comparison with a prototype Born inversion method and the more computationally intensive Hessian-based approach from \cite{Javaherian2}. A discussion of these results follows in Section \ref{sec:discussion}. The appendices provide additional theoretical details.

\section{Forward Problem}
\label{sec:greens}

This section outlines the forward problem underlying the inversion approach for reconstructing the spatially varying sound speed. Let \(\bx = \big[x^1, \dots, x^d\big]^\mathrm{T}\) denote a spatial position in \(\mathbb{R}^d\), where \(d\) represents the number of spatial dimensions, typically 2 or 3. While this study focuses on the case \(d = 2\), the findings aim to be extended to \(d = 3\). To avoid repetition, this section is kept concise, and readers are directed to Section 3 of \cite{Javaherian2} for further details.

\subsection{Problem setting}

The domain $\Omega \subset \mathbb{R}^2$ is an open, bounded set containing the spatially-varying sound speed distribution $c(\bx)$. The variation is represented by $\big[c_0 / c(\bx) - 1\big] \in C_0^\infty$, where $c_0$ is the uniform sound speed outside $\Omega$ (e.g., the sound speed in water). The boundary of $\Omega$ is a circular ring $\mathbb{S} \subset \mathbb{R}$ containing the emission and reception elements, denoted by $e$ (emitter) and $r$ (receiver). These elements are modeled as point sources and detectors, respectively.

Each emitter $e \in \{1, \dots, N_e\}$ is sequentially excited by a pulse acting as a source $s(t, e)$ over the time interval $t \in (0, T_s)$. The resulting acoustic pressure field propagates through the object immersed in water and is recorded by receivers $r \in \{1, \dots, N_r\}$ over the time interval $t \in (0, T)$, with $T \gg T_s$. For each emitter $e$, the time-series data recorded by a receiver $r$ is denoted by $\mathfrak{p}(t, r, e)$.

The inverse problem is to reconstruct the spatial distribution $c(\bx)$ from the measured pressure time-series $\mathfrak{p}(t, r, e)$ for all emitter-receiver pairs, given the source $s(t, e)$.

While the measurements are typically made in the time domain using a broadband excitation signal, the image reconstruction is performed in the frequency domain. To facilitate this, the following Fourier transform pair is defined between the time and temporal frequency domains:
\begin{align}
\hat{p}(\omega) = \mathscr{F}p(t) = \int_{-\infty}^{\infty}  dt \, p(t) e^{i\omega t}
\qquad  \mathscr{F}^{-1} \hat{p}(\omega) = \frac{1}{2\pi}\int_{-\infty}^{\infty} d\omega \, \hat{p}(\omega) e^{-i\omega t} .
\end{align}

\subsection{Lossy Helmholtz equation and complex wavevector}
The propagation of a single-frequency acoustic pressure field, $\hat{p}(\omega, \bx)$, in an absorbing medium is governed by the lossy Helmholtz equation:
\begin{align} 
\label{eq:second-order-lossy-omega}
\Big[ \tilde{k}(\bx)^2 + \nabla^2 \Big] \hat{p}(\omega, \bx, e) = -s(\omega, e),
\end{align}
where $\tilde{k}(\bx)$ is the complex wavenumber, which can be defined in different ways. In this study, the complex wavenumber is defined based on Szabo's absorption model, consistent with the absorption model used in the full-waveform solver for generating the synthetic ultrasound data benchmarked in Section \ref{sec:results} \cite{Treeby}. The associated complex wavenumber is given by \cite{Kelly}:

\begin{align}
\label{eq:szabo_dispersion_equation}
\begin{split}
    \tilde{k} & =  \frac{\omega}{c}  - \frac{\alpha_0 (-i)^{y+1} \omega^y}{ \cos(\pi y / 2) }\\
              &  = \frac{\omega}{c} + \alpha \Big[ \tan\left(\frac{\pi y}{2}\right) + i \Big] = k + i\alpha,
\end{split}
\end{align}

where we have used $(-i)^y = \cos(\pi y / 2) - i\sin(\pi y / 2)$. The imaginary part of $\tilde{k}$ represents the power-law absorption $\alpha = \alpha_0 \omega^y$, where $\alpha_0$ has units of $\text{Np} (\text{rad}/\text{s})^{-y} \, \text{m}^{-1}$, and $y$ is a non-integer power-law exponent, typically in the range $1 < y \leqslant 1.5$ for soft tissues \cite{Szabo, Liebler}. Furthermore, the real part $k$ satisfies:
\begin{align}  \label{eq:wave-number}
 k = \frac{\omega}{c} + \alpha_0 \, \tan\left(\frac{\pi y}{2}\right) \omega^y = \frac{\omega}{c_p(\omega)},
\end{align}
where $c_p(\omega)$ denotes the phase speed \cite{Kelly}.

Additionally, the corresponding complex wavevector is defined as $\tilde{\bk} = \bk + i \bk_i$, where:
\begin{align} \label{eq:complex-wavevector}
 |\bk| = k, \quad
 \bk_i = \frac{\alpha}{k} \bk,
\end{align}
with $|\cdot|$ denoting the Euclidean magnitude.

The Green's function solution to the wave equation \eqref{eq:second-order-lossy-omega} is given by:
\begin{align}   \label{eq:general-greens-source}
    \hat{p}(\omega,\bx) = \int g(\omega, \bx, \bx') s(\omega,\bx') \, d\bx',
\end{align}
where $g(\omega, \bx, \bx')$ satisfies the equation:
\begin{align}  \label{eq:greens-wave-equation}
\Big[ \tilde{k}(\bx)^2 + \nabla^2  \Big] g(\omega, \bx, \bx') =  -\delta(\bx - \bx').
\end{align}

For 2D settings in heterogeneous and absorbing media, the Green's function can be approximated as:
\begin{align}
\label{eq:2D_greens}
    g(\omega,\bx, \bx') & \approx A(\omega,\bx, \bx') \exp\left( i\Big[\phi(\omega,\bx, \bx') + \frac{\pi}{4}\Big] \right),
\end{align}
where $\phi$ is the phase, and $A = A_{abs} A_{geom}$ is the amplitude factor, incorporating contributions from absorption ($A_{abs}$) and geometric spreading ($A_{geom}$). Substituting \eqref{eq:2D_greens} into \eqref{eq:greens-wave-equation}, assuming $\alpha^2 \ll k^2$ and the high-frequency approximation $|\nabla^2 A / A| \ll k^2$, yields the Eikonal equation:
\begin{align} \label{eq:eikonal}
\nabla \phi \cdot \nabla \phi = k^2,
\end{align}
and the Transport equation:
\begin{align} \label{eq:transport}
    \nabla \cdot \big[ A_{geom}^2 \nabla \phi \big] = 0.
\end{align}
Further details on the derivation of these two equations are provided in Appendix A. Using the Eikonal equation \eqref{eq:eikonal}, rays are trajectories perpendicular to surfaces of constant phase and are tangent to the real part of the complex wavevector, $\bk$, satisfying:
\begin{align} \label{eq:wave-vector}
    \bk = \nabla \phi.
\end{align} 

Correspondingly, the accumulated phase along a ray trajectory $\mathcal{C}(\bx, \bx')$ is given by:
\begin{align} \label{eq:phase}
\phi(\omega, \bx,\bx') = \int_{\mathcal{C}(\bx, \bx')} d \bx  \cdot \bk ,
\end{align}
and the absorption contribution to the amplitude is given by:
\begin{align} \label{eq:absorption}
A_{abs}(\omega, \bx,\bx') = \exp\left( - \int_{\mathcal{C}(\bx,\bx')}  d \bx \cdot  \bk_i \right).
\end{align}

The geometric spreading term $A_{geom}$ is determined using the Transport equation \eqref{eq:transport} and satisfies:
\begin{align}  \label{eq:attenuation_geom}
    A_{geom}(\omega, \bx, \bx' ) = \bigg[ \frac{J(\omega, \bx') \rho(\bx) c(\bx)}{J(\omega, \bx) \rho(\bx') c(\bx')} \bigg]^{1/2},
\end{align}
where $\rho$ is the medium density (assumed homogeneous, $\rho(\bx) = \rho(\bx')$), and $J(\bx)$ is the ray-tube cross-sectional measure (Jacobian determinant) at \( \bx \). Using the initial point $\bx_0$ of a ray as the reference point, the geometric amplitude $A_{geom}(\bx, \bx_0)$ vanishes \cite{Cerveny2001}. Therefore, the reference point $\bx_{ref}$ is chosen near $\bx_0$, where $|\bx_{ref} - \bx_0| < \epsilon$, with $\epsilon$ being a small neighborhood radius in which the medium is homogeneous (only water). The geometric amplitude at the reference point is expressed as:
\begin{align} \label{eq:attenuation_geom_reference}
A_{\text{geom}}(\omega, \bx_{ref}, \bx_0) = \big[8 \pi \phi_0(\omega, \bx_{ref}, \bx_0) \big]^{-1/2}, 
\end{align}
where \( \phi_0 (\omega, \bx_{ref}, \bx_0) = k_0 |\bx_{ref} - \bx_0| \), and \( k_0 = \omega / c_0 \) represents the wavenumber in water.

The geometric amplitude at an arbitrary point along the ray is then given by:
\begin{align} \label{eq:attenuation_geom2}
A_{geom}(\omega, \bx, \bx_0) = A_{geom}(\omega, \bx_{ref}, \bx_0)  A_{geom}(\omega, \bx, \bx_{ref}), 
\end{align}
where $A_{geom}(\omega, \bx, \bx_{ref})$ is computed from \eqref{eq:attenuation_geom} by setting $\bx' = \bx_{ref}$.

\section{Inverse problem}  \label{sec:minimisation_approach}

\noindent
This section presents the proposed approach for solving the nonlinear inverse problem in quantitative ultrasound tomography (QUT), which aims to reconstruct the sound speed distribution \(c\) of an object from measured pressure time series $  \mathfrak{p}(t,r,e) $. These measurements are acquired through sequential excitations and recordings using an array of ultrasound transducers surrounding the object. 

{\rev{
In the proposed approach, the objective function is linearized once over the entire frequency range of the transducers. The resulting full-frequency linearized problem is then preconditioned so that the Hessian operator is transformed into the identity operator, thereby enabling its inversion in a single step while transferring its informative content to the gradient. This yields a full-frequency search direction in the form of a preconditioned negative gradient, which is evaluated sequentially from low to high frequencies. At each stage, the updated wavenumber map obtained from the current frequency set is used as the initial guess for computing the search direction (i.e., the preconditioned negative gradient) associated with the next frequency set.
}}

In the standard formulation proposed in \cite{Javaherian2}, the objective function is linearized separately for each frequency set. Consequently, solving each linearized subproblem requires evaluating the product of the inverse Hessian operator and the negative gradient of the objective function. This is accomplished iteratively by repeatedly computing products of the Hessian operator with perturbations of the sound speed until the perturbation that minimizes the linearized objective function is obtained \cite{Javaherian2}. In contrast, the present work introduces a preconditioning scheme that transforms the Hessian operator of the full-frequency linearized objective function into the identity operator while transferring its informative content to the gradient. As a result, the Hessian operator can be inverted in a single step, substantially reducing the computational cost of solving the resulting linearized inverse problem.

In \cite{Javaherian2}, the objective function was minimized with respect to the sound speed \( c \). In contrast, this study minimizes the objective function with respect to the squared slowness \( m = c^{-2} \), to facilitate the identity transformation of the Hessian operator. Accordingly, the objective function is formulated in terms of the Green’s function in the frequency domain:
\begin{align} \label{eq:objective-function}
    \mathcal{F}(m) = \frac{1}{2} \sum_{e,r}  \int    d \omega  \, g_{\text{res}}(m;\omega, r,e) ^*    g_{\text{res}}(m;\omega, r,e), 
\end{align}
where \( * \) denotes the complex conjugate. The term \(  g_{\text{res}} \) represents the \textit{residual}, defined as
\begin{align} \label{eq:residual}
     g_{\text{res}}(m;\omega,r,e) = g(m; \omega,r,e) - \mathfrak{g}(\omega,r,e),
\end{align}
where \( g(m; \omega,r,e) \) and \( \mathfrak{g}(\omega,r,e) \) denote the \textit{approximated} and \textit{measured} Green’s functions, respectively, at frequency \( \omega \), receiver position \( r \), and emitter position \( e \). The measured Green’s function \( \mathfrak{g}(\omega, r,e) \) is obtained by deconvolving the source \( s(\omega, e) \) from the measured pressure \( \mathfrak{p}(\omega, r,e) \) in the frequency domain. (An analogous formulation can be applied directly to an objective function based on pressure data, as in \cite{Javaherian2}.) The approximated Green's function is computed along the linked rays traced through the wavenumber field corresponding to the squared slowness $m$. 

\subsection{Linearized objective function}
\label{sec:objective function-linearized}

\noindent
A preconditioned full-frequency linearization of the objective function \eqref{eq:objective-function} about the squared slowness \(m\) yields
\begin{align}  
\begin{split}
\delta m(\bx) \approx \argmin_{\delta m} \, \frac{1}{2} \sum_{e,r}  \int  d \omega \, \bigg[  & \Big[ \delta g(m;\omega,r,e) \big[\delta m \big]   +   g_{\text{res}} (m;\omega,r,e) \Big]^* \mathcal{Q}(m;\omega,r,e,\bx)  \\
 &\Big[ \delta g (m;\omega,r,e) \big[\delta m \big]   +  g_{\text{res}} (m;\omega,r,e) \Big] \bigg] ,
\end{split}
\end{align}
where \( \mathcal{Q}(m;\omega, r,e,\bx) \) is a preconditioning operator. (The derivation of \( \mathcal{Q} \) will follow.) 
This linearized minimization subproblem seeks a perturbation \( \delta m \) such that the corresponding perturbation in the Green's function matches the negative residual. (For brevity, the dependence on $m$ is omitted hereafter unless otherwise stated.) Correspondingy, the perturbed Green’s function \( \delta g(\omega,r,e) [\delta m] \) is expressed as:
\begin{align} \label{eq:perturb1}
 \delta g(\omega,r,e) [\delta m]  = \int  d \bx' \,  \frac{\partial g}{\partial m} (\omega,r,e, \bx')  \delta m(\bx'),
\end{align}
where \( \frac{\partial g}{\partial m} (\omega,r,e,\bx')\) is the Fréchet derivative of the Green's function with respect to the squared slowness at \( \bx' \). This derivative satisfies:
\begin{align} \label{eq:derivative}
 \frac{\partial g(\omega,r,e)}{ \partial m} (\bx')=  g(\omega,r,\bx') \, \Upsilon_m(\omega,\bx')  \, g(\omega, \bx', e),  
\end{align}
where \( \Upsilon_m(\omega,\bx') \) denotes the scattering potential,
\begin{align} \label{eq:scattering-potential}
\Upsilon_m(\omega,\bx')
=
\Upsilon_c(\omega,\bx')
\frac{d c}{d m}(\bx')
=
\omega c(\bx') \tilde{k}(\bx'),
\end{align}
where, from Section~4.1 of \cite{Javaherian2},
\begin{align}
\Upsilon_c
=
-\frac{2\omega}{c^2}\tilde{k},
\end{align}
with \( \tilde{k} \) satisfying Eq.~\eqref{eq:szabo_dispersion_equation} for the corresponding values of \( c \) and \( \alpha \). (For further details, see Section~4.1 of \cite{Javaherian2}.) Recall that all quantities are evaluated on the wavenumber field $k$ associated with the squared slowness $m$.

\subsection{Preconditioned gradient and Hessian}
\label{sec:weighting}

The linearized problem is equivalent to solving:
\begin{align} \label{eq:linearised-linearised-subproblem}
     \nabla  \mathcal{F}(\bx)  + \int d \bx'  \, H (\bx, \bx') \, \delta m(\bx')  = 0,
\end{align}
where \( \nabla \mathcal{F} \) is the preconditioned functional gradient:
\begin{align}  
\label{eq:functional-gradient}
\begin{split}
    \nabla \mathcal{F}\big( \bx \big) &=  \sum_{e,r}  \int  d \omega \,   \left[ \frac{ \partial g}{ \partial m } (\omega,r,e, \bx)\right]^* \mathcal{Q} (\omega,r,e,\bx) \,  g_{\text{res}} (\omega,r,e).
\end{split}
\end{align}
The Hessian matrix's action, weighted similarly, is:
\begin{align}  \label{eq:hessian-action}
\int d\bx' \,  H (\bx, \bx' )  \delta m(\bx')  = \sum_{e,r}  \int  d \omega \,  \left[  \frac{\partial g} { \partial m } (\omega,r,e,\bx)\right]^* \mathcal{Q} (\omega,r,e,\bx) \, \delta g (\omega,r,e) \big[ \delta m \big]
\end{align}
Using Eqs.~\eqref{eq:perturb1} and \eqref{eq:derivative}, an explicit expression of the weighted Hessian is derived as:
\begin{align}
\label{eq:hessian}
H (\bx, \bx') = \sum_{e,r} \int d \omega \, \mathcal{Q}{(\omega, r, e, \bx)}  D (\omega, r, e, \bx, \bx')  e^{-\iu \Phi(\omega, r, e, \bx, \bx')} ,
\end{align}
where  \(D\) and \(\Phi\) are defined as:
\begin{align}
\begin{split}
D (\omega, r, e, \bx, \bx') &= \Big[ A(\omega, r, \bx) \Upsilon_m(\omega, \bx) A(\omega, \bx, e) \Big]^*
\Big[ A(\omega, r, \bx') \Upsilon_m(\omega, \bx') A(\omega, \bx', e) \Big], \\
\Phi (\omega, r, e, \bx, \bx') &= \phi(\omega, r, \bx) + \phi(\omega, \bx, e) -
\big[ \phi (\omega, r, \bx') + \phi (\omega, \bx', e) \big].
\end{split}
\end{align}

\subsection{Identity transformation of the Hessian}  \label{sec:approx-diag-hessian}
The transformation of the Hessian operator into the identity operator relies on two approximations applied to Eq.~\eqref{eq:hessian}, as proposed in \cite{Thierry, Jin, Lambare1, Lambare2}.
\rev{For every pair of points $\bx,\bx' \in \Omega$, we adopt the following approximations:}
\begin{align}
\label{eq:rayborn-approx1}
D(\omega, r, e, \bx, \bx') \approx  D(\omega, r, e, \bx', \bx'),
\end{align}
and
\begin{align}
\label{eq:rayborn-approx2}
\begin{split}
\Phi(\omega, r, e, \bx, \bx') &\approx \nabla
\left( \phi(\omega, r, \bx') + \phi (\omega, \bx', e) \right) \cdot (\bx - \bx') \\
&= \btk (\omega, r, e, \bx') \cdot (\bx - \bx'),
\end{split}
\end{align}
where \(\btk(\omega, r, e, \bx')\) represents the gradient of a two-way isochron—a curve in the scattering point space \(\bx'\) where the sum of the accumulated phase from the emitter position \(e\) to the scatterer \(\bx'\) and from \(\bx'\) to the receiver \(r\) is constant \cite{Jin, Lambare1}. {\rev{Appendix C discusses the conditions under which these two approximate formulas are valid.}}

Thus, \(\btk(\omega, r, e, \bx')\) is a vector passing through \(\bx'\) and normal to the two-way isochron, given by:
\[
\btk(\omega, r, e, \bx') = \bk{(\omega, \bx', e)} + \bk{(\omega, r, \bx')} = \bk{(\omega, \bx', e)} - \bk{(\omega, \bx', r)},
\]
with \(\bk{(\omega, \bx', e)} = \nabla\phi(\omega, \bx', e)\) and \(\bk{(\omega, \bx', r)} = \nabla \phi(\omega, \bx', r)\) representing the wavevectors of rays emanating from \(e\) and \(r\), respectively, at \(\bx'\). 

Let us now introduce polar coordinates for the vectors used in this study. In polar coordinates, the shift vector \( \bx_{\fd} := \bx - \bx' \) is represented by the pair \( (x_{\fd}, \phi) \), where \( x_{\fd} = |\bx - \bx'| \). The wavevector \( \bk \) is represented by the pair \( (k, \gamma) \), and the two-way wavevector \( \btk \) is represented by \( (|\btk|, \zeta) \). These satisfy
\begin{align} \label{eq:polar-coordinates-1}
\bx_{\fd}
=
x_{\fd}
\begin{bmatrix}
\cos(\phi) \\
\sin(\phi)
\end{bmatrix}
\hspace{0.5cm}
\bk
=
k
\begin{bmatrix}
\cos(\gamma)\\
\sin(\gamma)
\end{bmatrix}
\hspace{0.5cm}
\btk
=
|\btk|
\begin{bmatrix}
\cos(\zeta)\\
\sin(\zeta)
\end{bmatrix}
\end{align}

Now, by applying the approximations \eqref{eq:rayborn-approx1} and \eqref{eq:rayborn-approx2} to \eqref{eq:hessian}, the Hessian operator simplifies to:
\begin{align}
\label{eq:hessian2}
\begin{split}
     H (\bx,\bx') &= \sum_{e,r} \int d \omega \, \mathcal{Q}(\omega,r,e, \bx) \, e^{-\iu  \btk (\omega,r,e,\bx') \cdot (\bx-\bx') } \, D (\omega,r,e,\bx',\bx')    \\
     &= \int d \omega \int de  \int dr  \, \frac{1}{\Delta e \Delta r} \mathcal{Q}(\omega,r,e, \bx) \, e^{-\iu  \btk(\omega,r,e,\bx') \cdot (\bx-\bx') } \,
    D (\omega,r,e,\bx',\bx'). 
\end{split}
\end{align}
Assuming that rays do not exhibit caustics or other singularities, the integration variables are transformed using the one-to-one mapping \((\omega,r,e) \rightarrow ( |\btk |, \zeta,  \theta )\). Here, the scattering angle \( \theta(\omega, r, e, \bx') \), defined at an arbitrary point \( \bx' \) for the emitter–receiver pair \( (e, r) \), is given by:
\begin{align}
    \theta (\omega,r,e,\bx') = \left[\gamma(\omega, \bx', r) + \pi \right] - \gamma(\omega, \bx', e),
\end{align}
where \(\gamma(\omega, \bx', e)\) (resp. \(\gamma(\omega, \bx', r)\)) is the angle of the wavevector for a ray originating from \(e\) (resp. \(r\)). The additional $\pi$ accounts for the reversal in direction, as the scattered wave propagates from $\bx'$ to $r$. Figure \ref{figure-1} illustrates the ray parameterization at the scattering point $\bx'$, with the dependence of parameters on $\omega$ omitted for simplicity.

\begin{figure}[h]
\centering

\begin{tikzpicture}
    \draw[thick] (0,0) circle (3cm);
    \filldraw[blue!10] (0.1,0.2) ellipse (1.75cm and 2.25cm);
    \filldraw[blue] ({3*cos(135)},{3*sin(135)}) circle (1pt) node[above left] {\( e \)}; 
    \filldraw[red] ({3*cos(45)},{3*sin(45)}) circle (1pt) node[above right] {\( r \)};  
    \filldraw[black] (0.5,0.5) circle (1pt) node[left] {\( \boldsymbol{x}' \)};
    
    \draw[thin, ->, black] 
        ({3*cos(135)},{3*sin(135)}) 
        .. controls (-1,1.2) .. 
        (0.5,0.5);

    \draw[thin, ->, black] 
        (0.5,0.5)
        .. controls (1,0.8) .. 
        ({3*cos(45)},{3*sin(45)}) ;

    \draw[dotted, ->, green] (0.5,0.5) -- (1.3,0.9)  node[above right] {\(  \bk(r, \bx') \)};
     \draw[dotted, ->, red] (0.5,0.5) -- (-0.3,0.1)  node[below left] {\(  \bk(\bx', r) \)};
     
    \draw[dotted, ->, blue] (0.5,0.5) -- (1.3, 0.1) node[below right] {\( \bk(\bx', e) \)};

    \draw[dashed, ->, black] (0.5,0.5) -- (2.1, 0.5)  node[right] {\( \btk(r,e, \bx')  \)};

    \draw[thick, <->] (0.7,0.37) arc[start angle=-45,end angle=45,radius=0.2];
    \node at (0.95,0.55) {\( \theta \)};

\end{tikzpicture}
\caption{Ray parameterization at the scattering point $\bx'$. The wavevectors $\bk(\bx', e)$ (blue) and $\bk(\bx', r)$ (red) correspond to rays emanating from the emitter $e$ and receiver $r$, respectively, at $\bx'$. These wavevectors are expressed in polar coordinates as $\big(k(\bx'), \gamma(\bx', e)\big)$ and $\big(k(\bx'), \gamma(\bx', r)\big)$. The two-way wavevector $\btk(r,e, \bx')$ passes through $\bx'$ and represents the gradient of a two-way isochron, dedfined as $\btk(r,e, \bx') = \bk(\bx', e) + \bk(r, \bx')$. Its polar representation is given by $\big(|\btk(r,e, \bx')|, \zeta(r,e, \bx') \big)$. The direction-reversed wavevector $\bk(r, \bx')$ (green) is represented in polar coordinates as $\big(k(\bx'), \gamma(\bx', r) + \pi\big)$. Additionally, $\theta(r,e, \bx')$ denotes the scattering angle. The dependence on $\omega$ is omitted for brevity.}

\label{figure-1}
\end{figure}
The pair \((|\btk|, \zeta)\) represents the polar coordinates of the two-way wavevector \(\btk (\omega,r,e,\bx')\), which is normal to the two-way isochron, associated with the accumulated phase \(\phi(\omega,r ,\bx') + \phi(\omega,\bx',e)\). Specifically:
\begin{align}  \label{eq:btk-amp-angle}
|\btk | &= 2 k   \left| \cos\left(\frac{\theta}{2}\right) \right| ,  \nonumber \\
\zeta(\omega,r,e, \bx') &= \frac{1}{2} \Big[ \left[ \gamma(\omega, \bx', r) + \pi \right] + \gamma(\omega, \bx', e) \Big].
\end{align}

Applying the change of variables \((\omega,r,e) \rightarrow (|\btk|,  \theta, \zeta)\) in \eqref{eq:hessian2}, we obtain:
\begin{align}
\label{eq:hessian3}
\begin{split}
     H (\bx,\bx') = \int_0^{2\pi} d\theta  \int_0^ {+\infty} d |\bar{\mathbf{k}}|   \int_0^{2\pi}  d \zeta &
      \Bigg[ \frac{1}{\Delta e \Delta r}  \mathcal{Q}(\omega, r,e, \bx) \, e^{-\iu  \bar{\mathbf{\bk}}(\omega,r,e,\bx') \cdot (\bx-\bx') }  \\
     & D (\omega,r,e,\bx',\bx') 
     \left| \frac{\partial (\omega,r,e)}{\partial (|\bar{\mathbf{k}}|,\theta, \zeta)} \right| \Bigg].
\end{split}
\end{align}
Following \cite{Thierry,Jin,Lambare1,Lambare2}, we choose \(\mathcal{Q}\) to transform \(H (\bx,\bx')\) into an identity operator, allowing inversion in a single step. Specifically, with:
\begin{align}
\label{eq:preconditioner}
    \mathcal{Q}(\omega,r,e, \bx) = 
    \frac{\Delta e \Delta r }{(2 \pi)^2}
    \frac{ |\btk(\omega,r,e,\bx) | }{ D (\omega,r,e,\bx, \bx) } \left| \frac{\partial \left(|\btk |,\theta, \zeta \right)} {\partial (\omega,r,e)} \right| ,
\end{align}
the Hessian operator in \eqref{eq:hessian3} simplifies to:
\begin{align} 
\label{eq:hessian-diagonalised}
H(\bx,\bx') = \frac{1}{(2 \pi)^2}   \int_0^{2\pi} d\theta  \int_0^ {+\infty} d |\bar{\mathbf{k}}| \,   | \btk |   \int_0^{2\pi}  d \zeta  \,   e^{-\iu  \btk(\bx') \cdot (\bx-\bx') }  \approx 2 \pi \delta (\bx-\bx').
\end{align}

\subsection{The search direction formula}

In Eq. \eqref{eq:preconditioner}, the Jacobian determinant simplifies to:
\begin{align}
\left| \frac{\partial (|\btk|,\theta, \zeta )} {\partial (\omega,r,e)} \right| =  \left| \frac{\partial \gamma_e}{\partial e} \right| \left| \frac{\partial \gamma_r}{\partial r} \right| \left| \frac{\partial (|\btk|)} {\partial \omega} \right| ,
\end{align}
where \(  \gamma_e \) and \( \gamma_r \) denote the angle of rays initialized on emitter position \(e\) and receiver position \(r\), respectively. Also,
\begin{align}
 \frac{\partial (|\btk(\bx)|)} {\partial \omega}  = 2  \left| \cos\left(\frac{\theta(\bx)}{2}\right) \right| \left[ \frac{1}{c(\bx)} +  y \tan\left(\frac{\pi y}{2}\right) \omega^{y-1} \alpha_0(\bx) \right].
\end{align}

By substituting the weighting function in \eqref{eq:preconditioner} into the functional gradient \eqref{eq:functional-gradient}, and then inserting the result into the linearized equation \eqref{eq:linearised-linearised-subproblem} alongside \eqref{eq:hessian-diagonalised}, a full-frequency search direction is derived as:
\begin{align}  \label{eq:sound_speed_update}
    \delta m (\bx') \approx \Re \left\{ \sum_{e,r,\omega} 
    \Lambda(\omega,r,e,\bx')
  \    g_{\text{res}} (\omega, r, e) \right\},
\end{align}

\noindent
where $\Re \left\{ \cdot \right\}$ denotes the real part, and $\Lambda$is the backprojection operator at point $\bx'$, acting on $g_{\text{res}} (\omega, r, e)$. We employ a recursive scheme to compute this search direction. Specifically, it is evaluated sequentially over consecutive frequency intervals, each comprising a fixed number of discrete frequencies (two in the present implementation), proceeding from low to high frequencies, yielding:
\begin{align}
\label{eq:sound_speed_update2}
\begin{split}
    \Lambda(\omega^{(n)},r,e,\bx')   = & \frac{\Delta e \Delta r  \Delta \omega } {(2 \pi)^3}  \
    \bigg| \frac{\partial \gamma(\omega^{(n)}, \bx', e)}{\partial e}  \bigg| \bigg| \frac{\partial \gamma(\omega^{(n)}, \bx', r)}{\partial r} \bigg| \bigg| \frac{\partial \big( |\btk(\omega^{(n)}, r, e, \bx')| \big) } {\partial \omega } \bigg| \ \times\\
    &  \big| \btk (\omega^{(n)},r,e, \bx') \big|  \big[ \Upsilon_m(\omega^{(n)},\bx')  \big]^{-1}
     g_{\dagger}(\omega^{(n)},\bx',e)  \ g_{\dagger} (\omega^{(n)},\bx',r),
     \end{split}
\end{align}
where 
\begin{align}  \label{eq:greens-time-reversed}
g_{\dagger}(\omega^{(n)},\bx,\bx') = \big[ A(\omega^{(n)},\bx,\bx') \big]^{-1}  \exp{\Big(-\iu \big[\phi (\omega^{(n)},\bx,\bx') + \pi/4 \big] \Big) } 
 .
\end{align}
Here, \(n\) denotes the iteration index associated with the frequency intervals, progressing from low to high frequencies. The term $g_{\dagger}$ is referred to as the \textit{reciprocal} Green's function, where the dagger symbol ($\dagger$) signifies inverting both phase and amplitude. Specifically, for $g_{\dagger}(\omega^{(n)},\bx',r)   $, which acts on $ g_{\text{res}} (\omega^{(n)}, r, e) $, the reciprocity of the Green's function is utilized \cite{Pierce}.

The backprojection operator in Eq. \eqref{eq:sound_speed_update2} compensates for amplitude decay and phase shifts in the scattered waves by reversing both the incident and scattered Green's functions. The factors $|\partial \gamma(\omega^{(n)}, \bx', e) / \partial e| \Delta e $ and $|\partial \gamma(\omega^{(n)}, \bx', r)/ \partial r| \Delta r $ account for the angular nonuniformity and sparsity with which emitters and receivers are observed by a scattering point $\bx'$ within the medium, respectively. The Born approximation, which inherently relies on a low-frequency assumption, considers that waves originating from emitter $e$ and measured at receiver $r$ may be scattered by any point in the medium. In contrast, the backprojection operator derived in Eq. \eqref{eq:sound_speed_update2} assumes that high-frequency components dominate the scattered waves. The correction factor $|\btk| \, \big| \partial (|\btk|) / \partial \omega \big|$, which depends on the scattering angle $\theta$, prioritizes contributions from scatterers located near the ray linking the emitter--receiver pair $(e,r)$ over those farther away. 

At the end of each iteration \(n\), the squared slowness is updated according to
\[
m^{(n+1)} = m^{(n)} + \tau\,\delta m^{(n)},
\]
where \(\tau\) is a fixed step length. The updated sound speed \(c^{(n+1)}\) is then computed from \(m^{(n+1)}\).

\section{Ray tracing}  
\label{sec:ray-tracing}  

This section details the numerical implementation of ray-based quantitative ultrasound tomography (QUT) for heterogeneous and absorbing media. Specifically, it describes how ray theory, derived from high-frequency approximations and governed by the \textit{Eikonal} equation \eqref{eq:eikonal} and \textit{Transport} equation \eqref{eq:transport}, is used to compute the approximate Green’s function and its reciprocal counterpart.  

\subsection{Hamiltonian formulation of the Eikonal equation}  
The Eikonal equation can be expressed as a Hamiltonian system \cite{Virieux}:  
\begin{align} \label{eq:hamiltonian}  
    H(\bx, \bk)= \frac{1}{2} k^{-1}\big[ \bk \cdot \bk - k^2 \big],  
\end{align}  
where \( k \), the real part of \(\tilde{k}\), is dependent on the sound speed \(c\) via Eq. \eqref{eq:szabo_dispersion_equation}. Along a ray, \( H = 0 \), and the system satisfies the canonical equations:  
\begin{align}  \label{eq:canonical}  
\begin{split}  
\boldsymbol{\dot{x}} & = \nabla_{\bk} H, \\  
\boldsymbol{\dot{k}} &= -\nabla_{\bx} H.  
\end{split}  
\end{align}  

Correspondingly, the ray’s position \(\bx\) and wavevector \(\bk\) satisfy:  
\begin{align}  
\label{eq:ray-equation-reference}  
   \dot{\bx} = k^{-1} \bk,  \ \ \ \ \ \  
   \dot{\bk}  = \frac{1}{2} \nabla k \big[ k^{-2} \bk \cdot \bk+ 1 \big],  
\end{align}  
where \(k^{-2} |\bk|^2 = 1\) for \(H=0\).  

For each emitter-receiver pair \((e, r)\), the ray is described by the canonical vector \(\by(s) = \big[\bx(s), \bk(s)\big]^T\), where \(s\) denotes the arc length along the ray. The ray's initial position \(\bx(s_0) \) coincides with the emitter position \(e\) on the ring, while the initial wavevector \(\bk(s_0)\) is determined through \textit{ray linking} \cite{Javaherian1}, ensuring the ray intersects the receiver position \(r\). Further details on ray linking are provided later in this section.

\subsection{Ray Jacobian and paraxial equations} 
The amplitude factor \(A\) in Eq. \eqref{eq:2D_greens} includes contributions from geometrical spreading, denoted \(A_{geom}\), determined by the \textit{Transport} equation \eqref{eq:transport}. This equation describes changes in the ray tube area relative to a reference ray, relying on the \textit{ray Jacobian} \cite{Cerveny2001}. Previously, the ray Jacobian was computed using finite differences of auxiliary rays near the reference ray \cite{Javaherian2,Rullan}. Here, \textit{paraxial ray tracing}, also known as \textit{dynamic ray tracing}, is employed instead.  

Paraxial ray tracing involves solving an additional system of linear ordinary differential equations alongside the reference ray, eliminating the need for independent auxiliary rays. For each emitter-receiver pair, the paraxial ray perturbation vector \(\delta \by(s) = \big[\delta \bx(s), \delta \bk(s)\big]^T\) satisfies the system:  
\begin{align}  
\label{eq:paraxial-system1}  
\delta  \dot{\by} = \mathbf{B} \  \delta \by,  
\end{align}  
where  
\begin{align} \label{eq:paraxial-system}  
    \mathbf{B} = \begin{bmatrix}  
    \nabla_{\bx} \nabla_{\bk} H  & \ \nabla_{\bk} \nabla_{\bk} H\\  
    -\nabla_{\bx} \nabla_{\bx} H & \  -\nabla_{\bk} \nabla_{\bx} H  
    \end{bmatrix}.  
\end{align}

Correspondingly, the paraxial ray perturbation equations are given by
\begin{align}  
\label{eq:ray-equation-paraxial}  
\begin{split}  
\delta \dot{\bx} &= \big[\frac{\partial}{\partial \bx} \dot{\bx}\big] \delta \bx + \big[\frac{\partial }{\partial \bk}\dot{\bx} \big] \delta \bk, \\  
\delta \dot{\bk} &= \big[\frac{\partial }{\partial \bx}\dot{\bk}\big] \delta \bx + \big[\frac{\partial }{\partial \bk}\dot{\bk} \big] \delta \bk,  
\end{split}  
\end{align}  
where  
\begin{align}  
\begin{split}
    \frac{\partial}{\partial \bx}  \dot{\bx}= -k^{-2} \bk (\nabla k)^T, \ \ &\frac{\partial}{\partial \bk} \dot{\bx} = k^{-1} \mathbf{I}_{2\times2}, \\  
    \frac{\partial }{\partial \bx}\dot{\bk} = (\nabla^2 k)^T - k^{-1} \nabla k (\nabla k)^T, \ \ &\frac{\partial }{\partial \bk}\dot{\bk} = k^{-2} \nabla k \bk^T, 
\end{split}
\end{align}  
where \( \mathbf{I}_{2\times2} \) is the \(2 \times 2\) identity matrix.

\subsection{Numerical implementation}

Here, the paraxial ray tracing system defined by \eqref{eq:ray-equation-reference} and \eqref{eq:ray-equation-paraxial} is numerically implemented using Heun's method, a second-order variant of the Runge-Kutta (RK) scheme that balances computational efficiency and accuracy \cite{Butcher,Kreyszig}. An outline of Heun’s approach for solving the paraxial ray tracing equations is provided in Algorithm \ref{alg:ray_tracing}. In this implementation, normalization steps are applied to the wavevector $\bk$ as a safeguard to enhance the stability of the algorithm in cases where wavenumber maps exhibit sharp gradients. However, numerical experiments show that for soft tissues, such as breast tissue, where refractive index variations are within the range of $0.9$ to $1.1$, the normalization steps have negligible effects on ray trajectories and can thus be omitted if desired.
\noindent
The system \eqref{eq:paraxial-system1} represents a ray if the perturbation vector $\delta \by$ satisfies 
\begin{align} \label{eq:paraxial-condition}
    \delta H = \nabla_{\bk} H \cdot \delta \bk + \nabla_{\bx} H \cdot \delta \bx = 0.
\end{align}
Additionally, as per \eqref{eq:ray-equation-reference}, $\delta H$ remains constant along any solution of the paraxial system. Consequently, it is sufficient to ensure the condition \eqref{eq:paraxial-condition} is met at the initial point \cite{Virieux}.

\subsection{Initial conditions for ray tracing}
The initial wavevector is determined by the frequency $\omega$, the sound speed in water $c_0$, and the initial unit direction vector. Following \cite{Javaherian2}, the rays are initialized at the emitter position $e$ and connected to the receiver position $r$ using ray linking \cite{Javaherian1}. 

Ray linking, a type of shooting method, identifies a ray trajectory that provides a stationary path within a family of neighboring paths between $e$ and $r$. This is achieved by imposing a boundary condition on the ray's path such that the ray initialized at the emitter position $e$ intersects the receiver position $r$ after traversing the medium. For each frequency set in the QUT inverse problem, ray linking is performed for each emitter-receiver pair $(e,r)$ separately. This involves iteratively adjusting the initial unit direction of the ray originating at $e$ using an optimization algorithm to ensure that the ray intercepts the detection surface (ring) at the receiver position $r$ within a specified tolerance \cite{Cerveny2001,Javaherian1}. 

Once the initial wavevector $\bk(s_0)$ is determined, the condition \eqref{eq:paraxial-condition}, along with the enforcement of $\delta \bx(s_0) = 0$, requires the initial perturbation \(\delta \bk(s_0)\) to satisfy 
\begin{align}
    \bk(s_0) \cdot \delta \bk(s_0) = 0.
\end{align}

\begin{algorithm}
    \caption{Paraxial ray tracing for the linked ray initialized on $e$ and intercepted by $r$  using Heun's method}
    \label{alg:ray_tracing}
    \begin{algorithmic}[1]
        \State \textbf{input:} $(e, r)$, $ k(\bx)$
            \Comment{Input emitter and receiver positions and wavenumber map}
            \State \textbf{initialize:} $\bx(s_0) , \bk(s_0)$ \Comment{Set initial position, and compute the initial wavevector through ray linking }\\
            $\delta \bx=0, \ \delta \bk$ satisfying $\delta \bk \cdot \bk = 0$,\Comment{Set initial conditions: paraxial ray (amplitude) }
        \While { $\bx(s) \ \text{is inside} \ \Omega$ } \\
         \State $ --- $
         \Comment{\textbf{Update reference ray}}
        \State $\bk \leftarrow k \   \bk / \mid \bk \mid $
           \Comment{Normalize the ray direction}
            \State $q_{\bx} = \bk / k$
            \Comment{Compute the update variables   } 
            \State $q_{\bk} = \nabla k(\bx)$  
           \State $\bk' \leftarrow \bk + \Delta s  \ q_{\bk} $
      \Comment{Update the auxiliary ray direction}
        \State $k' \leftarrow k (\bx + \Delta s \  q_{\bx})$
        \Comment{Update the auxiliary wavenumber}
               \State $\bk' \leftarrow   k'  \bk' / \mid \bk' \mid $
        \Comment{Normalize the auxiliary ray direction}  
             \State $q_{\bx}' = \bk' / k' $
              \Comment{Compute the auxiliary update variables }
             \State $q_{\bk}' = \nabla k(\bx + \Delta s  \ q_{\bx})$   
             \State $\bx \leftarrow \bx + \Delta s \big[  q_{\bx} + q_{\bx}' 
 \big]/ | q_{\bx} + q_{\bx}'   |   $
                \Comment{Update the ray position}
                 \State $\bk \leftarrow \bk + \Delta s \big[  q_{\bk} + q_{\bk}'\big] /2 $
                  \Comment{Update the ray direction}
                  \\
         \State $ ---$
         \Comment{\textbf{Update paraxial ray: geometrical amplitude}}  
             \State $q_{\delta \bx} = \big[- k^{-2} \bk \nabla k^T \big] \delta \bx + \big[ k^{-1} \big] \delta \bk$
            \Comment{Compute the update variables  }  
            \State $q_{\delta \bk} = \big[ (\nabla^2 k)^T - k^{-1} \nabla k \nabla k^T  \big] \delta \bx + \big[ k^{-2} \nabla k \bk^T \big] \delta \bk$ 
           \State $\delta \bk' \leftarrow \delta \bk+ \Delta s  \ q_{\delta \bk} $
      \Comment{Update the auxiliary ray direction perturbation}
        \State $\delta \bx' \leftarrow \delta \bx + \Delta s q_{\delta \bx}$
              \Comment{Update the auxiliary ray position perturbation}
             \State $q_{\delta \bx}' = \big[-  k'^{-2} \bk' \nabla k'^T \big] \delta \bx' + \big[  k'^{-1} \big] \delta \bk'$
              \Comment{Compute the auxiliary update variables}
             \State $q_{\delta \bk}'  = \big[ (\nabla^2 k')^T - k'^{-1} \nabla k' \nabla k'^T  \big] \delta \bx' + \big[ k'^{-2} \nabla k' \bk'^T\big] \delta \bk' $ 
             \State $\delta \bx \leftarrow \delta \bx + \Delta s\big[q_{\delta \bx} + q_{\delta \bx}'\big] /2  $
                \Comment{Update the ray position perturbation}
             \State $\delta \bk \leftarrow \delta \bk + \Delta s \big[q_{\delta \bk} + q_{\delta \bk}'\big]/2$
                \Comment{Update the ray direction perturbation}\\
        \EndWhile
    \end{algorithmic}
\end{algorithm}

\subsection{Interpolation}
\textit{Grid-to-ray interpolation:} The squared slowness map is defined on grid points. To implement Algorithm \ref{alg:ray_tracing}, this map must be interpolated onto the sampled points along the rays. In this study, grid-to-ray interpolation is performed using B-spline interpolation, which provides smooth and continuous values for both $\nabla k$ and $\nabla^2 k$ at any arbitrary (off-grid) point. For further details, refer to \cite{Javaherian2}.  

\textit{Ray-to-grid interpolation:} The Green's function parameters at the grid points are computed by interpolating the corresponding quantities from the linked rays onto the computational grid via trilinear interpolation, as described in \cite{Javaherian2}.

\section{Ray coordinates}
\label{sec:Discretised-approximation}

This section describes the approximation and discretization of the Green's function in the residual defined by Eq.~\eqref{eq:residual} and its reciprocal counterparts in the backprojection operator \eqref{eq:sound_speed_update2}. These quantities are evaluated along linked rays to compute the update directions associated with successive frequency sets. The derivations will focus on $g(\omega, r, e)$ and $g_{\dagger}(\omega, \bx, e)$; however, the formulae for $g_{\dagger}(\omega, \bx, e)$ can be similarly applied to $g_{\dagger}(\omega, \bx, r)$ by interchanging $e$ and $r$. In general, for the 2D case, rays are parameterized using two coordinates: one specifying the initial direction (angle) of the ray and another representing a monotonic parameter along the ray \cite{Cerveny2001}. Here, the ray parameters are chosen as the initial angle, $\gamma^0$, and the arc length, $s$.

\begin{definition} \label{def:ray-parameterisation}
The trajectory of a ray linking an emission point $e$ to a reception point $r$ is discretized by sampled arc lengths $s_i$, where $i \in \{0, \dots, M_{(e,r)}\}$. The sampled points begin at $s_0$, where $\bx(s_0)$ coincides with the emitter position $e$, and end at $s_{M_{(e,r)}}$, where $\bx(s_{M_{(e,r)}})$ coincides with the receiver position $r$. The sampled arc lengths $s_i$ are defined as:
\begin{align}  
\label{eq:ray_points_sm}
s_i = 
\begin{cases}
i \Delta s,                                 & \text{for } i \in \{0, \dots, M_{(e,r)} - 1\}, \\
(i-1) \Delta s + \Delta s',                & \text{for } i = M_{(e,r)},
\end{cases}
\end{align}
where the second case ensures that the last point of the ray coincides with the reception point $r$, and $\Delta s' = s_{M_{(e,r)}} - s_{M_{(e,r)}-1}$, satisfying $\Delta s' \leq \Delta s$ \cite{Javaherian1}.
\end{definition}

\begin{definition} \label{def:raylinked-coordinates}
For each emitter position $e$, the Green's function parameters are approximated along a set of linked rays defined by $f_{(k,r,e)} = 0$. Each ray is parameterized in space as $\bx(s_i, \gamma_{(r,e)}^0)$, representing the position at arc length $s_i$ along the ray linking the emitter $e$ to the receiver $r$, with the ray’s initial polar direction (angle) denoted by $\gamma_{(r,e)}^0$.
\end{definition}

\noindent
Using the coordinates defined in Definition \ref{def:raylinked-coordinates}, the Green's function $g(\omega, \bx, e)$ and its reciprocal variant $g_{\dagger}(\omega, \bx, e)$ are discretized at points sampled along the rays linking emitter $e$ to all receivers $r$ as:
\begin{align}  \label{eq:greens-heterogeneous-sampled1}
g \big( \omega, \bx(s_i, \gamma_{(r,e)}^0), e \big) 
\approx A \big( \omega, \bx(s_i, \gamma_{(r,e)}^0), e \big) 
\exp \Big( \iu \big[ \phi \big( \omega, \bx(s_i,  \gamma_{(r,e)}^0), e \big) + \pi / 4 \big] \Big),
\end{align}
and
\begin{align}  \label{eq:greens-heterogeneous-sampled2}
g_{\dagger} \big( \omega, \bx(s_i, \gamma_{(r,e)}^0 ), e \big) 
\approx \Big[ A \big( \omega, \bx(s_i,  \gamma_{(r,e)}^0), e \big) \Big]^{-1}
\exp \Big( -\iu \big[ \phi \big( \omega, \bx(s_i,  \gamma_{(r,e)}^0), e \big) + \pi / 4 \big] \Big),
\end{align}
wher $\bx(s_0,  \gamma_{(r,e)}^0)$ coincides \(e\) for all \( r\) (cf. Eqs.~\eqref{eq:2D_greens} and \eqref{eq:greens-time-reversed}).

\noindent
In Eqs.~\eqref{eq:greens-heterogeneous-sampled1} and \eqref{eq:greens-heterogeneous-sampled2}, the accumulated phase $\phi$ is computed as:
\begin{align}  \label{eq:phase-discretised}
\phi \big( \omega, \bx(s_i, \gamma_{(r,e)}^0), e \big) 
= \int_{s_0}^{s_i} k \big( \bx(s, \gamma_{(r,e)}^0) \big) \, ds 
- \frac{\pi}{2} \mathcal{K}(s_i, \gamma_{(r,e)}^0),
\end{align}
where $\mathcal{K}(s_i, \gamma_{(r,e)}^0)$ is the cumulative count of caustics (points where the ray Jacobian changes sign), each contributing a $\pi/2$ phase shift \cite{Cerveny2001}.

\noindent
The amplitude factor $A$ has contributions from absorption, $A_{\text{abs}}$, discretized as:
\begin{align}  \label{eq:amplitude-abs-discretised}
A_{\text{abs}} \big( \omega, \bx(s_i,\gamma_{(r,e)}^0), e \big) 
= \exp \left( -\int_{s_0}^{s_i} \alpha \big( \bx(s, \gamma_{(r,e)}^0 ) \big) \, ds \right),
\end{align}
and from geometrical spreading, $A_{\text{geom}}$, given by:
\begin{align}  \label{eq:amplitude-geom-discretised}
A_{\text{geom}} \big( \omega, \bx(s_i, \gamma_{(r,e)}^0), e \big) 
= \bigg[ \frac{c \big(\bx(s_i, \gamma_{(r,e)}^0) \big)}{c \big(\bx(s_1, \gamma_{(r,e)}^0) \big)} 
\frac{J(s_1, \gamma_{(r,e)}^0)}{J(s_i, \gamma_{(r,e)}^0)} \bigg]^{1/2} 
A_{\text{geom}} \big( \omega, \bx(s_1, \gamma_{(r,e)}^0), e \big),
\end{align}
where $J(s_i, \gamma_{(r,e)}^0) = \det \Xi(s_i, \gamma_{(r,e)}^0)$ is the ray's Jacobian, and $\Xi = \partial \bx / \partial \eta$ is the transformation matrix from ray coordinates $\eta = [\eta_1, \eta_2]^T$ (with $\eta_1 = \gamma_{(r,e)}^0$ and $\eta_2 = s$) to Cartesian coordinates $\bx = [\bx_1, \bx_2]^T$ \cite{Cerveny2001}. The rays' relative Jacobian is approximated by computing the relative \(\delta \bx\) at \(s_1\) with respect to the current point \(s_i\) along each ray, using the paraxial ray-tracing system introduced in Section~\ref{sec:ray-tracing}. The computation is initialized with a wavevector perturbation \(\delta \bk(s_0)\) satisfying Eq.~\eqref{eq:paraxial-condition} and a vanishing position perturbation, \(\delta \bx(s_0) = 0\). Here, $s_1$ is the arc length of the first sampled point after the initial point on the ray. This point is chosen as the reference point, at which the amplitude is evaluated analytically using Eq.~\eqref{eq:attenuation_geom_reference}.

\section{Numerical Results} \label{sec:results}

This section describes numerical experiments demonstrating the effectiveness of the proposed ray-based inversion approach for low-cost computation of a high-resolution image of the sound speed distribution inside the breast. 

\subsection{Data simulation} \label{sec:data-simulation}

\begin{figure} 
    \centering
	\subfigure[]{\includegraphics[width=0.45\textwidth]{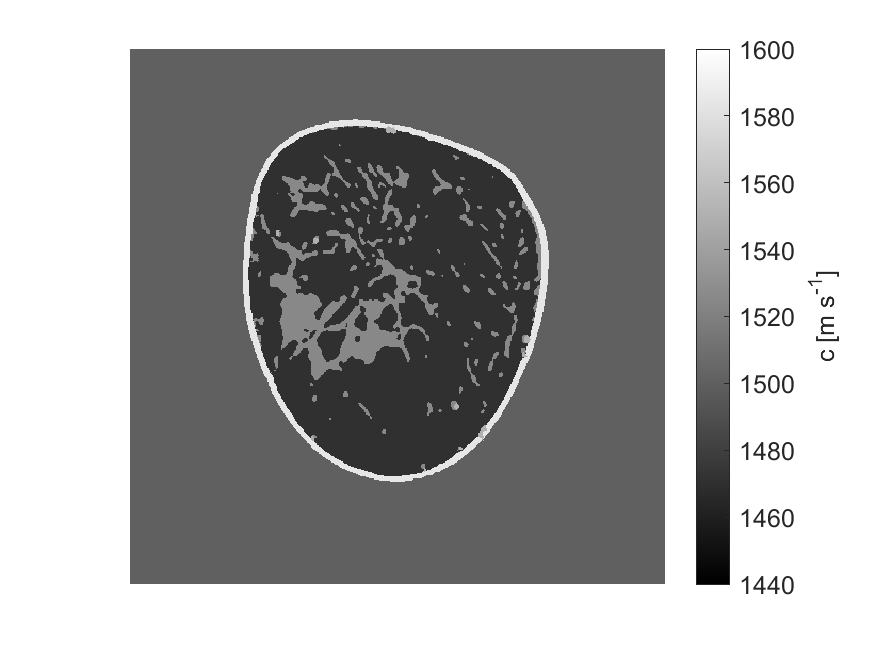}
     \label{fig:1a}  }
      \subfigure[]{\includegraphics[width=0.438\textwidth]{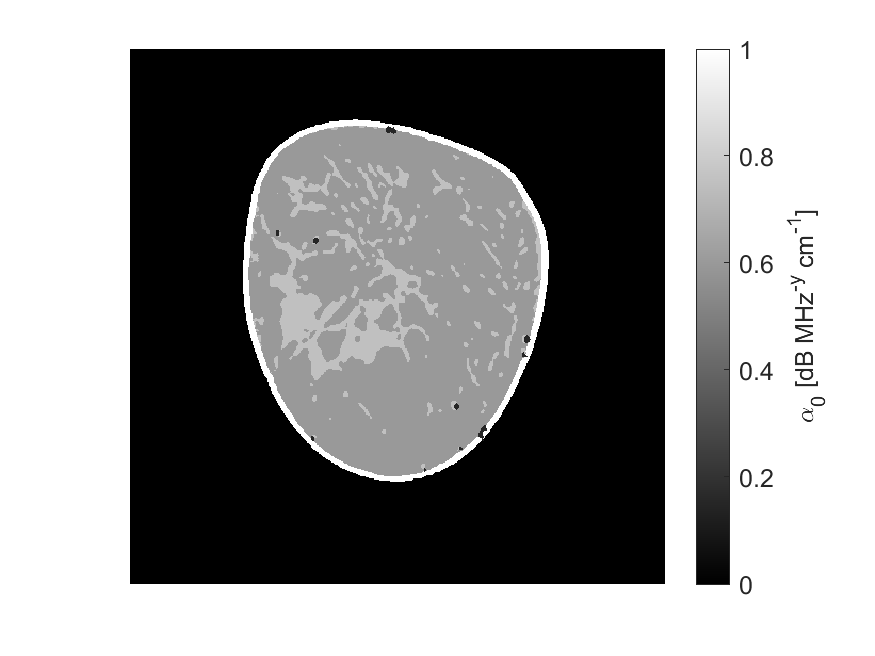}
     \label{fig:1b}   }
     	\subfigure[]{\includegraphics[width=0.45\textwidth]{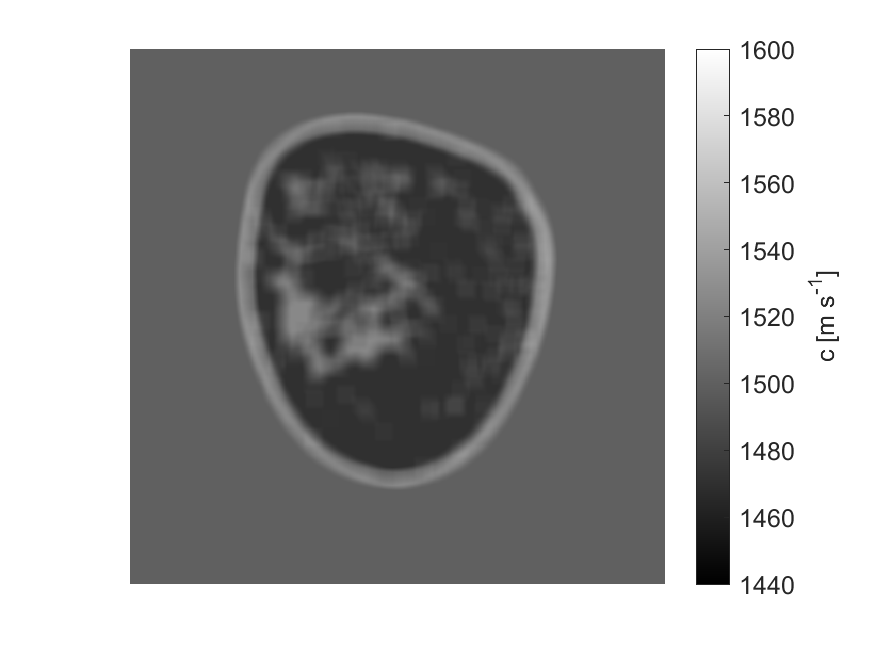}
     \label{fig:1c}  }
      \subfigure[]{\includegraphics[width=0.438\textwidth]{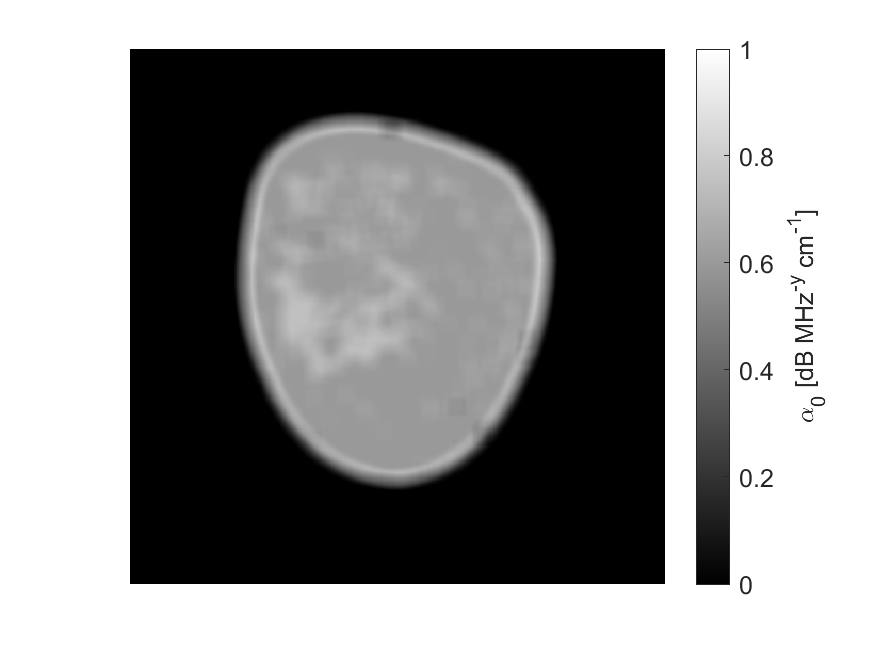}
     \label{fig:1d}   }
	\caption{Phantom used for simulation of synthetic UST data using the full-wave approach: (a) sound speed \([ \mathrm{m}  \mathrm{s}^{-1} ]\), (b) absorption coefficient \([ \mathrm{dB} \  \mathrm{MHz}^{-y} \ \mathrm{cm}^{-1} ]\), (c) smoothed sound speed \([ \mathrm{m}  \mathrm{s}^{-1} ]\), (d) smoothed absorption coefficient \([ \mathrm{dB} \ \mathrm{MHz}^{-y}  \ \mathrm{cm}^{-1} ]\). The maps are shown on a grid consisting of \(502 \times 502\) points (used for the wave simulation). The maps in (c) and (d) are smoothed by applying an averaging window of size 17 points to the original acoustic maps in (a) and (b), respectively. The original (nonsmoothed) acoustic maps were used for simulating the data for image reconstruction. The smoothed maps were used for simulating the data for benchmarking against the ray approximation to the heterogeneous Green's function, as described in Section~\ref{sec:results-greens} . The power law exponent was assumed to be \(y = 1.4\) and homogeneous.
}
\end{figure}

An imaging system consisting of 64 emitters and 256 receivers uniformly distributed along a circular ring with a radius of $R = 9.5~\mathrm{cm}$ was simulated. A horizontal slice of a 3D digital phantom, freely available from \cite{Lou}, was used to mimic the acoustic properties of the breast. The sound speed was set within the range of $1470$–$1580~\mathrm{m/s}$, the absorption coefficient $\alpha_0$ within $0$–$1~\mathrm{dB}  \ \mathrm{MHz}^{-y} \  \mathrm{cm}^{-1}$, and the power-law exponent $y$ was fixed at $1.4$. (In contrast to the full-wave solver used as the benchmark, the proposed ray-Born approach presented in this study can be straightforwardly extended to include spatially varying \( y\).) Figures \ref{fig:1a} and \ref{fig:1b} depict the sound speed and absorption coefficient maps of the breast phantom, respectively. For water, the sound speed and absorption coefficient were set to $1500~\mathrm{m/s}$ and $0$, respectively. 

The computational domain comprised a grid of $502 \times 502$ points covering the region $[-10.04, +10.00] \times [-10.04, +10.00]~\mathrm{cm}^2$, with a uniform grid spacing of $\Delta x = 0.4~\mathrm{mm}$. Based on this sound speed distribution and grid spacing, the maximum frequency supported by the grid, $f_{\text{max}}$, was determined to be $1.84~\mathrm{MHz}$.

\textit{Simulating time series data.} A time-domain \textit{k-space pseudo-spectral} method was employed to simulate the acoustic pressure time series data on the detection ring, generated by each time-varying source \(s\) \cite{Treeby,Bojarski,Mast,Tabei}. This numerical approach has been implemented in the open-source \textit{k-Wave} toolbox \cite{Treeby}, which solves a system of three coupled first-order wave equations equivalent to Szabo's second-order wave equation.\footnote{The scripts provided in \cite{Javaherian-toolbox} implement point sources in terms of \(s\), as defined in Eq.~\eqref{eq:second-order-lossy-omega}, consistent with the assumptions adopted in this study.} Szabo's wave equation is a time-domain variant of Eq.~\eqref{eq:second-order-lossy-omega}, with the complex wavenumber defined by Eq.~\eqref{eq:szabo_dispersion_equation} \cite{Kelly,Szabo,Liebler}.

The emitters and receivers were modeled as point transducers distributed along the circular detection ring. The interpolation of the pressure field between the computational grid and these transducers was performed using regularized Dirac delta distributions. Each emitter was individually driven by an excitation pulse, and the resulting acoustic pressure time series were recorded on the receivers at 6466 time points with a sampling rate of \(39.6~\mathrm{MHz}\) (\(25.25~\mathrm{ns}\) time spacing). This procedure was repeated for all emitters. 

\begin{figure} 
   \centering
	\subfigure[]{\includegraphics[width=0.45\textwidth]{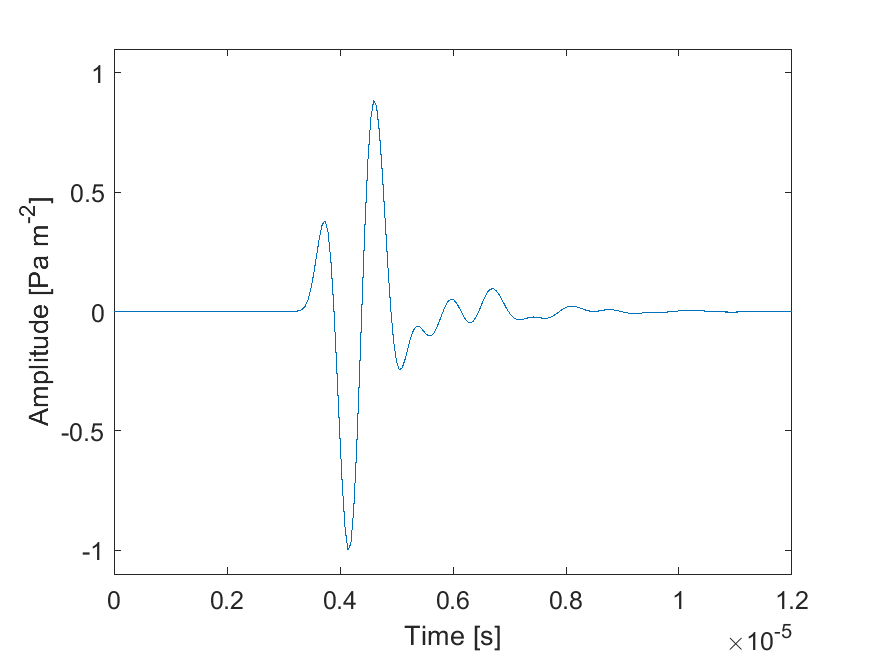}
   \label{fig:2a}   }
   \subfigure[]{\includegraphics[width=0.45\textwidth]{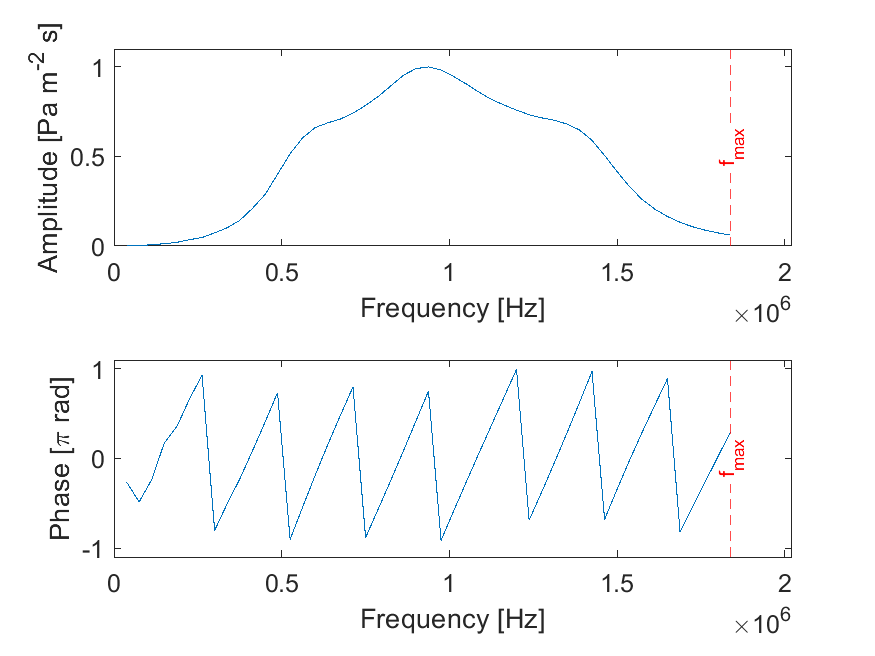}
    \label{fig:2b}  }
	\caption{Acoustic source used for all excitations (emitters): (a) time domain, (b) frequency domain: normalized amplitude and phase. \(f_{\text{max}}\) indicates the maximum frequency supported by the grid used for the wave simulations.
}\label{fig:excitation-pulse}
\end{figure}

Figure~\ref{fig:2a} presents the normalized amplitude of the acoustic source in the time domain, while Figure~\ref{fig:2b} illustrates the normalized amplitude and phase components of the acoustic source in the frequency domain. This signal was consistently used as the acoustic source for all excitations.

\subsection{Numerical validation of the ray approximation to the Green's function} \label{sec:results-greens}

This section compares the approximate Green's function solution to Szabo's wave equation with a full-wave solution computed using the k-space pseudo-spectral method \cite{Mast,Tabei,Treeby} in a heterogeneous and absorbing medium with spatially smooth variations in acoustic properties. The approximate heterogeneous Green's function was introduced in Section~\ref{sec:greens}, and its numerical approximation and discretization were described in Sections~\ref{sec:ray-tracing} and \ref{sec:Discretised-approximation}, respectively. Further details on the derivation of the Green’s function solution to Szabo’s wave equation are provided in Appendix A of the current manuscript and in \cite{Javaherian2}.

For the full-wave simulation, emitters 1 and 20 were excited by the acoustic pulse shown in Figure~\ref{fig:excitation-pulse}, and the induced pressure propagated through the breast immersed in water and was recorded on all 256 receivers. The Green's function, \(g(\omega, r, e)\), depends on the accumulated parameters and geometrical spreading along the linked rays, but does not account for scattering effects. Therefore, to enable a fair comparison with the full-wave simulation, scattering effects were reduced by applying an averaging window of 17 grid points to the sound speed and absorption coefficient maps. The smoothed maps are shown in Figures~\ref{fig:1c} and \ref{fig:1d}, respectively. The reason for selecting this particular smoothing window size is discussed below.

For the ray approximation, the Green's function was computed along the linked rays and then incorporated into the Green's formula, Eq.~\eqref{eq:general-greens-source}, to approximate the pressure time series on the receivers after exciting emitters 1~and~20. Note that the spatial integral in Eq.~\eqref{eq:general-greens-source} is omitted when the source is restricted to a point. As will be described in Section~\ref{sec:image-reconstruction}, an inverse crime is avoided during the image reconstruction process by employing two distinct computational grids for data simulation and reconstruction \cite{Wirgin}.
(Although this precaution may not be strictly necessary in the present study, since inherently different approaches are used for simulation and reconstruction.) 
Correspondingly, for the ray-based image reconstruction, a computational grid of \(204 \times 204\) points with a grid spacing of \(1~\mathrm{mm}\) was used. Consequently, the ray-based approximation of the Green's function used for comparison in this section was also computed on the same grid.

To implement the ray approximation, the smoothed sound speed and absorption coefficient maps from the full-wave simulation were interpolated onto the reconstruction grid. The interpolated wavenumber map was further smoothed using an averaging window of 7 grid points to mitigate interpolation artifacts. The ray theory presented in this study is based on a high-frequency approximation. At lower frequencies, a smoothing window must be applied to the wavenumber field on which the ray tracing is performed. In this study, the optimal size of the smoothing window (always chosen as an odd integer) was determined for each frequency, assuming the wavenumber field is discretized on a $1~\mathrm{mm}$ grid. For example, our numerical results indicate that ray tracing on a $1~\mathrm{mm}$ grid at $1~\mathrm{MHz}$ requires a smoothing window of size~7. For the $0.4~\mathrm{mm}$ grid used in the full-wave simulations to generate the benchmark pressure time series, the window size was increased inversely with the grid spacing, yielding a smoothing window of size~17, as described above. It should be noted that the smoothing window was applied to the full-wave simulation grid only for this specific experiment to ensure consistency with the ray-based model. For the full-wave simulated pressure data used for benchmarking the image reconstruction, no smoothing was applied.

The pressure fields generated by emitters 1 and 20 were approximated on all 256 receivers at a single frequency of \(1~\mathrm{MHz}\) for three cases: (1) only water, (2) a non-absorbing breast immersed in water, and (3) an absorbing breast immersed in water.

\begin{figure} 
   \centering
	\subfigure[]{\includegraphics[width=0.45\textwidth]{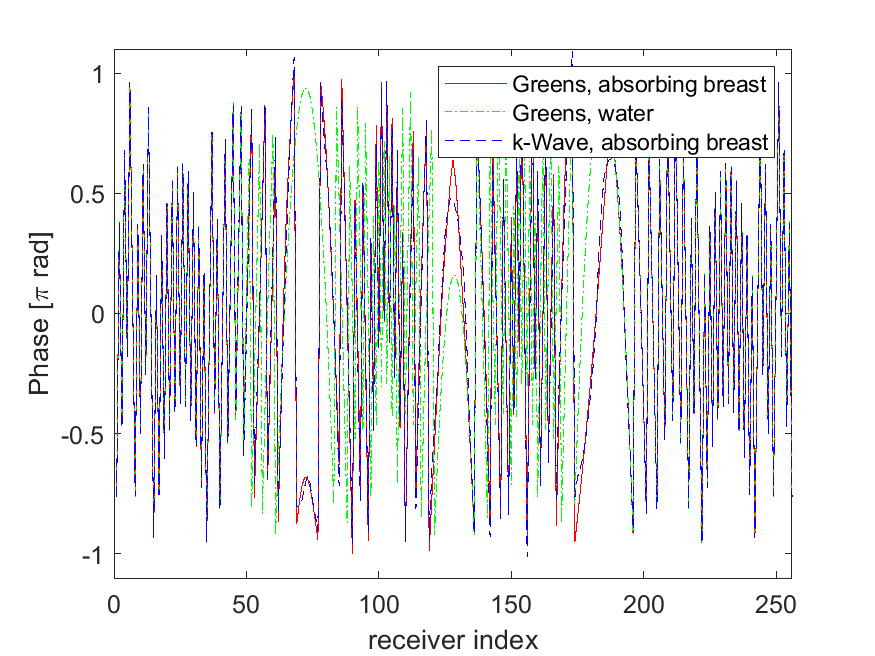}
   \label{fig:3a}   }
   \subfigure[]{\includegraphics[width=0.45\textwidth]{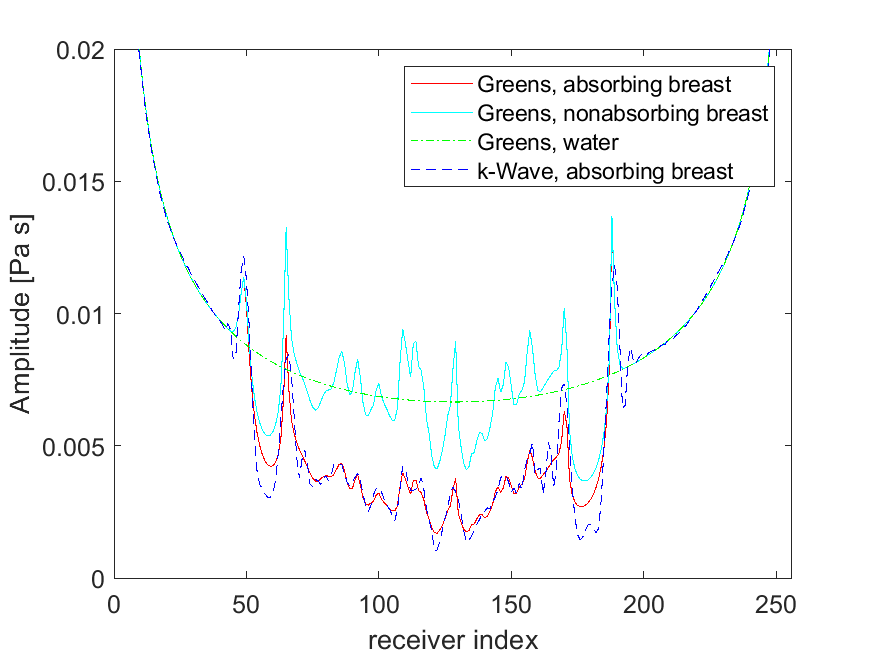}
    \label{fig:3b}  }
    	\subfigure[]{\includegraphics[width=0.45\textwidth]{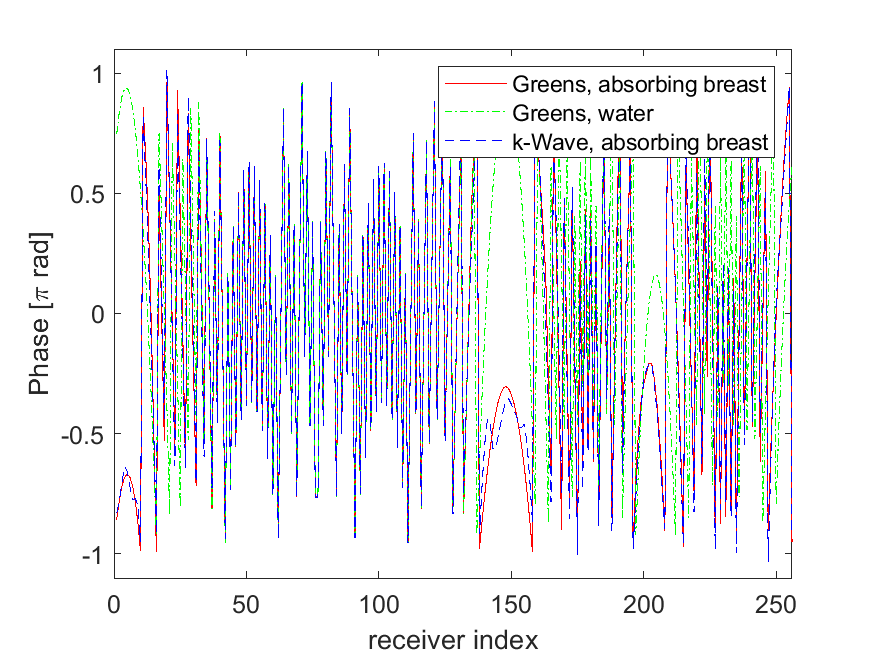} \label{fig:3c}   }
   \subfigure[]{\includegraphics[width=0.45\textwidth]{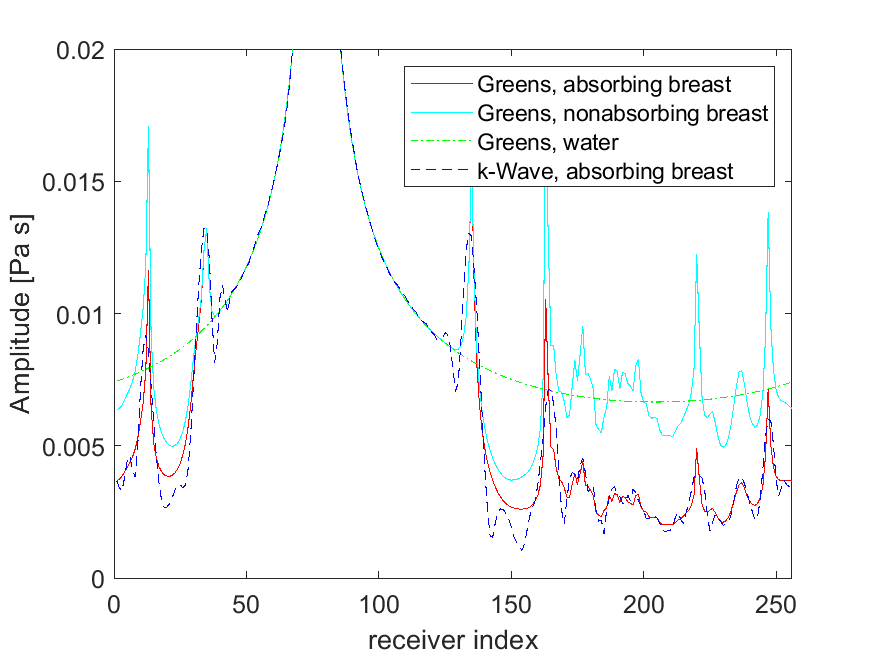}
    \label{fig:3d}  }
	\caption{Pressure time series at all 256 receivers for a single frequency of 1~MHz following the excitation of emitters~1 and~20: 
(a) phase for emitter~1, (b) amplitude for emitter~1, 
(c) phase for emitter~20, and (d) amplitude for emitter~20. 
For the full-wave simulation, point sources in terms of $s$ were used.}\label{fig:3}
\end{figure}

Figures~\ref{fig:3a} and~\ref{fig:3c} show the phase of the pressure time series at all receivers following the excitations of emitters~1 and~20, respectively. (Phases were wrapped to \([- \pi, \pi]\).) The green plot represents the phases analytically computed using the homogeneous Green's function assuming a water-only medium, whereas the red plot shows the phases computed from the ray-based approximation for the absorbing breast, based on the discretized formula~\eqref{eq:phase-discretised}. As the benchmark for comparison, the blue plot represents the phases obtained from the full-wave simulation for the absorbing breast. For receiver indices corresponding to rays that traverse the breast region rather than water alone, the ray approximation and full-wave simulation exhibit strong agreement, whereas the homogeneous Green's function shows substantial discrepancies.

Figures~\ref{fig:3b} and~\ref{fig:3d} illustrate the amplitude of the pressure time series at all receivers following the excitations of emitters~1 and~20, respectively. 
The green plot shows amplitudes computed using the homogeneous Green's function under a water-only assumption. 
Amplitudes for a non-absorbing breast, attenuated solely by geometrical spreading, were computed via Eq.~\eqref{eq:amplitude-geom-discretised} and are shown in light blue. 
(For Eq.~\eqref{eq:amplitude-geom-discretised}, the relative Jacobian was approximated using paraxial ray tracing, as described in Section~\ref{sec:ray-tracing}.) 
The red plot represents amplitudes computed using the ray-based approximation for the absorbing breast, incorporating both geometrical spreading (light blue) and accumulated absorption via Eq.~\eqref{eq:amplitude-abs-discretised}. 
For comparison, the dark blue plot shows amplitudes obtained from the full-wave simulation. 
For the absorbing breast, the ray approximation accounting for both geometrical spreading and accumulated absorption closely matches the full-wave results, whereas the homogeneous Green's function or the ray-based heterogeneous Green's function neglecting absorption exhibits significant discrepancies.

\subsection{Image reconstruction} \label{sec:image-reconstruction}

This subsection outlines the image reconstruction process from ultrasound data simulated using the full-waveform approach and presents the reconstructed images. Synthetic data were simulated using the \textit{k}-space pseudo-spectral approach, as described in Section~\ref{sec:data-simulation}. Two datasets were generated: the first simulated in water alone, and the second simulated using the digital breast phantom immersed in water. The breast-in-water data employed the sound speed and absorption coefficient maps shown in Figures~\ref{fig:1a} and \ref{fig:1b}, respectively. For each dataset, emitters were sequentially excited by the pulse shown in Figure~\ref{fig:2a}, and the resulting pressure time series were recorded on all receivers. This process was repeated for all emitters. 

The simulation of the UST dataset for the digital breast phantom immersed in water, using the k-Wave MATLAB toolbox \cite{Treeby} on a single 6-core Intel Xeon E5-2620 v2 2.1 GHz CPU, required approximately 4 hours. Additive white Gaussian noise (AWGN) was introduced to the simulated time series at varying levels to achieve signal-to-noise ratios (SNRs) of 40~dB, 30~dB, and 25~dB based on the peak amplitudes, for both datasets.

To mitigate the inverse crime related to temporal discretization, the time series simulated by the full-waveform approach were downsampled by a factor of 2 before ray-based reconstruction. The reconstruction grid consisted of $204 \times 204$ points, spanning the region $[-10.1650, +10.1350] \times [-10.1650, +10.1350]~\mathrm{cm}^2$, with a uniform grid spacing of $\Delta x = 1~\mathrm{mm}$. Image reconstruction was confined to grid points inside a binary mask with a radius equal to 90\% of the detection ring's radius, while sound speed values outside this mask were set to that of water ($1500~\mathrm{ms}^{-1}$).

The Green’s function \( g(\omega, r, e) \), utilized in the residual \( g_{\text{res}} \), was approximated directly at the receiver locations through ray linking. Since the sound speed is reconstructed at the grid points, the reciprocal Green's functions, \(g_{\dagger}(\omega,\bx,e)\), were first approximated and reciprocated along the linked rays initialized at the emitter position \(e\), and their parameters were subsequently interpolated onto the computational grid. This interpolation involved applying triangulation to the sampled points along the linked rays and using trilinear interpolation. 

Furthermore, the Green's functions, \(g(\omega,\bx,r)\), and their reciprocal counterparts were computed without additional ray tracing by reversing the accumulated parameters of the forward Green's functions, \(g(\omega,r,e)\), as described in \cite{Javaherian2}. To compute the geometrical components of the amplitudes for \( g(\omega, \bx, r) \), an additional paraxial ray was traced along each reversed ray. Finally, the reciprocal Green's functions, \( g_{\dagger}(\omega, \bx, r) \), were derived from the Green's functions, \( g(\omega, \bx, r) \), by taking the reciprocals of their amplitudes and phases.

\subsubsection{Inversion approach using Time-of-Flights (initial guess)} 
Because the objective function in Eq. \eqref{eq:objective-function} is highly nonlinear, an initial guess is often employed \cite{Sandhu,Wiskin2}. In this work, an image reconstruction approach based on direct time-of-flights (TOFs) extracted from the measured (simulated) pressure time series was utilized to provide an initial guess. The TOF-based inversion approach iteratively minimizes the L2 norm of the discrepancy between TOFs modeled via the ray tracing algorithm and TOFs extracted from the measured (simulated) pressure time series using a first-arrival picking algorithm \cite{Javaherian1,Li3}.  

To mitigate the effects of measurement errors on the picked first-arrivals, a difference inversion approach was employed. In this method, the difference in slowness (\(1/c\)) between the breast-in-water and only-water cases is computed from the discrepancy in first-arrival times obtained from the respective data sets. The minimization procedure iteratively linearizes the associated objective function and solves the resulting linear subproblems using a Radon-type technique.  

In the present study, each linearized subproblem was solved using the Simultaneous Algebraic Reconstruction Technique (SART), which accounts for the nonuniform ray density resulting from the curved trajectories of the rays \cite{Anderson1}. In each linearization step, the sound speed update was smoothed using an averaging window of 7 grid points, specifically for the purpose of ray tracing. However, the integration of accumulated TOFs along the ray trajectories was performed using the nonsmoothed sound speed updates. From the perspective of the linearization, this means that the system matrix was constructed using the smoothed sound speed update, whereas it acts on the unsmoothed update to preserve the accuracy of the accumulated TOFs.

Ray linking was carried out using the Secant method. For each emitter--receiver pair, the linked ray (i.e., the optimal ray obtained after ray linking) associated with each linearized subproblem served as the initial guess for the ray-linking procedure in the subsequent linearized subproblem \cite{Javaherian2,Javaherian1}. In the present study, the TOF-based algorithm was terminated at an early stage—after only three linearizations—to balance reconstruction accuracy against artifacts arising from errors in the picked first-arrival times \cite{Javaherian2}.

\subsubsection{Inversion approaches using ray approximation to heterogeneous Green's function}
Below, the procedure for implementing the inversion approach based on the Green's function is explained. To evaluate the effectiveness of the proposed Hessian-free inversion approach, the method used in \cite{Javaherian2} serves as the benchmark. The goal of the inversion is to determine the squared slowness map, $m$, which minimizes the objective function \eqref{eq:objective-function}. This problem is solved by discretizing the objective function at a range of frequencies within the frequency band of the transducers, and performing minimization from low to high frequencies until the computed update direction associated with a frequency set becomes smaller than a specified tolerance. To compute the residual \eqref{eq:residual}, the measured Green's function, \(\mathfrak{g}(\omega,r,e)\), is obtained by deconvolving the excitation pulse from the measured pressure time series, which are simulated here using the full-wave approach. In addition, each Green's function, \(g(m;\omega,r,e)\), appearing in the residual \eqref{eq:residual} is approximated at the terminal points of the linked rays, which are computed recursively from low to high frequency intervals using the most recently updated sound speed.

\textit{Ray linking.} For the Green's function-based approaches, the ray linking was performed on the updates of the sound speed and wavenumber maps for each of emitter-receiver pairs separately. For each update of the wavenumber map and each emitter-receiver pair, the ray linking was done by iteratively updating the initial unit direction of the ray using a Secant method \cite{Javaherian2,Javaherian1}. The initial guess for the unknown initial unit direction of the ray was set the obtained optimal initial direction after ray linking for the last previous update of the wavenumber map, and the ray's trajectory was iteratively computed using the first part of Algorithm \ref{alg:ray_tracing} until the interception point of the ray by the detection ring matches the receiver position within a tolerance. 

The parameters of the Green’s functions \( g \) and reciprocal Green’s functions \( g_{\dagger} \) (the latter included in the backprojection formula for the Hessian-free approach) were computed along the linked rays, following the expressions provided in Section~\ref{sec:Discretised-approximation}. 
To compute the geometrical component of the amplitudes, the linked rays were used as reference rays, while paraxial rays were traced in their vicinity by implementing the second part of Algorithm~\ref{alg:ray_tracing}. For ray tracing and ray linking, an averaging window of variable size—13, 11, 9, and 7 grid points, decreasing from low to high frequencies—was applied to the updated wavenumber maps. However, integration along the linked rays and the approximation of the Green’s function parameters using the formulae in Section~\ref{sec:Discretised-approximation} were performed on the non-smoothed updates.

\textit{a) Hessian-free ray-Born inversion.}  
The inversion approach described in section \ref{sec:minimisation_approach} was implemented at 180 discretized frequencies in the range $f \in \left\{0.2, ..., 1.72 \right\} $MHz. Image reconstruction progressed sequentially from low to high frequencies, with each update of the squared slowness map computed at two consecutive frequencies using Eq. \ref{eq:sound_speed_update} ($n \in \left\{1, ..., 90 \right\} $). For each search direction $n$, the Green's functions $g(\omega,r,e)$ included in the residual \eqref{eq:residual}, and the reciprocal Green's functions $g_{\dagger}(\omega,\bx, e)$ and $g_{\dagger}(\omega,\bx,r)$ included in the backprojection formula \eqref{eq:sound_speed_update2}, were computed along the forward and backward rays, respectively. The step length was heuristically chosen as $\tau = 1.2 \times 10^{-1}$ for all linearization steps. 

The dominant computational cost in determining the search direction associated with each frequency set arose from ray tracing, since the update direction was computed in a single step using Eqs.~\eqref{eq:sound_speed_update} and \eqref{eq:sound_speed_update2}.

\textit{b) Hessian-based ray-Born inversion (Gauss-Newton).}  
Using the Hessian-based approach, instead of linearizing the objective function over the entire frequency range, the objective function is linearized recursively from low to high frequencies, with $\mathcal{Q} = 1$ for all $(\omega, r, e, \bx)$. Each resulting frequency-dependent linearized subproblem $n$ \eqref{eq:linearised-linearised-subproblem} is solved by first forming the gradient $\nabla \mathcal{F}^{(n)}$ and then determining the update direction through iterative computations of the actions of the Hessian operator $H^{(n)}$ on perturbations to the squared slowness map. This iterative procedure, using inner iterations, determines the optimal perturbation vector and effectively produces a Gauss–Newton search direction for each frequency set. Each linearized subproblem in this study was solved using 10–15 inner iterations, with early termination of these iterations providing a regularizing effect on the solution. Details of the algorithm can be found in \cite{Javaherian2}.

To benchmark the proposed Hessian-free ray-Born inversion algorithm, the Hessian-based inversion approach described in \cite{Javaherian2} was implemented at 120 discretized frequencies within the range \( f \in [0.2, 1.2]~\mathrm{MHz} \). (Reconstructed images begin to degrade slightly at frequencies above \( 1.2~\mathrm{MHz} \).) Similar to the Hessian-free approach, the image reconstruction proceeded from low to high frequencies, with each update of the sound speed computed at two consecutive discretized frequencies ($n \in \left\{1, ..., 60 \right\}$). The step length was heuristically set to $\tau = 2.75 \times 10^4$ for all linearisations. Using the developed MATLAB code on the CPU described in the second paragraph of section \ref{sec:image-reconstruction}, the computational time for solving each linearized subproblem $n$ and computing the update direction was approximately an order of magnitude higher than the corresponding computational cost for the proposed Hessian-free approach. 

As described in section \ref{sec:ray-tracing}, the geometrical amplitudes were computed by solving a paraxial system of equations rather than using auxiliary rays, as previously done in \cite{Javaherian2,Rullan} (see Algorithm \ref{alg:ray_tracing}).

\subsubsection{Prototype Born inversion based on a water-only assumption}
To benchmark the ray-based approximation to the Green's function for sound-speed image reconstruction, a prototype Born approximation was implemented at 50 discretized frequencies within the range \( f \in [0.2, 0.62]~\mathrm{MHz} \). (Reconstructed images begin to degrade slightly at frequencies above \( 0.62~\mathrm{MHz} \).) 
Similar to the ray-based approaches, the reconstruction proceeded from low to high frequencies, with each sound-speed update computed at two consecutive discretized frequencies (\( n \in \{1, \dots, 25\} \)). 
The image reconstruction followed the same procedure as the Hessian-based approach, using 8–12 inner iterations. However, in this case, the Green’s functions were computed analytically under a water-only assumption, eliminating the need for ray tracing.

\subsection{Reconstructed images}
This section presents the sound speed images reconstructed using the proposed Hessian-free ray-Born inversion approach and compares them with images reconstructed using the Hessian-based inversion approach described in \cite{Javaherian2} and the prototype Born inversion approach based on a water-only assumption for approximating the Green's function. As noted earlier, the proposed Hessian-free method is approximately an order of magnitude faster than the Hessian-based one due to inverting a full-frequency Hessian matrix in a single step.

The reconstructed images are evaluated using the Relative Error (RE), defined as:
\begin{align}
    RE_{\text{image}} = \frac{\| c_{\text{image}} -c_{\text{phantom}}  \|_2}{\|c_{\text{water}} -c_{\text{phantom}}  \|_2} \times 100,
\end{align}
where $c_{\text{phantom}}$ represents the sound speed map of the digital breast phantom interpolated onto the reconstruction grid, and $c_{\text{image}}$ is the reconstructed sound speed image. These quantities were calculated only on grid points within the binary mask used for image reconstruction. For grid points outside this mask, the sound speed was set to $c_{\text{water}} = 1500 \ \mathrm{ms}^{-1}$. (cf. Section \ref{sec:image-reconstruction}.) Reconstructed sound speed images are presented for synthetic datasets with signal-to-noise ratios (SNRs) of 40~dB, 30~dB, and 25~dB.

\subsubsection{Reconstructed images from UST data with high SNR}
\begin{figure}[h!]
   \centering
	\subfigure[]{\includegraphics[width=0.32\textwidth]{Fig2a.png}
   \label{fig:4a}   }
   \subfigure[]{\includegraphics[width=0.32\textwidth]{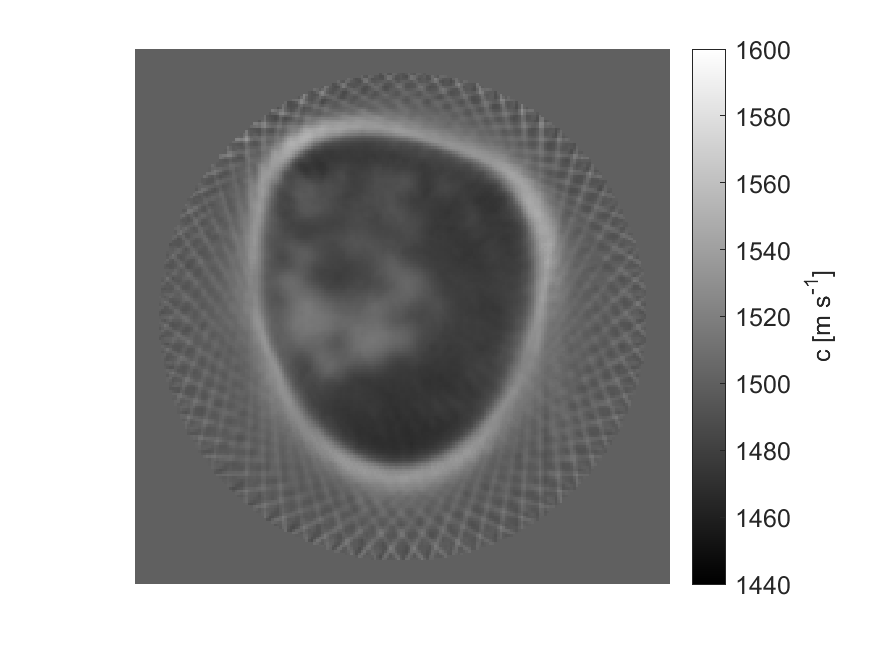}
    \label{fig:4b}  }\\
       \subfigure[]{\includegraphics[width=0.32\textwidth]{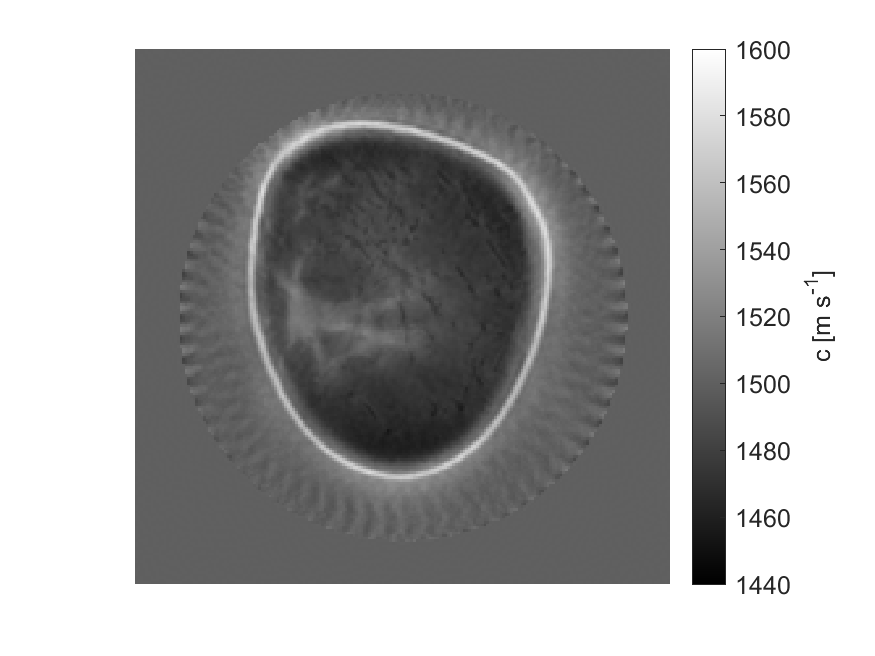}
    \label{fig:4c}  }
     \subfigure[]{\includegraphics[width=0.32\textwidth]{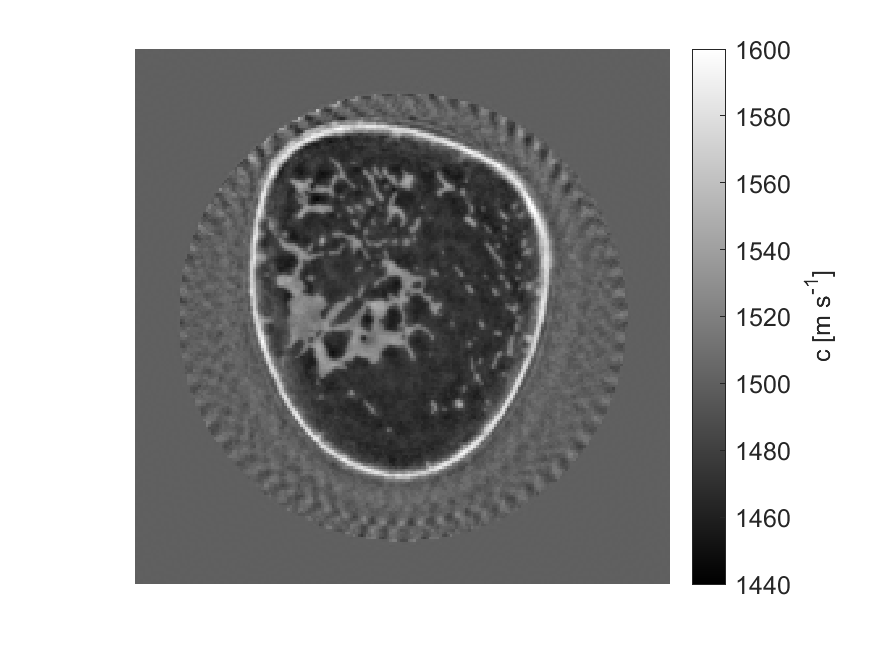}
    \label{fig:4d}  }
    \subfigure[]{\includegraphics[width=0.32\textwidth]{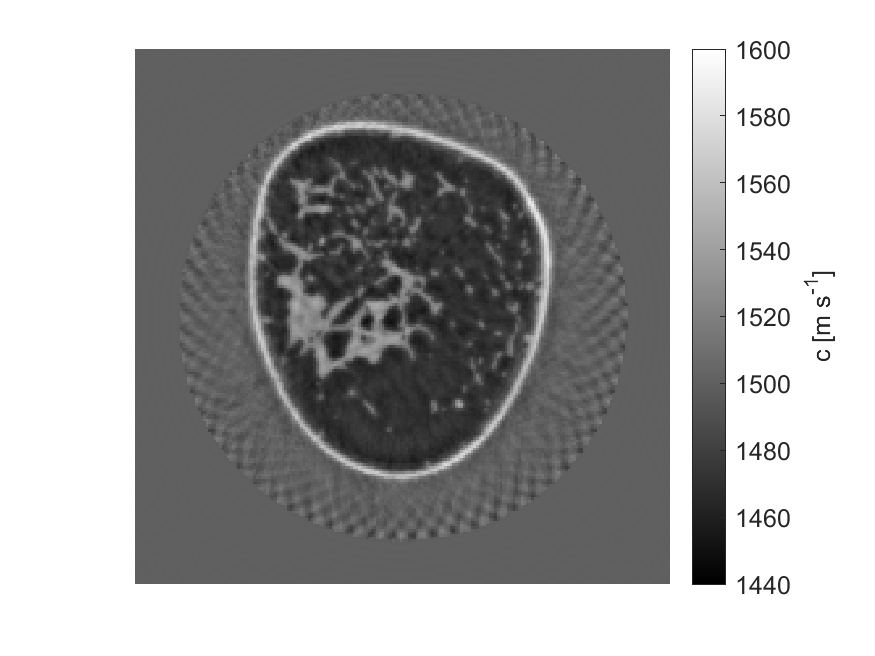}
    \label{fig:4e}  }\\
     \subfigure[]{\includegraphics[width=0.32\textwidth]{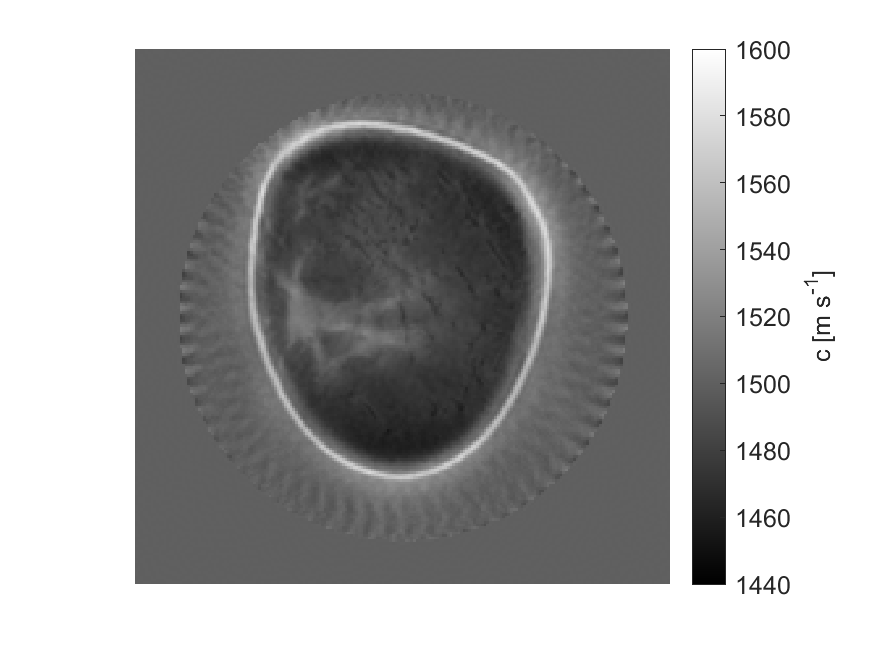}
    \label{fig:4f}  }
        \subfigure[]{\includegraphics[width=0.32\textwidth]{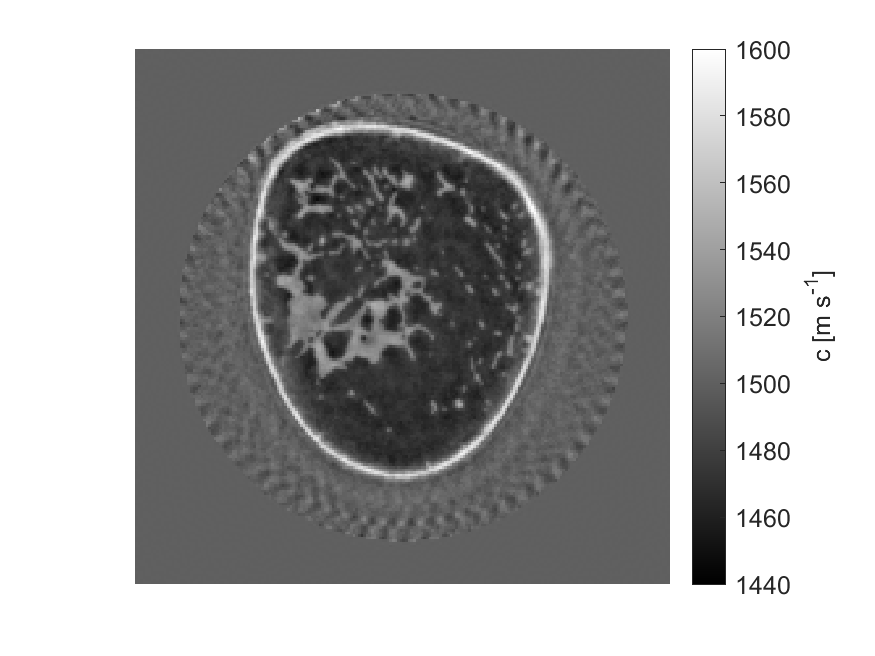}
    \label{fig:4g}  }
     \subfigure[]{\includegraphics[width=0.32\textwidth]{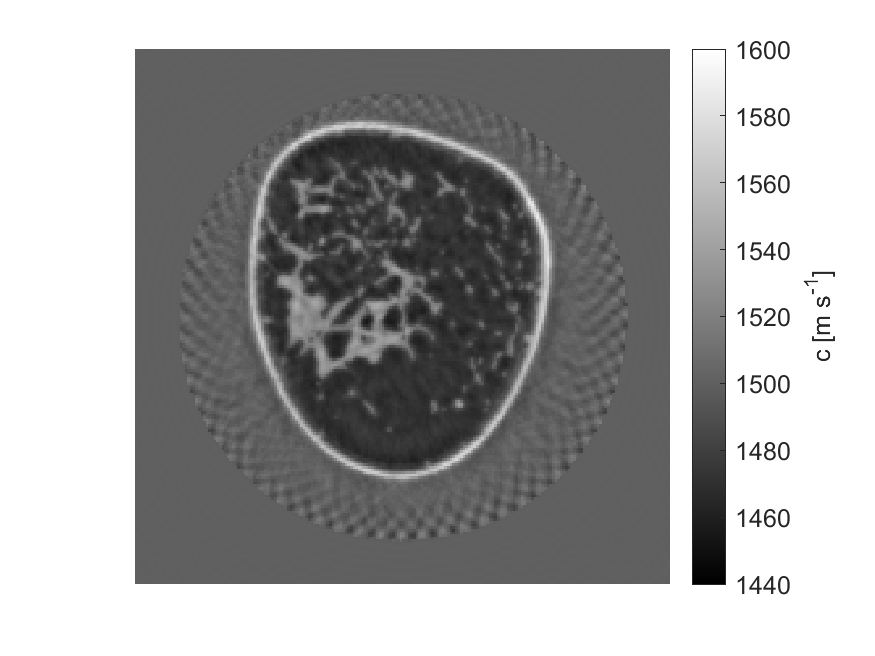}
    \label{fig:4h}  }\\
        \subfigure[]{\includegraphics[width=0.32\textwidth]{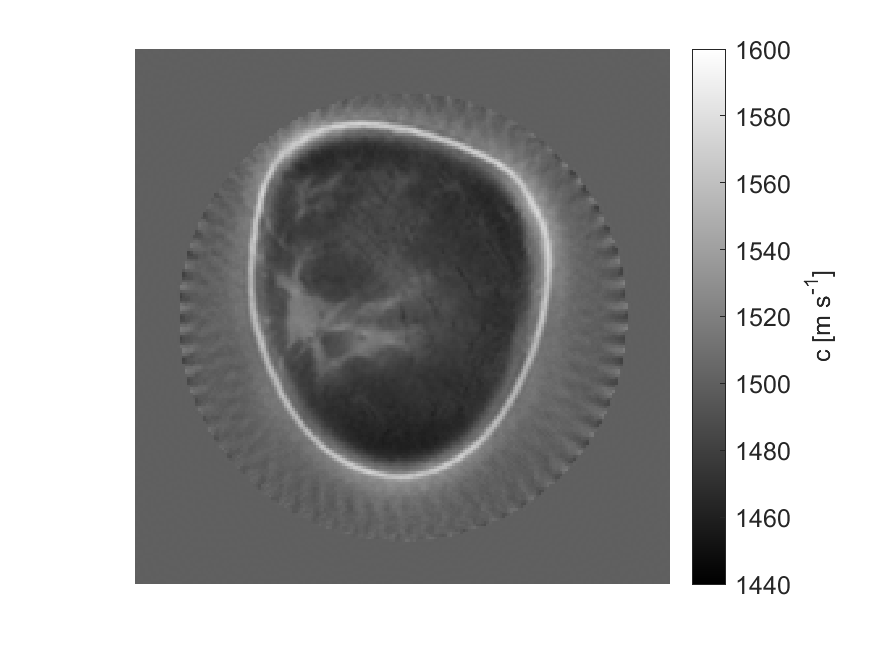}
    \label{fig:4i}  }
     \subfigure[]{\includegraphics[width=0.32\textwidth]{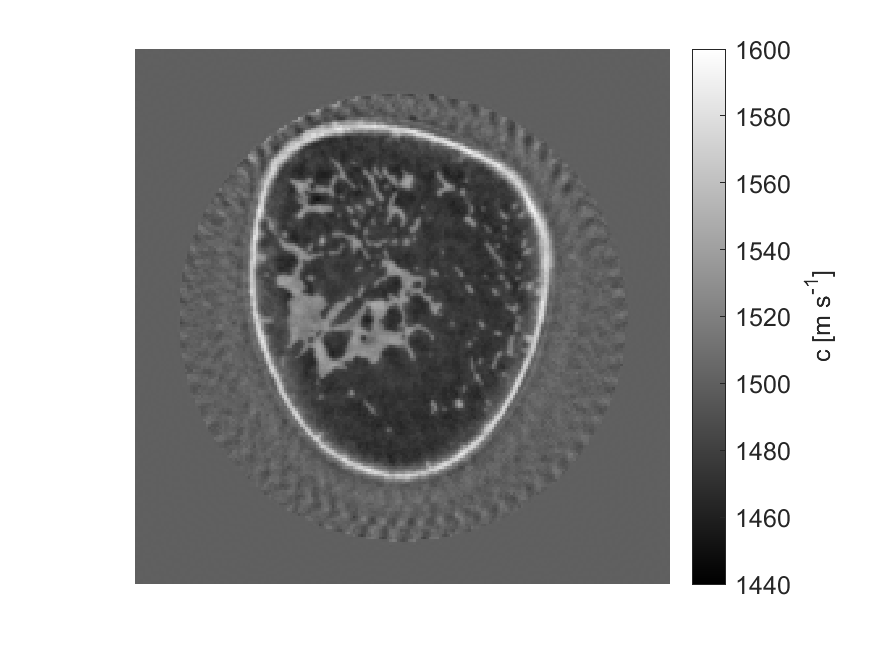}
    \label{fig:4j}  }
       \subfigure[]{\includegraphics[width=0.32\textwidth]{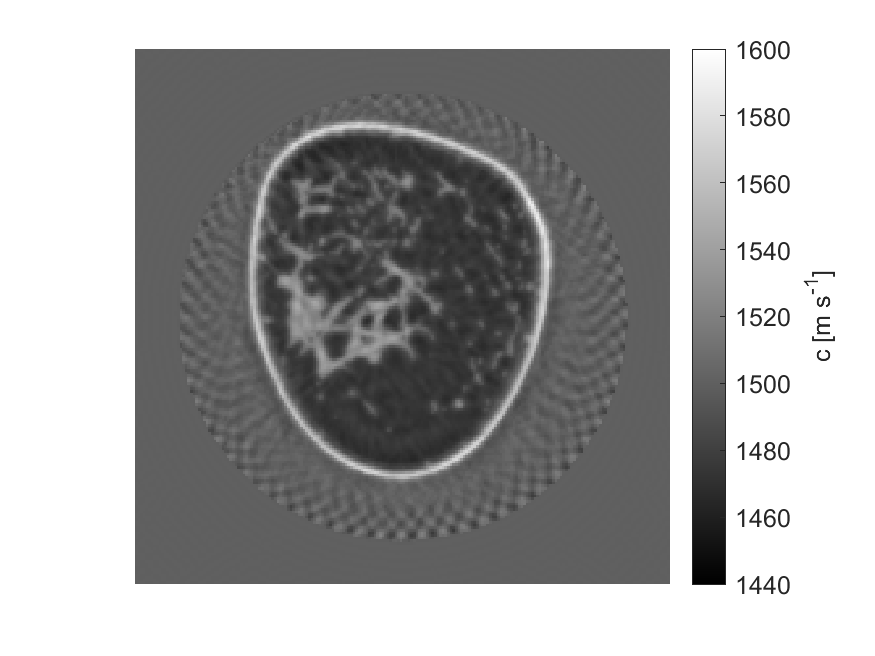}
    \label{fig:4k}  }
    \caption{(a) Ground truth. Reconstructed sound speed images from synthetic data with 40~dB SNR: (b) Time-of-flight-based approach (initial guess). Assuming the true $\alpha_0$ (Figure \ref{fig:1b}): (c) Born, (d) Hessian-based ray-Born, (e) Hessian-free ray-Born. Assuming $\alpha_0 = 0$: (f) Born, (g) Hessian-based ray-Born, (h) Hessian-free ray-Born. Assuming $\alpha_0 = 0.5 \ \mathrm{dB \ MHz}^{-y} \mathrm{cm}^{-1}$ (homogeneous within the breast): (i) Born, (j) Hessian-based ray-Born, (k) Hessian-free ray-Born.}\label{fig:reconstructed-images4}
\end{figure}

This subsection focuses on the reconstructed sound speed images from synthetic ultrasound tomography data with an SNR of 40~dB, along with their evaluation in terms of RE. Figure \ref{fig:4a} displays the sound speed map of the digital breast phantom (ground truth), while Figure \ref{fig:4b} shows the image reconstructed from time-of-flights (TOFs) extracted using a modified Akaike Information Criterion (AIC) approach \cite{Li3}. 

As discussed in Section \ref{sec:greens}, the proposed forward model, which uses the ray approximation to the heterogeneous Green's function, accounts for acoustic absorption and dispersion following a frequency power law. However, in practical scenarios, the absorption coefficient $\alpha_0$ map and the exponent power $y$ are typically unknown. In this study, the exponent power was assumed known and set to $y=1.4$. Reconstructions were performed under three assumptions: (1) a known $\alpha_0$ map (Figure \ref{fig:1b}), (2) zero $\alpha_0$, and (3) a homogeneous $\alpha_0$ within the breast. In practical scenarios, a rough homogeneous estimate of $\alpha = \alpha_0 \omega^y$ can be derived from the mean logarithmic relative amplitudes of the measured time series for water-only to breast-in-water datasets. (For case 3, the homogeneous absorption coefficient within the breast was heuristically set to $0.5~\mathrm{dB\,MHz}^{-\mathrm{y}}\,\mathrm{cm}^{-1}$.)

Figures~\ref{fig:4c},~\ref{fig:4d},~and~\ref{fig:4e} show the reconstructed sound speed images obtained using the prototype Born inversion, and the Hessian-based and Hessian-free ray-Born inversion approaches, respectively, under the assumption of a known heterogeneous $\alpha_0$ map. Recall that in the prototype Born inversion approach, the true absorption map and the sound speed updates are included only in the scattering potential, while the Green's functions are approximated using the water-only assumption. In contrast, in the ray-Born approaches, the absorption map and the sound speed updates are incorporated in both the scattering potential and the Green's functions.

Similarly, Figures~\ref{fig:4f},~\ref{fig:4g},~and~\ref{fig:4h} present reconstructions obtained under the assumption $\alpha_0 = 0$, neglecting both acoustic absorption and dispersion. Figures~\ref{fig:4i},~\ref{fig:4j},~and~\ref{fig:4k} show reconstructions assuming a homogeneous absorption coefficient of $\alpha_0 = 0.5~\mathrm{dB\,MHz}^{-y}\mathrm{cm}^{-1}$ within the breast. The corresponding relative error (RE) values for these reconstructions are summarized in Table~\ref{tab:RE}.

As seen in the figures, the prototype Born inversion approach fails to reconstruct an accurate image. In contrast, the Hessian-free ray-Born inversion approach produces more accurate reconstructions in terms of RE compared to the Hessian-based method, which implicitly and iteratively inverts the full Hessian matrix.

For the ray-Born reconstructions, assuming $\alpha_0 = 0$ results in less accurate images with more artifacts. In contrast, assuming a homogeneous $\alpha_0$ within the breast significantly improves image quality and reduces RE relative to the zero $\alpha_0$ assumption. Notably, while the reconstructed sound speed images degrade when assuming $\alpha_0 = 0$, assuming a homogeneous $\alpha_0$ within the breast yields images with RE values comparable to those obtained using the heterogeneous true $\alpha_0$, particularly for the Hessian-free ray-Born inversion approach.

\begin{table}[ht]
\centering
\caption{Relative Error (RE) values (\%) for reconstructed images at different SNR levels and assumptions for $\alpha_0$.}
\label{tab:RE}
\begin{tabular}{cccccc}
\hline
\multirow{2}{*}{SNR (dB)} & \multirow{2}{*}{\(\alpha_0\) Assumption} & \multicolumn{4}{c}{Reconstruction Approach (RE \%)}  \\ \cline{3-6} 
                          &                                         & Time-of-Flight  & Born       & Hessian-based & Hessian-free  \\ \hline
\multirow{4}{*}{40}       & True                                    &    77.97        & 69.27      &   45.86      &  42.12         \\ \cline{2-6} 
                          & Zero                                    &    -            & 67.82      &   47.60      &  47.95         \\ \cline{2-6} 
                          & Homogeneous                             &    -            & 69.09      &   45.98      &  43.05         \\ \hline
\multirow{4}{*}{30}       & True                                    &    78.92        & 67.34      &   48.18      &  44.03        \\ \cline{2-6} 
                          & Zero                                    &       -         & 66.06      &   51.32      &  50.53         \\ \cline{2-6} 
                          & Homogeneous                             &       -         & 67.12      &   48.86      &   45.13        \\ \hline
\multirow{4}{*}{25}       & True                                    &   81.67         & 68.79      &   53.31      &  46.21         \\ \cline{2-6} 
                          & Zero                                    &      -          & 67.51      &   56.08      &  52.55         \\ \cline{2-6} 
                          & Homogeneous                             &                 & 68.56      &   53.90      &  47.34         \\ \hline
\end{tabular}
\end{table}

\subsubsection{Reconstructed images from UST data with medium and low SNR}
\begin{figure}[h!]
   \centering
	\subfigure[]{\includegraphics[width=0.32\textwidth]{Fig2a.png}
   \label{fig:5a}   }
   \subfigure[]{\includegraphics[width=0.32\textwidth]{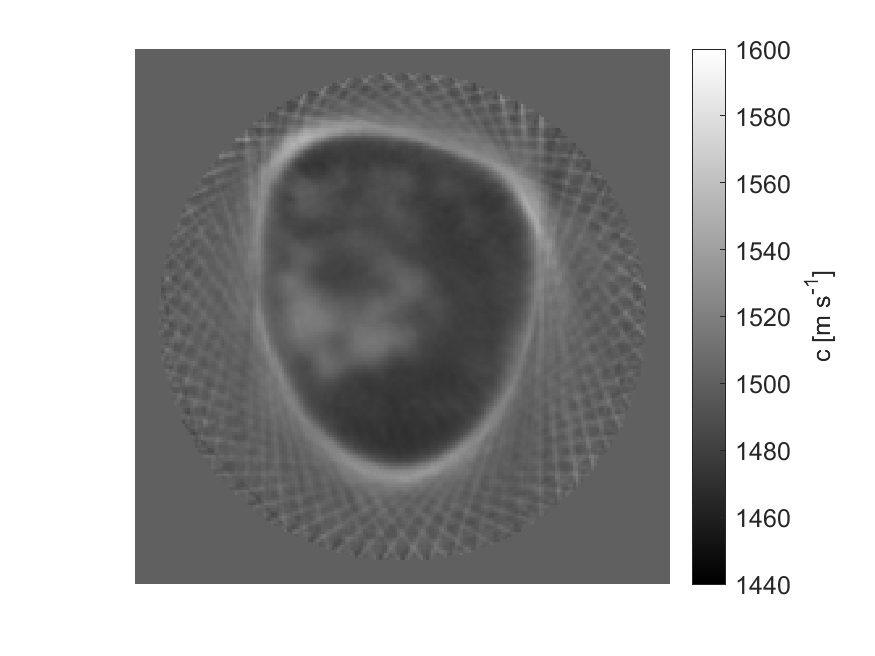}
    \label{fig:5b}  }\\
       \subfigure[]{\includegraphics[width=0.32\textwidth]{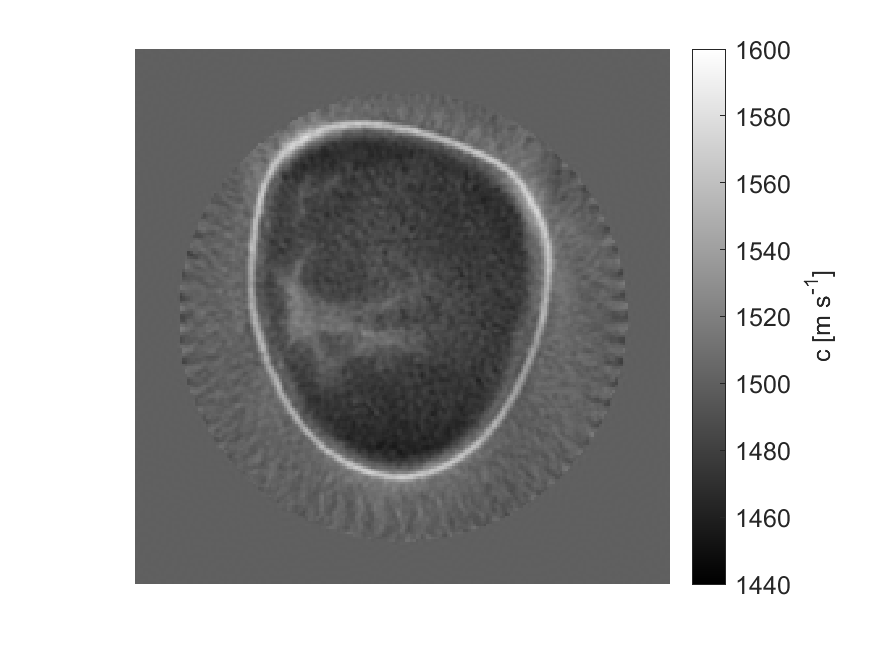}
    \label{fig:5c}  }
     \subfigure[]{\includegraphics[width=0.32\textwidth]{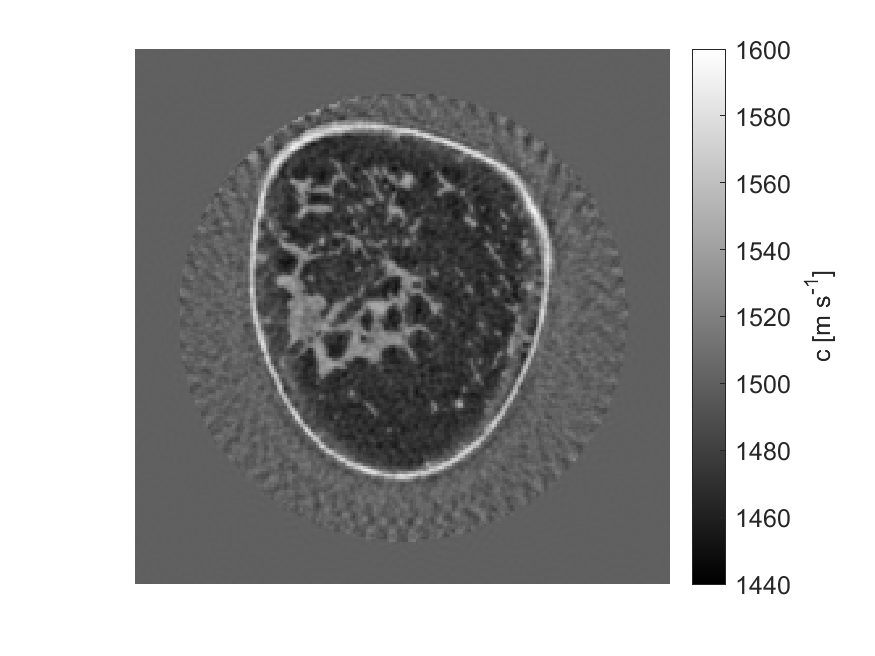}
    \label{fig:5d}  }
    \subfigure[]{\includegraphics[width=0.32\textwidth]{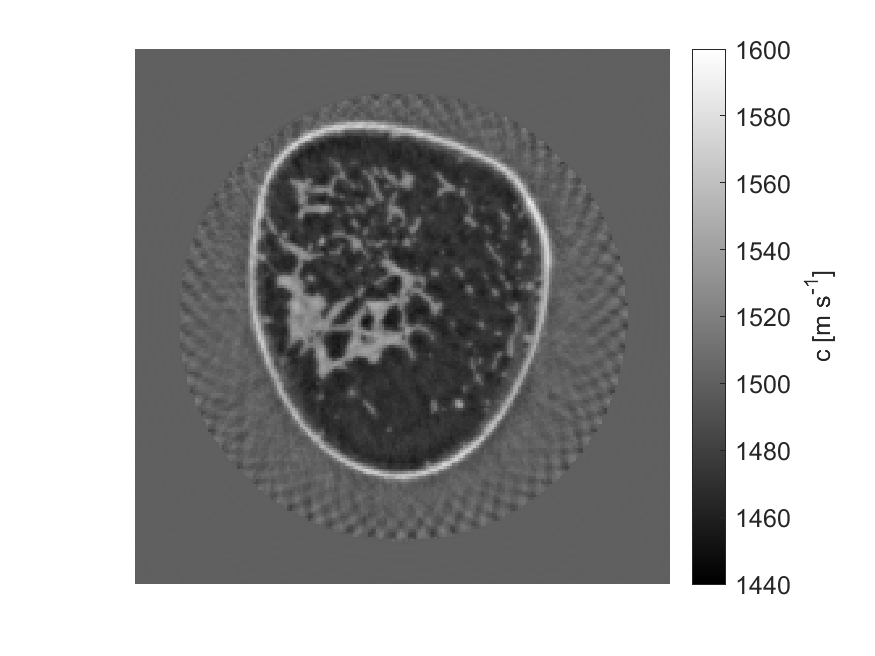}
    \label{fig:5e}  }\\
     \subfigure[]{\includegraphics[width=0.32\textwidth]{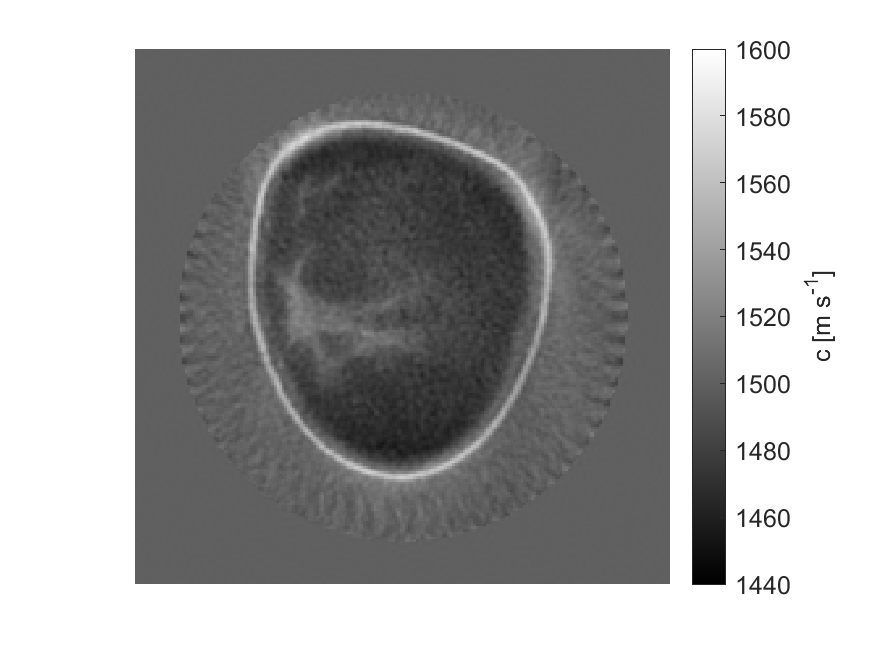}
    \label{fig:5f}  }
        \subfigure[]{\includegraphics[width=0.32\textwidth]{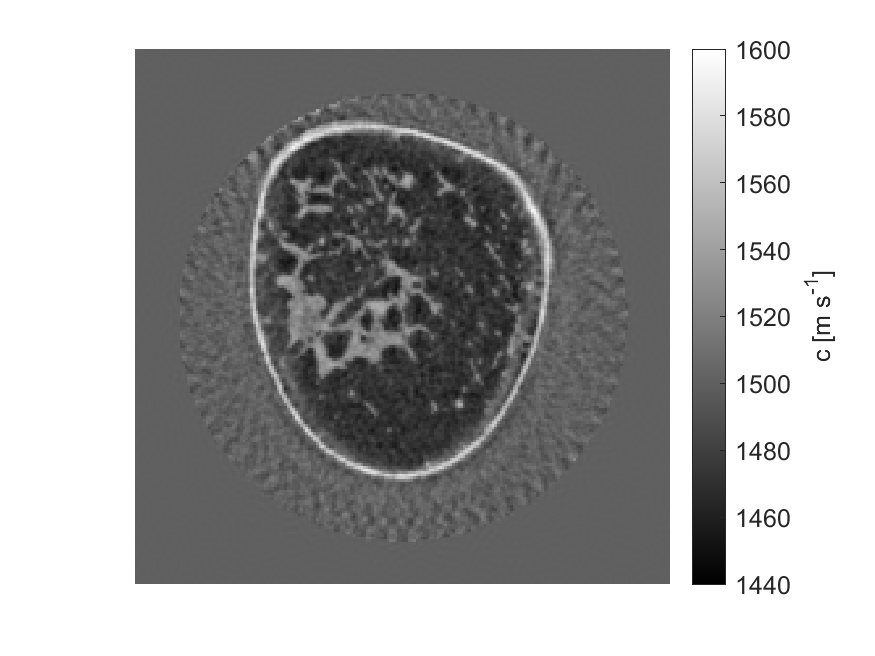}
    \label{fig:5g}  }
     \subfigure[]{\includegraphics[width=0.32\textwidth]{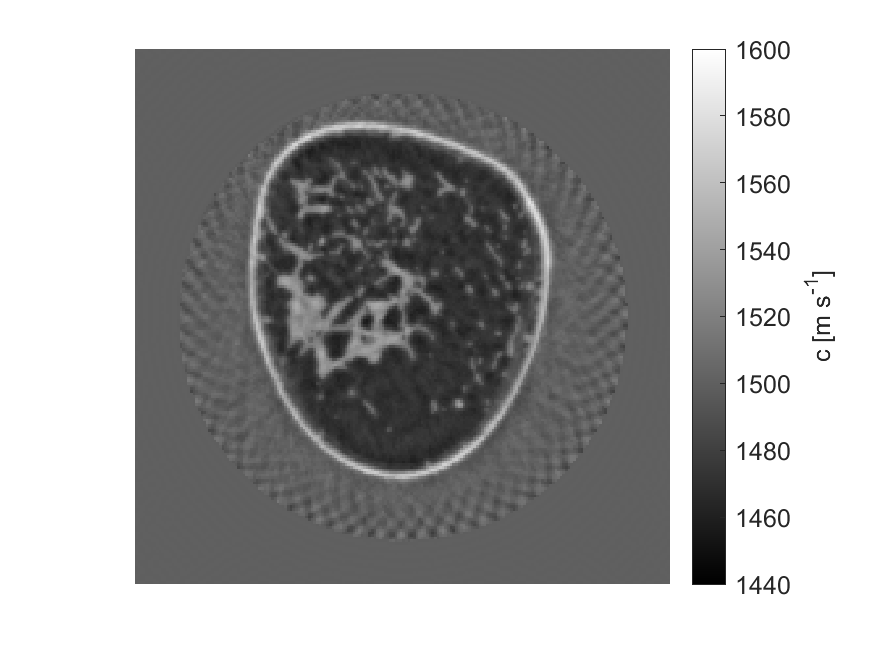}
    \label{fig:5h}  }\\
        \subfigure[]{\includegraphics[width=0.32\textwidth]{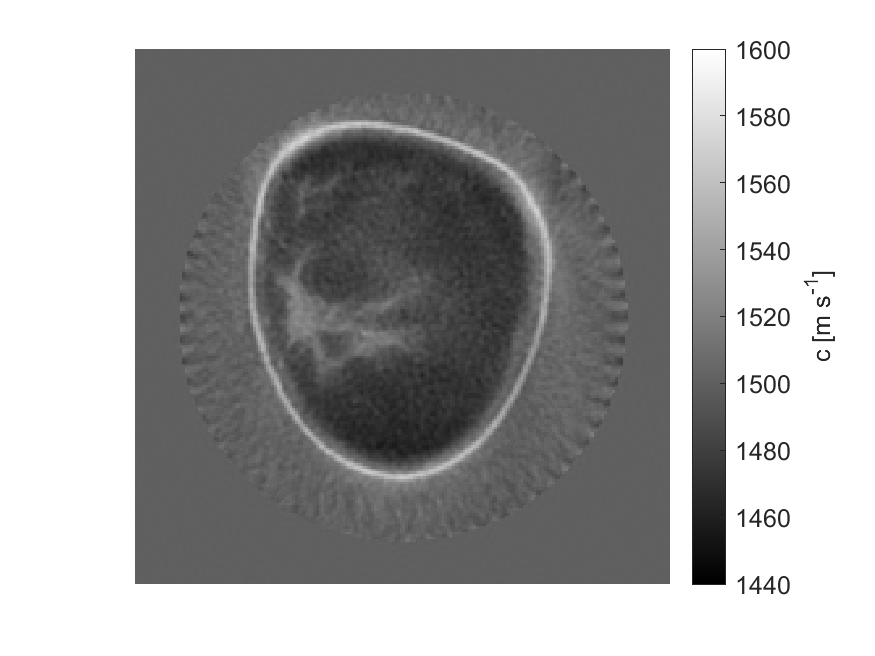}
    \label{fig:5i}  }
     \subfigure[]{\includegraphics[width=0.32\textwidth]{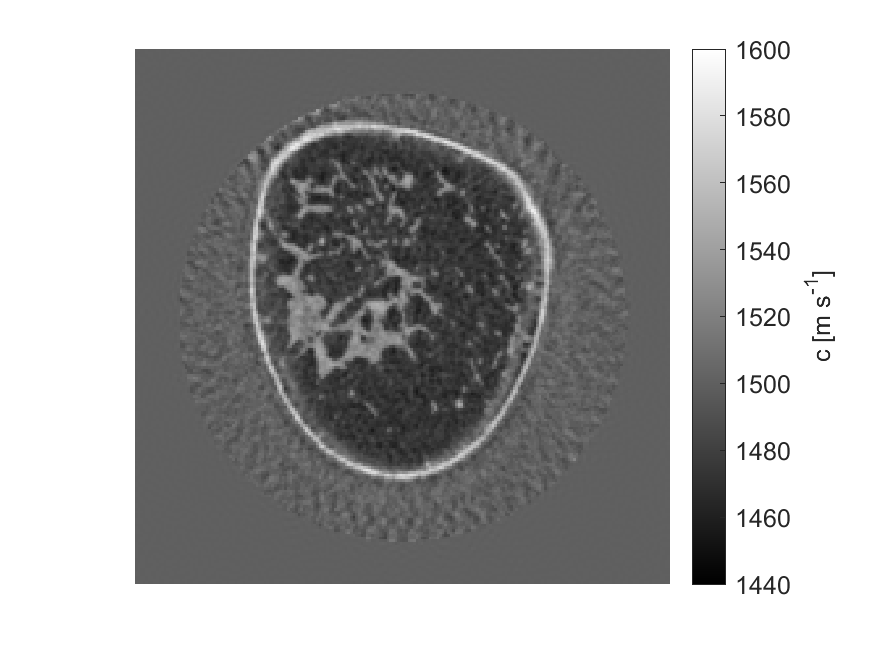}
    \label{fig:5j}  }
       \subfigure[]{\includegraphics[width=0.32\textwidth]{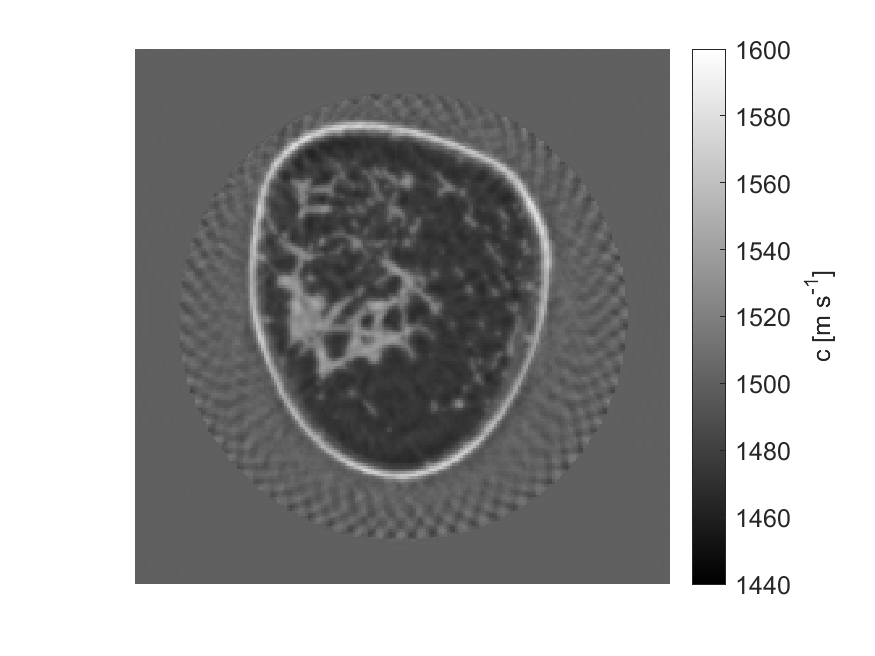}
    \label{fig:5k}  }
   \caption{(a) Ground truth. Reconstructed sound speed images from synthetic data with 30~dB SNR: (b) Time-of-flight-based approach (initial guess). Assuming the true $\alpha_0$ (Figure \ref{fig:1b}): (c) Born, (d) Hessian-based ray-Born, (e) Hessian-free ray-Born. Assuming $\alpha_0 = 0$: (f) Born, (g) Hessian-based ray-Born, (h) Hessian-free ray-Born. Assuming $\alpha_0 = 0.5 \ \mathrm{dB \ MHz}^{-y} \mathrm{cm}^{-1}$ (homogeneous within the breast): (i) Born, (j) Hessian-based ray-Born, (k) Hessian-free ray-Born. 
     }\label{fig:reconstructed-images5}
\end{figure}

\begin{figure}[h!]
   \centering
	\subfigure[]{\includegraphics[width=0.32\textwidth]{Fig2a.png}
   \label{fig:6a}   }
   \subfigure[]{\includegraphics[width=0.32\textwidth]{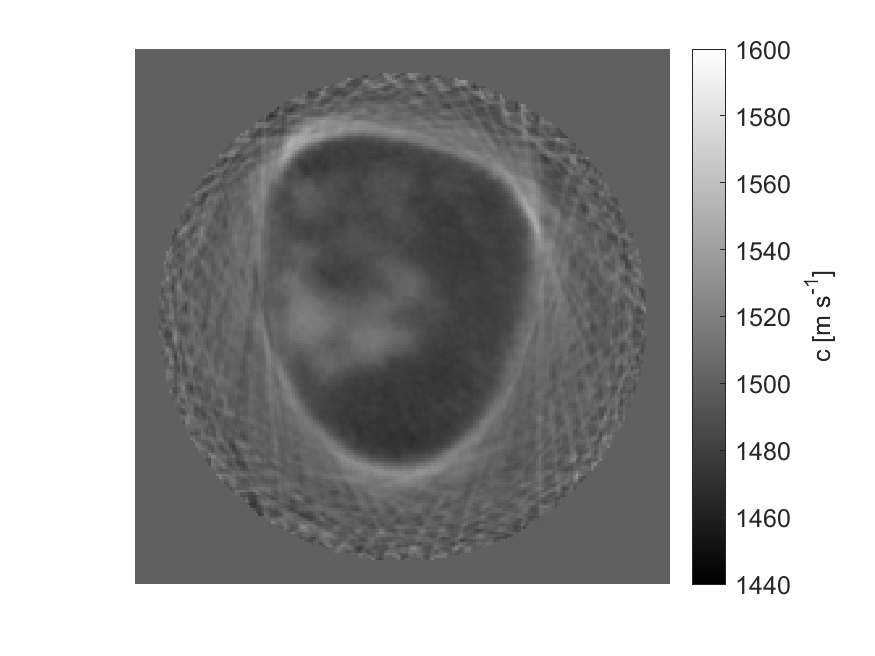}
    \label{fig:6b}  }\\
       \subfigure[]{\includegraphics[width=0.32\textwidth]{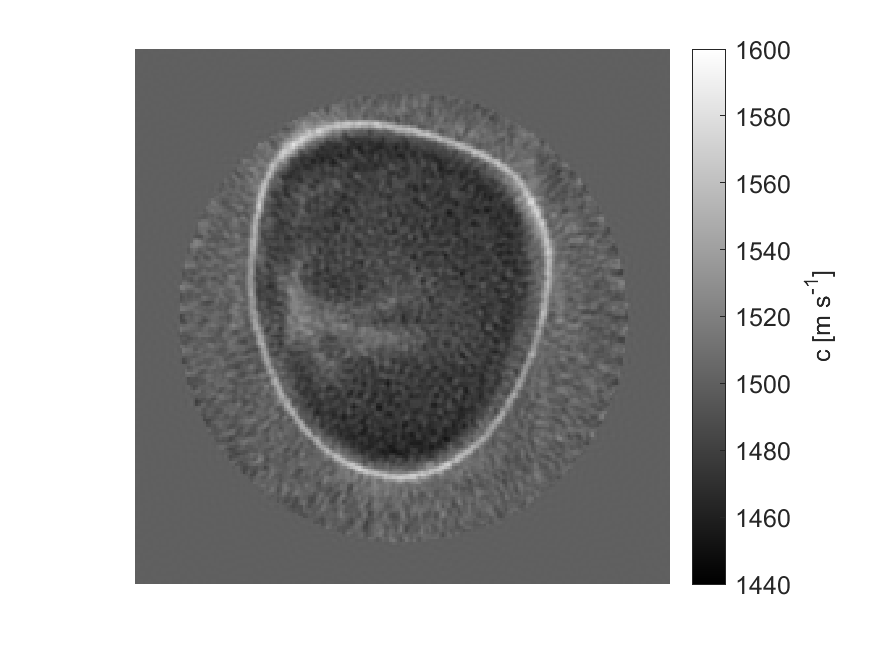}
    \label{fig:6c}  }
     \subfigure[]{\includegraphics[width=0.32\textwidth]{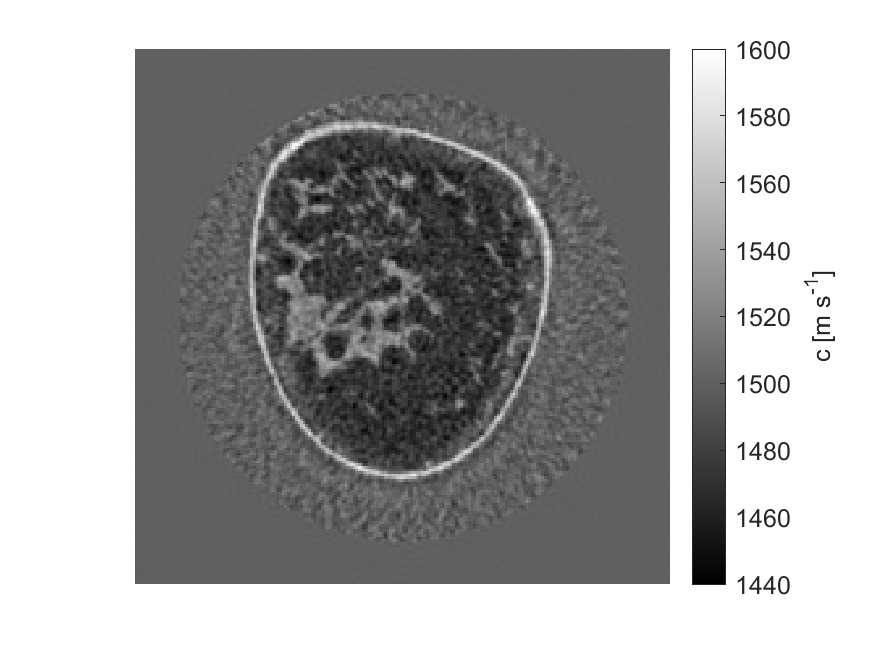}
    \label{fig:6d}  }
    \subfigure[]{\includegraphics[width=0.32\textwidth]{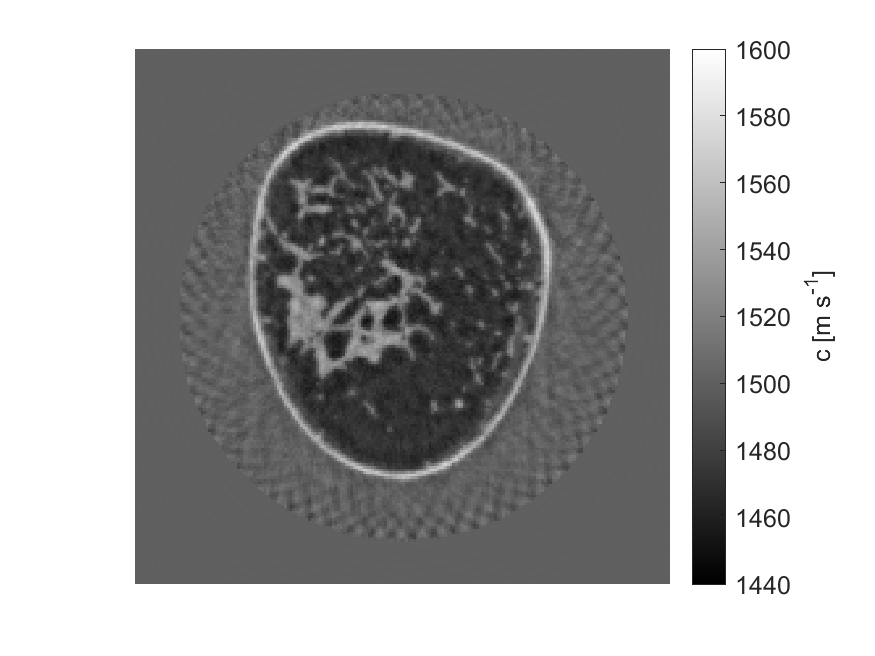}
    \label{fig:6e}  }\\
     \subfigure[]{\includegraphics[width=0.32\textwidth]{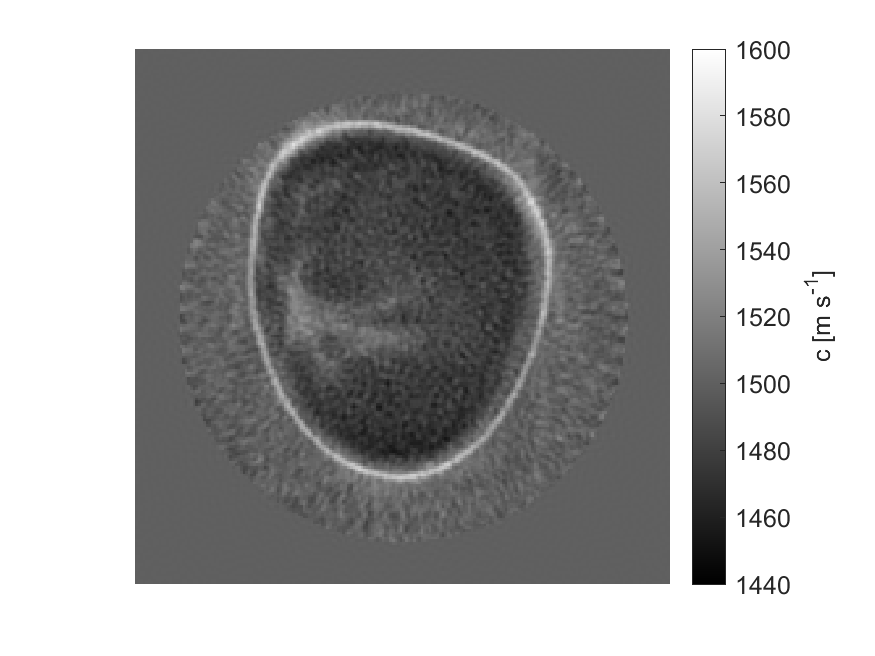}
    \label{fig:6f}  }
        \subfigure[]{\includegraphics[width=0.32\textwidth]{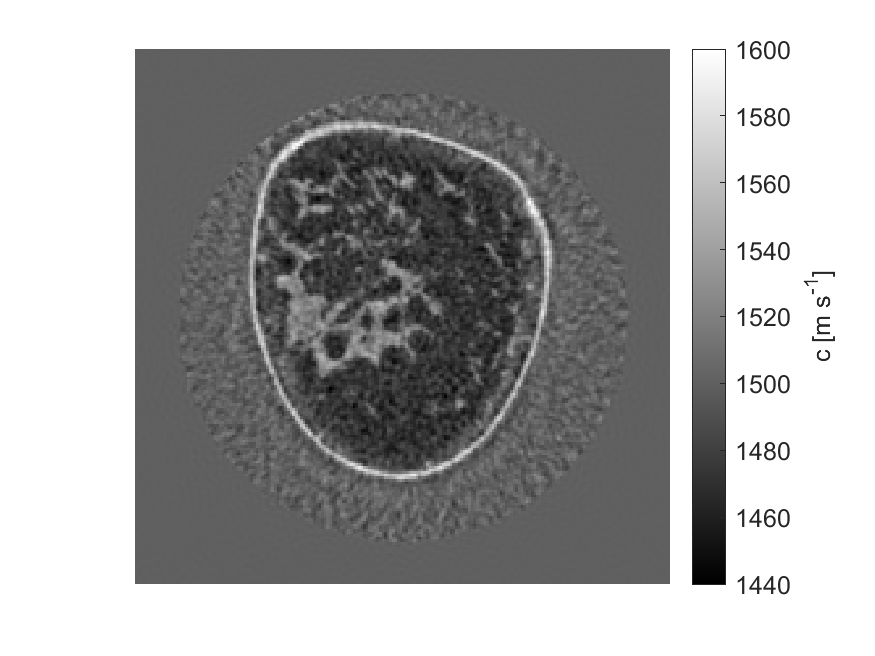}
    \label{fig:6g}  }
     \subfigure[]{\includegraphics[width=0.32\textwidth]{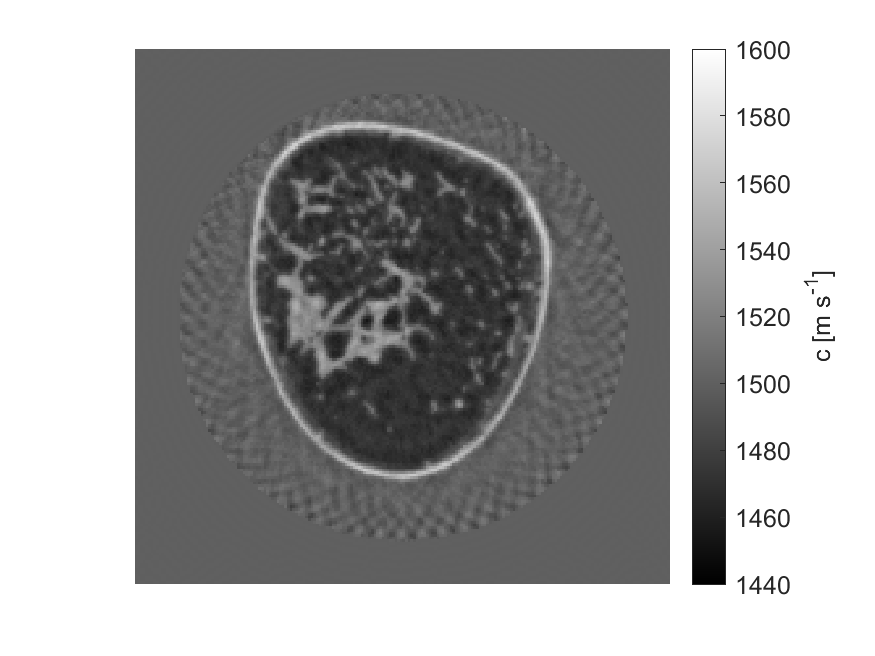}
    \label{fig:6h}  }\\
        \subfigure[]{\includegraphics[width=0.32\textwidth]{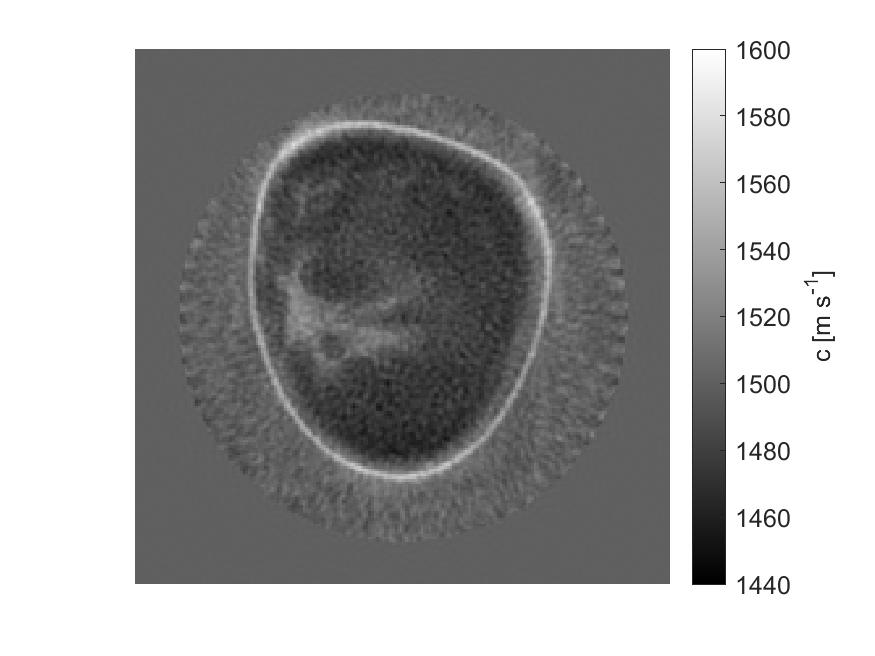}
    \label{fig:6i}  }
     \subfigure[]{\includegraphics[width=0.32\textwidth]{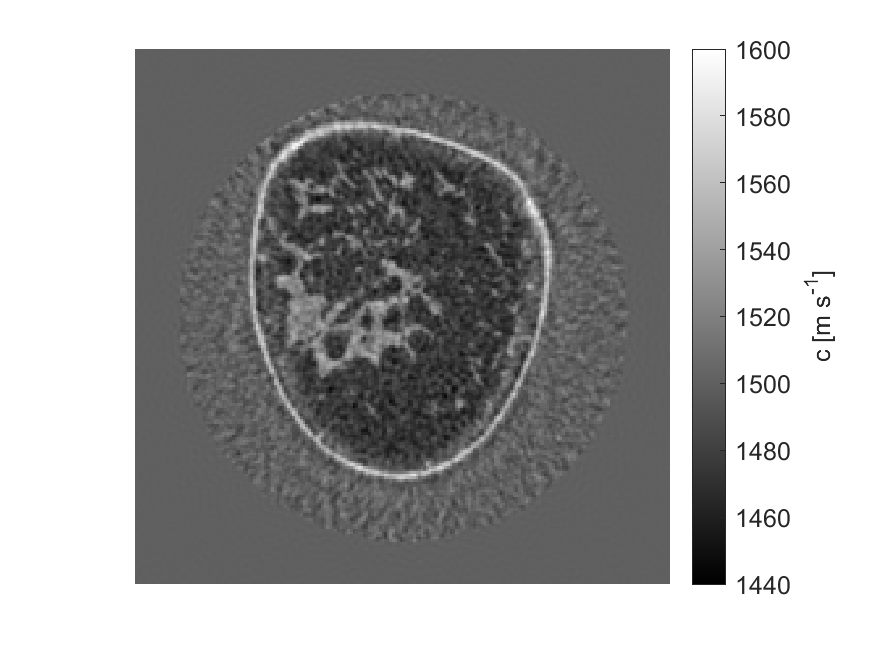}
    \label{fig:6j}  }
       \subfigure[]{\includegraphics[width=0.32\textwidth]{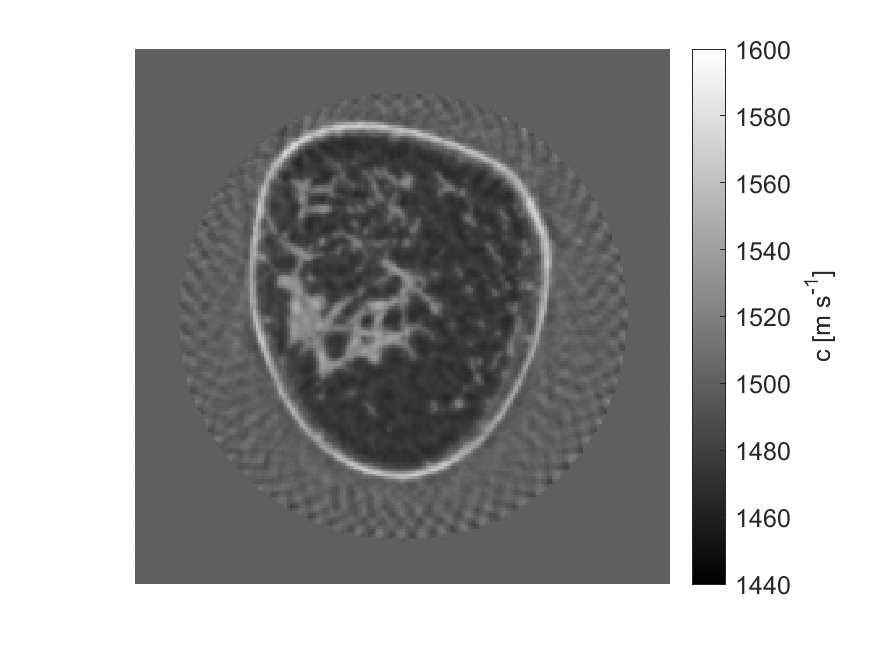}
    \label{fig:6k}  }
 \caption{(a) Ground truth. Reconstructed sound speed images from synthetic data with 25~dB SNR: (b) Time-of-flight-based approach (initial guess). Assuming the true $\alpha_0$ (Figure \ref{fig:1b}): (c) Born, (d) Hessian-based ray-Born, (e) Hessian-free ray-Born. Assuming $\alpha_0 = 0$: (f) Born, (g) Hessian-based ray-Born, (h) Hessian-free ray-Born. Assuming $\alpha_0 = 0.5 \ \mathrm{dB \ MHz}^{-y} \mathrm{cm}^{-1}$ (homogeneous within the breast): (i) Born, (j) Hessian-based ray-Born, (k) Hessian-free ray-Born.
     }\label{fig:reconstructed-images6}
\end{figure}

This section presents the sound speed images reconstructed from synthetic ultrasound data with 30~dB and 25~dB SNR, shown in figures \ref{fig:reconstructed-images5} and \ref{fig:reconstructed-images6}, respectively. Similar to the high SNR case, the reconstructions were performed using the three assumptions for $\alpha_0$, and the results are displayed in the same format as figure \ref{fig:reconstructed-images4}. The corresponding RE values are summarized in Table \ref{tab:RE}.

As observed in these figures, the images reconstructed using the Hessian-based inversion approach degraded more rapidly with decreasing SNR compared to those reconstructed using the Hessian-free approach, exhibiting more artifacts. This suggests that the Hessian-free method demonstrates improved robustness against variations in SNR. The RE values in Table \ref{tab:RE} further support this observation, showing that the Hessian-free approach maintains better stability under noise. 

For all these cases, the prototype Born inversion approach, which uses a water-only assumption to approximate the Green's functions, fails to reconstruct accurate images. One contributing factor to this degradation may be the avoidance of an inverse crime \cite{Wirgin}, as inherently different methods and grid spacings were used for simulating the synthetic data and performing the image reconstruction.

\subsection{Reconstructed sound speed profiles}
\begin{figure}
   \centering
\subfigure[]{\includegraphics[width=0.35\textwidth]{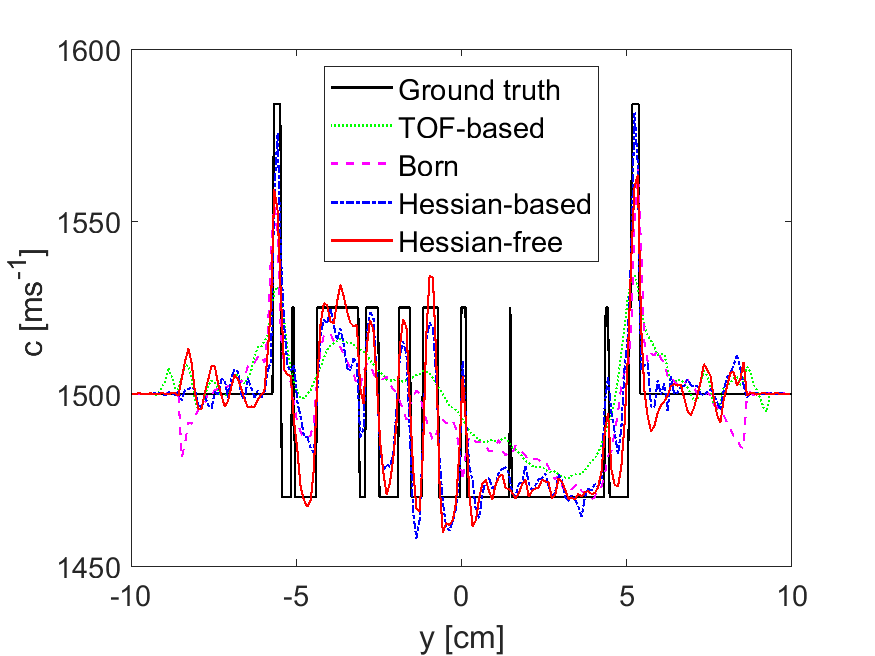}
\label{fig:7a} }
\subfigure[]{\includegraphics[width=0.35\textwidth]{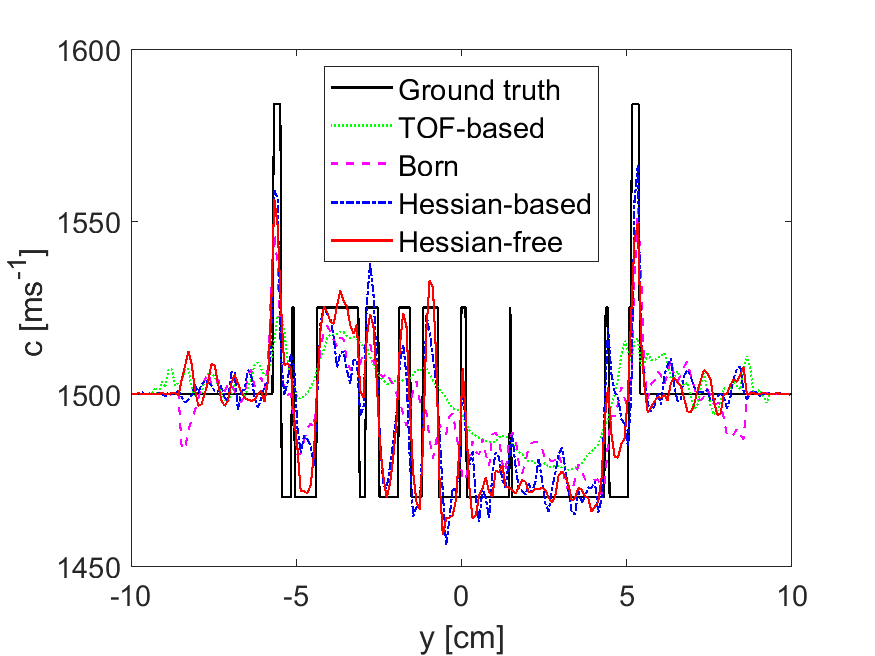}
\label{fig:7b} }
\subfigure[]{\includegraphics[width=0.35\textwidth]{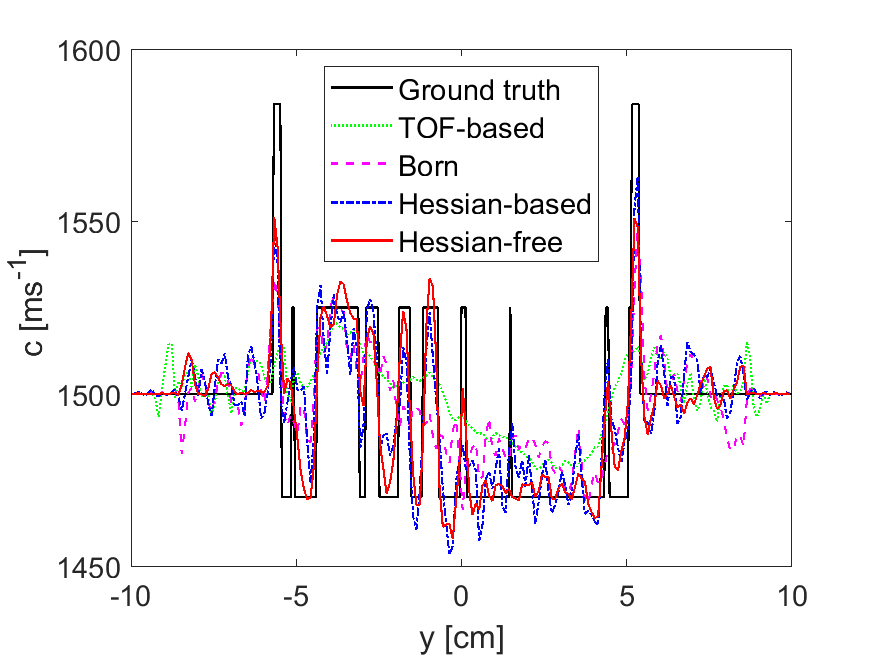}
\label{fig:7c}} \\
\subfigure[]{\includegraphics[width=0.35\textwidth]{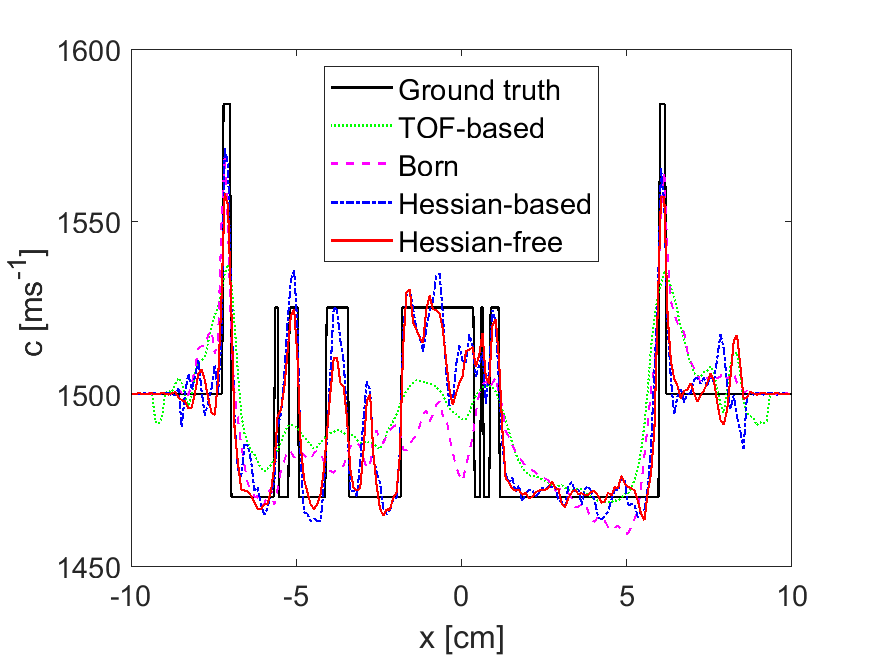}
\label{fig:7d} }
\subfigure[]{\includegraphics[width=0.35\textwidth]{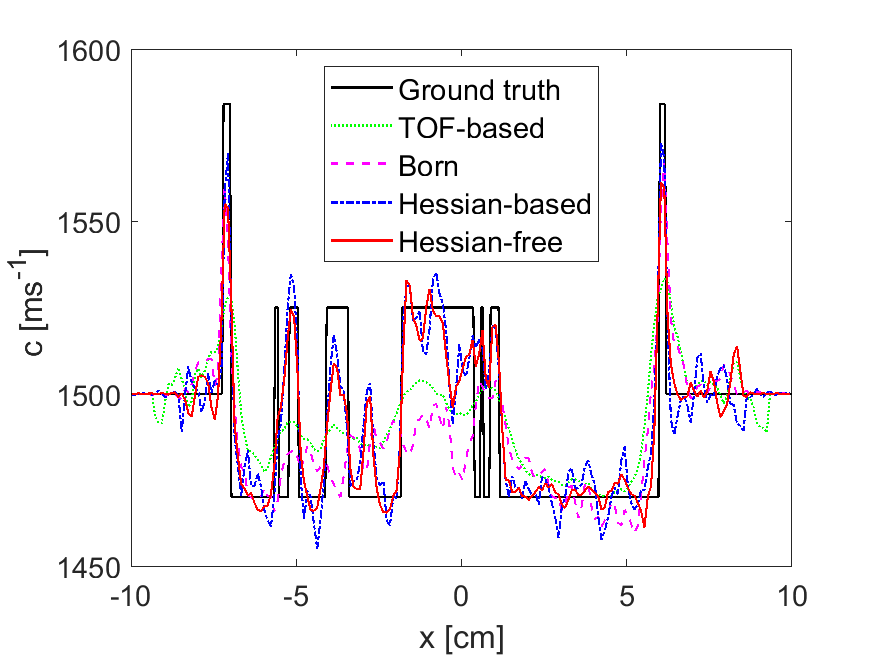}
\label{fig:7e} }
\subfigure[]{\includegraphics[width=0.35\textwidth]{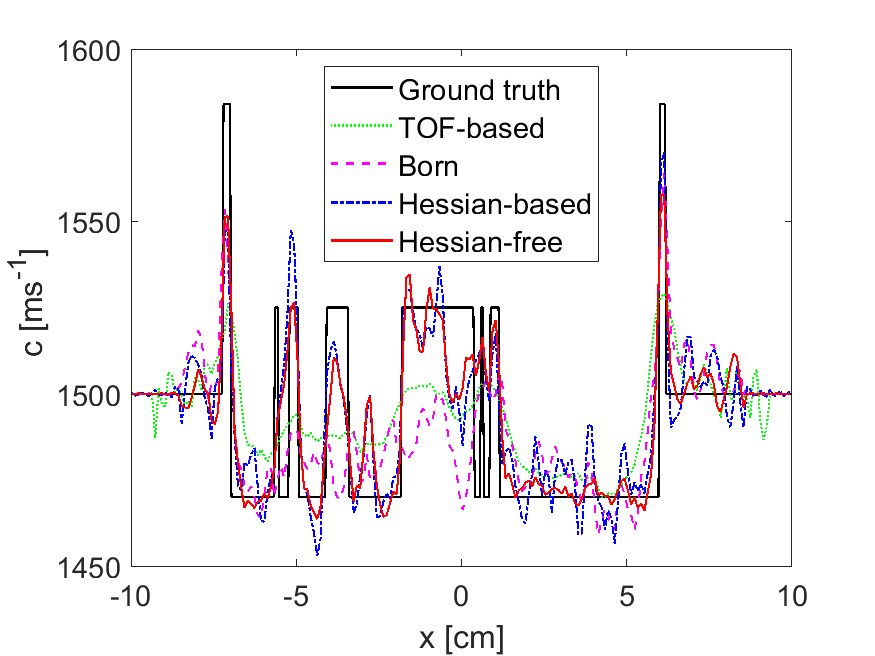}
\label{fig:7f}}   
 \end{figure}

Figures~\ref{fig:7a},~\ref{fig:7b}, and~\ref{fig:7c} show the reconstructed sound speed profiles along \(x = 0\) for ultrasound data with SNR values of 40~dB, 30~dB, and 25~dB, respectively. The true sound speed values for the digital breast phantom are plotted in black, while the profiles obtained using the TOF-based and Born inversion approaches are shown in green and magenta, respectively. Sound speed profiles reconstructed using the Hessian-based method are presented in blue, and those reconstructed using the proposed Hessian-free inversion approach are shown in red. To highlight the avoidance of inverse crime, the true phantom values are sampled on the grid used for data simulation rather than the reconstruction grid.

\begin{figure}
   \centering
\subfigure[]{\includegraphics[width=0.35\textwidth]{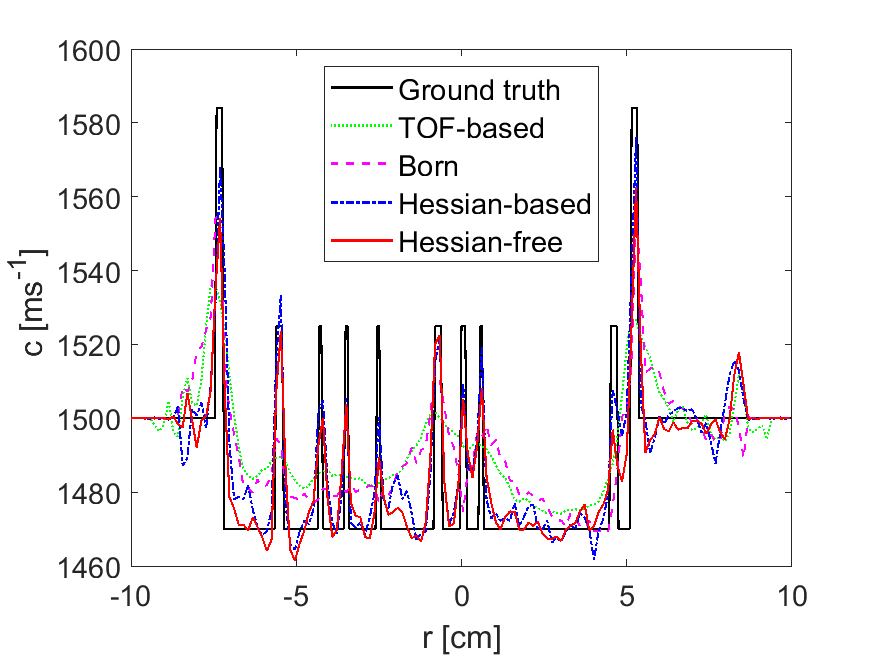}
\label{fig:7g} }
\subfigure[]{\includegraphics[width=0.35\textwidth]{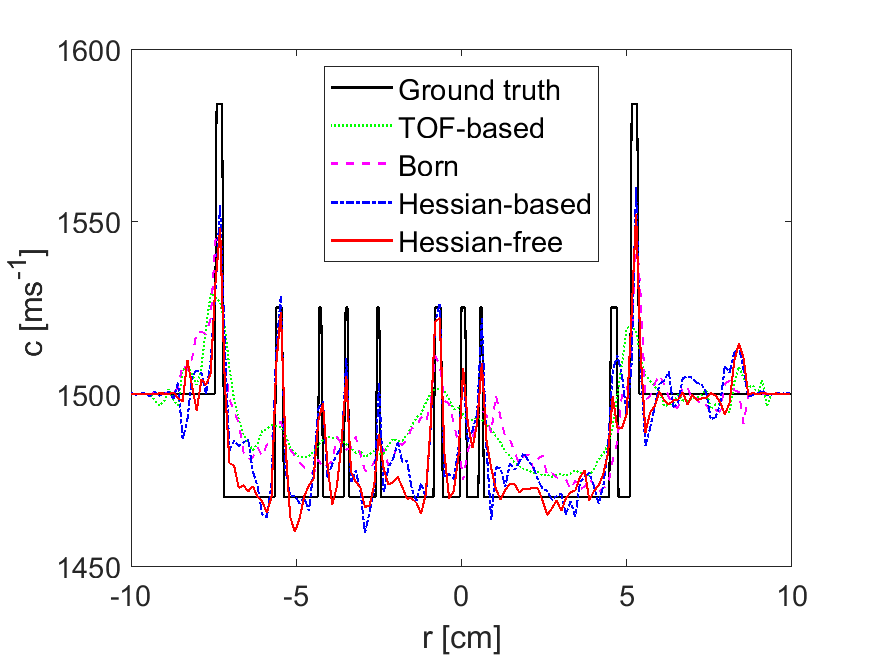}
\label{fig:7h} }
\subfigure[]{\includegraphics[width=0.35\textwidth]{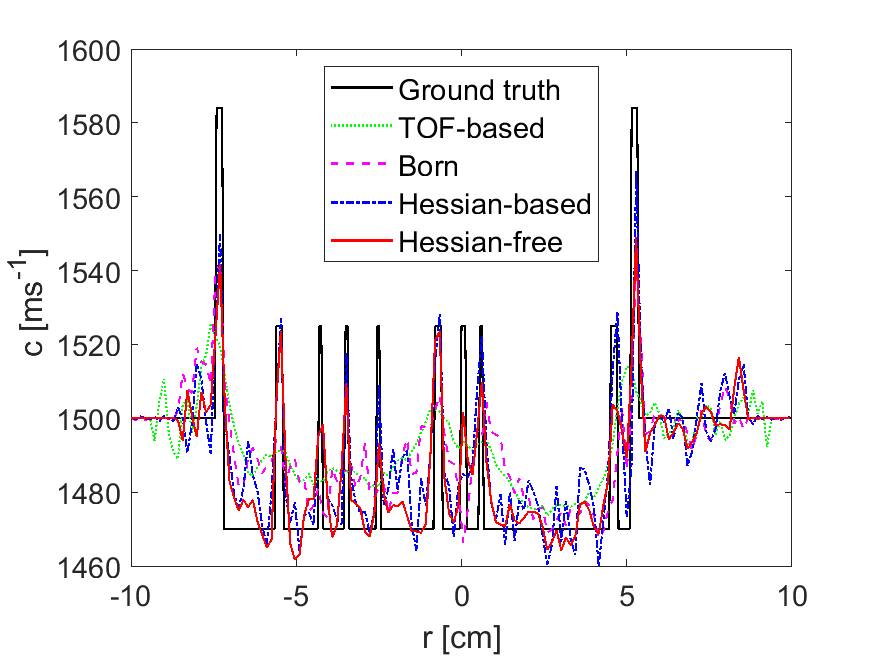}
\label{fig:7i}} \\
\subfigure[]{\includegraphics[width=0.35\textwidth]{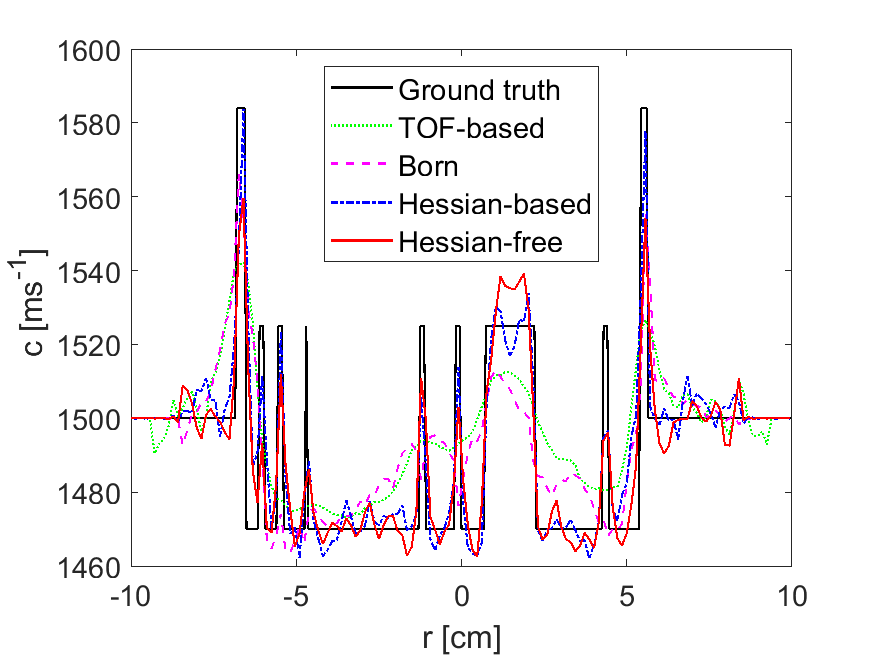}
\label{fig:7j} }
\subfigure[]{\includegraphics[width=0.35\textwidth]{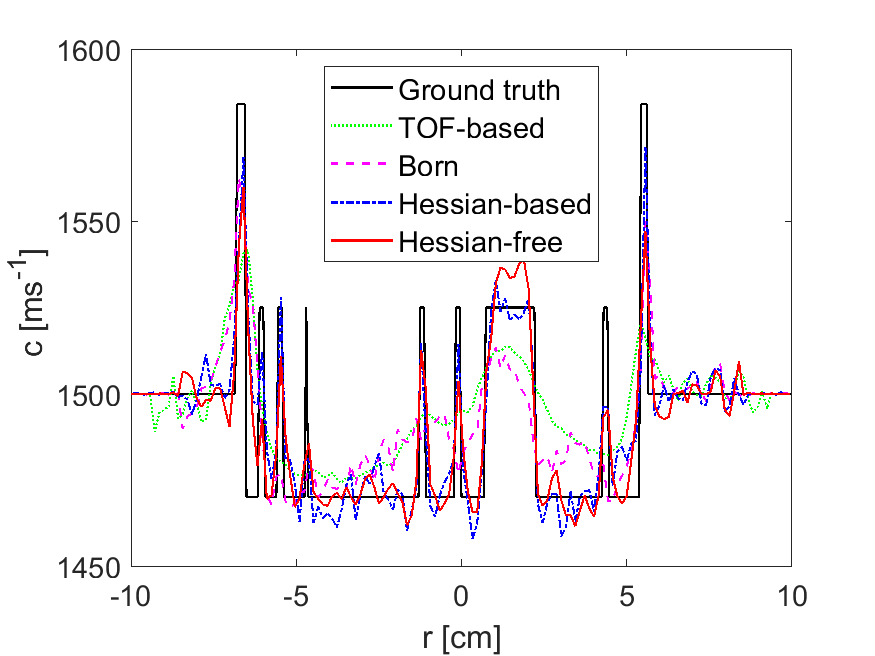}
\label{fig:7k} }
\subfigure[]{\includegraphics[width=0.35\textwidth]{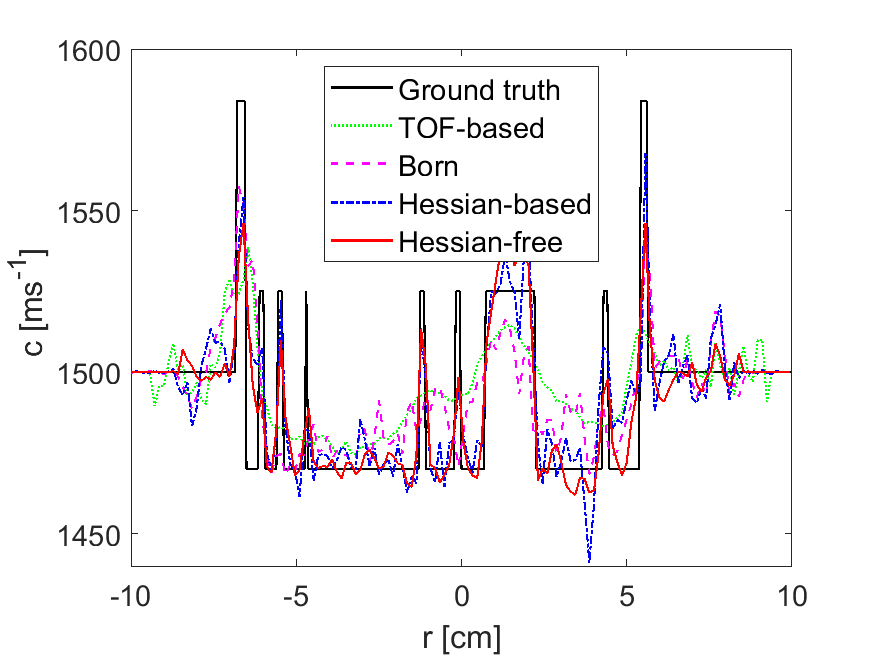}
\label{fig:7l}}  
   \caption{Sound speed profiles: along \(x = 0\) for synthetic data with SNR values of (a) 40~dB, (b) 30~dB, and (c) 25~dB,  along \(y = 0\) for synthetic data with SNR values of (d) 40~dB, (e) 30~dB, and (f) 25~dB, along the first diagonal for synthetic data with SNR values of (g) 40~dB, (h) 30~dB, and (i) 25~dB, along the second diagonal for synthetic data with SNR values of (j) 40~dB, (k) 30~dB, and (l) 25~dB.}
\end{figure}

Similarly, Figures~\ref{fig:7d},~\ref{fig:7e}, and~\ref{fig:7f} show the sound speed profiles along \(y = 0\); Figures~\ref{fig:7g},~\ref{fig:7h}, and~\ref{fig:7i} present the sound speed profiles along the first diagonal; and Figures~\ref{fig:7j},~\ref{fig:7k}, and~\ref{fig:7l} illustrate the sound speed profiles along the second diagonal of the sound speed phantom and reconstruction grids.

As seen in these figures, the ray-Born Hessian-based and Hessian-free approaches effectively reconstruct sharp sound speed variations associated with small anomalies, whereas the Born inversion approach shows little improvement over the time-of-flight-based reconstruction used as the initial guess.

Additionally, the Hessian-free approach demonstrates superior robustness to data noise compared with the Hessian-based approach. This improved robustness is attributed to the elimination of the ill-conditioning of the Hessian operator introduced by the approximations underlying its identity transformation (i.e., Eqs.~\eqref{eq:rayborn-approx1} and~\eqref{eq:rayborn-approx2}).

Furthermore, it is important to note that the computational cost of the proposed Hessian-free inversion approach is nearly an order of magnitude lower than that of the Hessian-based method, as each frequency subproblem is solved in a single step.

\clearpage

\section{Discussion and Conclusion} \label{sec:discussion}

This manuscript proposed a robust and computationally efficient approach for quantitative reconstruction of the sound speed from transmission ultrasound time series. It demonstrated the ability to generate high-resolution and quantitatively accurate images of the sound speed in soft tissues. The proposed image reconstruction method was validated using a realistic digital breast phantom \cite{Lou}. Similar to the inversion approach proposed in \cite{Javaherian2}, acoustic wave propagation was modeled using Green's formula, accounting for aberrations in the phase and amplitude of the Green's function caused by heterogeneity, refraction, geometrical spreading, and acoustic absorption and dispersion.
Correspondingly, amplitude losses due to geometrical spreading were approximated by computing the relative changes in the rays' Jacobian with respect to a reference point for each linked ray via solving the paraxial system of equations as described in Section \ref{sec:ray-tracing}. The proposed ray-based inversion method achieved high-resolution imaging of the breast phantom by including first-order scattered waves in the reconstruction. In practical settings, higher-order scattered waves are often attenuated due to acoustic absorption or buried in noise. 

The Hessian-free ray-Born inversion approach proposed in the present study and the Hessian-based ray-Born inversion proposed in \cite{Javaherian2} both showed significant improvements over a prototype Born inversion, which neglects medium heterogeneity in modeling the Green's function and accounts for it only in the scattering potential \cite{Devaney2,Devaney3,Beydoun,Borup,Mast0,Simonetti,Martiartu,Fan,Fauk,Guo}. As illustrated in Figure \ref{fig:3}, neglecting acoustic heterogeneity in approximating the Green's function results in substantial errors in phase and amplitude approximations.

The proposed approaches are computationally less expensive than full-waveform inversion methods, which utilize the complete pressure time-series data for image reconstruction \cite{Sandhu,Ali}. To further reduce the computational cost, this study introduced a Hessian-free inversion method that is approximately an order of magnitude less expensive than the Hessian-based approach proposed in \cite{Javaherian2}. This improvement is achieved by linearizing the objective function once over the entire frequency range and transforming the associated Hessian operator into the identity operator through a preconditioning scheme. The informative content of the Hessian operator is thereby transferred to the gradient, yielding a single-step full-frequency search direction, as derived in Eq.~\eqref{eq:sound_speed_update}. This search direction is evaluated recursively using a frequency-continuation scheme, progressing from low to high frequencies through the backprojection operator derived in Eq.~\eqref{eq:sound_speed_update2}.

The identity transformation of the Hessian operator relies on the approximations introduced in Eqs.~\eqref{eq:rayborn-approx1} and~\eqref{eq:rayborn-approx2}. As a result, the proposed formulation avoids the implicit inversion of the resulting ill-conditioned Hessian operator, thereby improving robustness to noise and producing reconstructed images with fewer artifacts than the Hessian-based approach for ultrasound data with SNR values of 30~dB and 25~dB. Furthermore, the Hessian-free inversion method exhibited reduced sensitivity to variations in the reconstruction parameters and the initial guess, further demonstrating its robustness.

The proposed forward model incorporates acoustic absorption and dispersion by solving a lossy Helmholtz equation based on Szabo's wave model \cite{Kelly,Szabo,Liebler}. Since absorption coefficient maps are typically unknown in practice, this study demonstrated that assuming a homogeneous absorption coefficient within the breast produced images comparable to those obtained using a true absorption map, whereas neglecting absorption entirely resulted in less accurate reconstructions.

{\rev{
The identity transformation of the Hessian operator relies on the assumption of an infinite frequency range in the two-dimensional Fourier integral representation of the Dirac delta distribution. However, because the transducers have an inherently band-limited response, the spectral support of \(|\btk|\) is finite in practice, leading to only an approximate evaluation of this Fourier integral. As shown in Appendix~D, the exact full-frequency preconditioned Hessian operator, which accounts for the finite bandwidth of the operating transducers, is a smoothing operator that converges to the Dirac delta distribution as the frequency coverage tends to infinity. A detailed analysis of the derived preconditioned Hessian operator is provided in Appendix~D. Consequently, replacing the exact finite-frequency Hessian operator by its infinite-frequency approximation eliminates the need for Hessian inversion and substantially improves reconstruction stability by reducing sensitivity to data noise, variations in the initial guess, and reconstruction parameters. The analysis presented in Appendix~D further suggests that employing broader-band transducers could improve reconstruction  accuracy, particularly when using the Hessian-free approach.}}

Considering that this study employed distinct methodologies for simulating synthetic ultrasound tomography (UT) data and for image reconstruction, the numerical results demonstrate the potential of the proposed framework for effective translation to practical two- and three-dimensional applications. Furthermore, our recent parallel study confirmed the successful adaptation of this approach to experimental settings, representing the first translation of acoustic inverse scattering methods to clinical applications \cite{Javaherian-experimental}

The primary motivation for developing this image reconstruction approach and the ray-based methods in \cite{Javaherian2,Javaherian1} lies in their potential application to volumetric reconstruction from 3D QUT data. Acoustic waves naturally propagate in three dimensions rather than being confined to a horizontal plane. While full-waveform approaches offer numerous well-known advantages, they face limitations in terms of computational cost and the ability to handle transducer directivity in specific 3D configurations, particularly when transducers are of finite size. These challenges can be addressed more effectively using ray-based methods.

\section*{Data and Code Availability Statement}
The MATLAB codes supporting the findings of this study, as well as those reported in \cite{Javaherian2} and \cite{Javaherian1} are publicly accessible via the GitHub repository referenced in \cite{Javaherian-toolbox}. An introduction to this GitHub repository, along with numerical experiments evaluating the accuracy of the ray-tracing algorithms provided in the toolbox, has been presented in \cite{Javaherian-validation}.

The original simulated ultrasound data \cite{Treeby}, used as a benchmark in this study, can be reproduced using the example scripts provided in \cite{Javaherian-toolbox} or downloaded directly from the Zenodo repository referenced in \cite{Ashkan-zenodo}. Users should add the downloaded files to the appropriate paths to utilize the data.

\section*{Applicability}
Results demonstrating the implementation of the image-reconstruction approach proposed in this study on in-vitro and in-vivo ultrasound datasets released by the University of Rochester Medical Center are reported in \cite{Javaherian-experimental} and are also included in the GitHub repository \cite{Javaherian-toolbox}. The datasets themselves can be accessed via the link provided in \cite{Ali2} and should be added to the appropriate path in the toolbox \cite{Javaherian-toolbox}.

\section*{Declaration of Interests}
The author declares no known competing financial interests or personal relationships that could have influenced the work reported in this paper.

\section*{Acknowledgments}
This work was predominantly conducted at the Department of Medical Physics and Biomedical Engineering, University College London, UK, during the author’s employment there. The project was partially funded by the UK EPSRC Grant under Project Reference: EP/T014369/1. Subsequent refinements and revisions were completed at the University of Tehran.

\section*{Appendix A: Eikonal and Transport equations}

This appendix outlines the derivation of the Eikonal equation \eqref{eq:eikonal} and the Transport equation \eqref{eq:transport}. Substituting the Green's function \eqref{eq:2D_greens} into the wave equation \eqref{eq:greens-wave-equation}, along with the complex wavenumber defined by \eqref{eq:wave-number}, yields the following system of equations:
\begin{align}
    \begin{split}
        k^2 A + \nabla^2 A - A \nabla \phi \cdot \nabla \phi &= 0, \\
        2k \alpha A + 2 \nabla A \cdot \nabla \phi + A \nabla^2 \phi &= 0,
    \end{split}
\end{align}
where the first equation corresponds to the real part, and the second equation corresponds to the imaginary part \cite{Javaherian2}. Here, we have used the approximation $\tilde{k}^2 \approx k^2 + 2 \alpha k \iu$. 

Applying the high-frequency approximation, $|\nabla^2 A / A| \ll k^2$, the first equation simplifies to the dispersive Eikonal equation \eqref{eq:eikonal}, which determines the rays' trajectories as tangents to the complex wavevector $\bk = \nabla \phi$. For the second equation, introducing $A = A_{abs} A_{geom}$, where $A_{abs}$ satisfies \eqref{eq:absorption}, leads to:
\begin{align} \label{eq:transport2}
    2k \alpha A_{geom} + 2 \nabla A_{geom} \cdot \nabla \phi - 2 A_{geom} \frac{\alpha}{k} \bk \cdot \nabla \phi + A_{geom} \nabla^2 \phi = 0,
\end{align}
where the relation $\nabla A_{abs} = -\left[\alpha /k\right] \bk A_{abs} $ has been applied \cite{Javaherian2}. Substituting $\bk = \nabla \phi$ into the above equation yields the Transport equation \eqref{eq:transport}.

\section*{Appendix B: Homogeneous Green's function}
This appendix presents the Green's functions for homogeneous media. For a 2D homogeneous medium, the Green's function is given as \cite{Javaherian2}:
\begin{align}  \label{greens-2d}
g_{0, 2D}(\omega, \bx, \bx') \approx \big[ 8 \pi \phi_0(\omega, \bx, \bx') \big]^{-1/2} 
\exp \Big( \iu \big[ \phi_0(\omega, \bx, \bx') + \pi/4 \big] \Big).
\end{align}

\noindent
For a 3D homogeneous medium, the Green's function is expressed as \cite{Javaherian2}:
\begin{align}  \label{greens-3d}
g_{0, 3D}(\omega, \bx, \bx') = \big[ 4 \pi | \bx - \bx' | \big]^{-1} 
\exp \Big( \iu \phi_0(\omega, \bx, \bx') \Big).
\end{align}

\noindent
Here, $ \phi_0(\omega, \bx, \bx') = k_0 | \bx - \bx' |$ represents the accumulated phase, with $k_0 = \omega/c_0$ denoting the wavenumber in water.

\section*{Appendix C: Conditions relying approximate formulas~\eqref{eq:rayborn-approx1} and~\eqref{eq:rayborn-approx2} }

This section provides further details on the two approximate formulas,
Eqs.~\eqref{eq:rayborn-approx1} and~\eqref{eq:rayborn-approx2}. In particular,
we derive the conditions under which these approximations are valid.

We begin with the approximation in Eq.~\eqref{eq:rayborn-approx2}, which
assumes that, for every point $\bx$ in a neighborhood of $\bx'$, the two-way
accumulated phase can be approximated by its first-order Taylor expansion about
$\bx'$, while neglecting the second- and higher-order terms. To facilitate the
analysis, we express the gradients $\nabla k$, $\nabla \theta$, and
$\nabla \zeta$ in polar coordinates by the pairs
$(|\nabla k|,\varphi_k)$, $(|\nabla \theta|,\varphi_{\theta})$, and
$(|\nabla \zeta|,\varphi_{\zeta})$, respectively, as
\begin{align}   \label{eq:polar-coordinates-2}
\nabla k
=
|\nabla k|
\begin{bmatrix}
\cos(\varphi_k)\\
\sin(\varphi_k)
\end{bmatrix},
\qquad
\nabla \theta
=
|\nabla \theta|
\begin{bmatrix}
\cos(\varphi_{\theta})\\
\sin(\varphi_{\theta})
\end{bmatrix},
\qquad
\nabla \zeta
=
|\nabla \zeta|
\begin{bmatrix}
\cos(\varphi_{\zeta})\\
\sin(\varphi_{\zeta})
\end{bmatrix}.
\end{align}

To quantify the approximation error, we retain the leading neglected term, namely the second-order term, in Eq.~\eqref{eq:rayborn-approx2}, yielding
\begin{align} \label{eq:phi-second-order}
 \Phi +\Delta \Phi
 =
 \btk \cdot \bx_{\fd}
 +\frac{1}{2}\bx_{\fd}^{T}\nabla\btk\,\bx_{\fd}
 +O(x_{\fd}^{3}),
\end{align}
where the arguments have been omitted for brevity. Here, $\Delta \Phi$ denotes the error in the approximated phase $\Phi$ due to neglecting the second- and higher-order terms. Substituting the polar-coordinate representations in Eqs.~\eqref{eq:polar-coordinates-1} and~\eqref{eq:polar-coordinates-2}, together with Eq.~\eqref{eq:btk-amp-angle}, into Eq.~\eqref{eq:phi-second-order} and performing straightforward algebra yields
\begin{align} \label{eq:phi-second-order-polar}
\Phi + \Delta \Phi
=
\pm x_{\fd}
\Big[
G\cos(\zeta-\phi)
-
F\sin(\zeta-\phi)
\Big],
\end{align}
where the upper and lower signs correspond to
$0 \leq \theta \leq \pi$ and
$\pi < \theta \leq 2\pi$, respectively, reflecting the sign of
$\cos(\theta/2)$. Here,
\begin{align} \label{eq:A}
G
=
G_1\cos\!\left(\frac{\theta}{2}\right)
-
G_2\sin\!\left(\frac{\theta}{2}\right),
\end{align}
with
\begin{align} \label{eq:A12}
\begin{split}
G_1
&=
2k
+
x_{\fd}
|\nabla k|
\cos(\varphi_k-\phi),
\\
G_2
&=
x_{\fd}
\frac{k}{2}
|\nabla\theta|
\cos(\varphi_{\theta}-\phi),
\end{split}
\end{align}
and
\begin{align} \label{eq:B}
F
=
x_{\fd}
k
\cos\!\left(\frac{\theta}{2}\right)
|\nabla\zeta|
\cos(\varphi_{\zeta}-\phi).
\end{align}

Equation~\eqref{eq:A} can be rewritten as
\begin{align}   \label{eq:A1}
G
=
\left(G_1^2+G_2^2\right)^{\frac{1}{2}}
\cos\!\left(
\frac{\theta}{2}
+
\varphi_{\delta_1}
\right),
\qquad
\varphi_{\delta_1}
=
\operatorname{atan2}(G_2,G_1).
\end{align}

Similarly, Eq.~\eqref{eq:phi-second-order-polar} can be rewritten as
\begin{align}  \label{eq:phi-second-order-final}
\Phi+\Delta \Phi
=
\pm x_{\fd}
\left(G^{2}+F^{2}\right)^{\frac{1}{2}}
\cos\!\left(
\zeta-\phi+\varphi_{\delta}
\right),
\qquad
\varphi_{\delta}
=
\operatorname{atan2}(F,G),
\end{align}
where
\begin{align}   \label{eq:phi-second-order-dominator}
G^{2}+F^{2}
={}&
\Big[
4k^{2}
+
4x_{\fd}k|\nabla k|
\cos(\varphi_k-\phi)
+
x_{\fd}^{2}
|\nabla k|^{2}
\cos^{2}(\varphi_k-\phi)
\nonumber\\
&
+
\frac14
x_{\fd}^{2}
k^{2}
|\nabla\theta|^{2}
\cos^{2}(\varphi_{\theta}-\phi)
\Big]
\cos^{2}\!\left(
\frac{\theta}{2}
+
\varphi_{\delta_1}
\right)
\nonumber\\
&
+
x_{\fd}^{2}
k^{2}
|\nabla\zeta|^{2}
\cos^{2}(\varphi_{\zeta}-\phi)
\cos^{2}\!\left(
\frac{\theta}{2}
\right).
\end{align}

Under the conditions
\begin{align}
\begin{split}
x_{\fd} \leq 2R
&\ll
\frac{k}
{|\nabla k| |\cos(\phi-\varphi_k)|},
\\
x_{\fd} \leq 2R
&\ll
\frac{4}{|\nabla\theta|
|\cos(\phi-\varphi_{\theta})|},
\end{split}
\end{align}
the last three terms in the bracket of Eq.~\eqref{eq:phi-second-order-dominator} become negligible compared with the leading term \(4k^2\). Consequently,
\(G_1 \approx 2k > 0 \) and \(|G_2| \ll G_1\), implying
\(\varphi_{\delta_1}=\operatorname{atan2}(G_2,G_1)\approx0\), and yielding $G \approx  2k \cos(\tfrac{\theta}{2})$. Here, \(R\) denotes the radius of the circular array on which the emitters and receivers are located.

Furthermore, if
\begin{align}
x_{\fd}
\leq
2R
\ll
\frac{2}{|\nabla\zeta| |\cos(\phi-\varphi_{\zeta})|},
\end{align}
the last term in Eq.~\eqref{eq:phi-second-order-dominator} is also negligible, yielding
\(|F| \ll |G| \) and therefore 
\begin{align}
\begin{split}
\varphi_{\delta}=\operatorname{atan2}(F,G)\approx
\begin{cases}
0,   &0 \leq \theta < \pi \\
\pi, & \pi < \theta \leq 2\pi
\end{cases}.
\end{split}
\end{align}

Consequently, Eq.~\eqref{eq:phi-second-order-final} reduces to the approximate phase expression in Eq.~\eqref{eq:rayborn-approx2},
\begin{align}
\Phi
\approx
2k
\left|  \cos\!\left(\frac{\theta}{2}\right) \right|
x_{\fd} \cos(\zeta-\phi).
\end{align}

Additionally, for deriving the conditions under which the approximation in
Eq.~\eqref{eq:rayborn-approx1} is valid, we introduce the following
polar representation of the gradient of the real-valued amplitude:
\begin{align}   \label{eq:polar-coordinates-A}
\nabla A
=
|\nabla A|
\begin{bmatrix}
\cos(\varphi_A)\\
\sin(\varphi_A)
\end{bmatrix},
\end{align}
where $\varphi_A=\operatorname{atan2}(\nabla_yA,\nabla_xA)$ denotes the
orientation of $\nabla A$.

Accordingly, for every fixed point $\bx'$ and every point in the data
space $(\omega,e,r)$, the approximation in
Eq.~\eqref{eq:rayborn-approx1} holds provided that
\begin{align}\label{eq:f2}
x_{\fd}
&\leq
2R
\ll
\frac{A(\omega,\bx',e)}
{|\nabla A(\omega,\bx',e)|\,|\cos(\phi-\varphi_{A}(\omega,\bx',e))|},
\nonumber\\
x_{\fd}
&\leq
2R
\ll
\frac{A(\omega,\bx',r)}
{|\nabla A(\omega,\bx',r)|\,|\cos(\phi-\varphi_{A}(\omega,\bx',r))|},
\nonumber\\
x_{\fd}
&\leq
2R
\ll
\frac{|\Upsilon_m(\omega,\bx')|}
{|\nabla\Upsilon_m(\omega,\bx')|}.
\end{align}

{\rev{
\section*{Appendix D: A frequency analysis of the preconditioned Hessian}
\label{sec:exact-weighted-hessian}

The identity transformation of the Hessian operator in Eq.~\eqref{eq:hessian-diagonalised} was derived under the assumption of an infinite frequency range, whereas practical ultrasound transducers operate over a finite frequency bandwidth. In this section, the effect of a finite bandwidth on the derived preconditioned Hessian operator is investigated. To this end, the wavenumber is assumed to be bounded within the finite interval \([0,K(\bx)]\), where \(K(\bx)\) is a spatially varying upper bound.

Recall the polar coordinates represenations in Eq.~\eqref{eq:polar-coordinates-1}. The vector \( \bx_{\fd} = \bx - \bx' \) is represented by the pair \( (x_{\fd}, \phi) \). From Eq.~\eqref{eq:btk-amp-angle}, we also use \( \big(2k  |\cos(\tfrac{\theta}{2})|, \zeta \big) \) as the polar representation of \( \bar{\mathbf{k}} \). For the analysis that follows, we use \(J_0\) and \(J_1\) to denote the Bessel functions of the first kind of orders zero and one, respectively  \cite{Abramowitz}.

The Hessian matrix can then be written in the form of a two-dimensional Fourier integral representation:
\begin{align}
H(\bx,\bx') = \frac{1}{(2 \pi)^2}
\int_{0}^{2 \pi} d \theta
\int_{0}^{2K \cos(\tfrac{\theta}{2})} d |\bar{\mathbf{k}}| \, |\bar{\mathbf{k}}|
\int_{0}^{2 \pi} d \zeta \,
e^{-\iu  |\bar{\mathbf{k}}| x_{\fd} \cos(\zeta - \phi)}.
\end{align}
Now, introducing the variable
\begin{align} \label{eq:a-theta}
a(\theta) = 2  \left| \cos(\tfrac{\theta}{2})\right|,
\end{align}
satisfying $ |\bar{\mathbf{k}}|=ka $, and using the identity \cite{Abramowitz}
\begin{align} \label{eq:zeta-integral1}
\frac{1}{2 \pi} \int_{0}^{2 \pi} d \zeta \, e^{-\iu a k x_{\fd} \cos(\zeta - \phi)} = J_0(a k  x_{\fd}),
\end{align}
and performing the change of variables \( |\bar{\mathbf{k}}| \rightarrow k \), we obtain
\begin{align} \label{eq:hessian-exact1}
H(\bx,\bx') = \frac{1}{2 \pi}
\int_{0}^{K} dk \, k
\int_{0}^{2 \pi} d \theta \, a(\theta)^2 \, J_0\!\big(a(\theta) k x_{\fd}\big).
\end{align}

Using the definition of \( a \) in Eq. \eqref{eq:a-theta}, the integral over \( \theta \) becomes
\begin{align}   \label{eq:theta-integral-1}
\int_{0}^{2 \pi} d \theta \, 4 \cos^2\!\big(\tfrac{\theta}{2}\big)\,
J_0\!\big(2  k x_{\fd} \cos(\tfrac{\theta}{2})\big), 
\end{align}
where we have used the fact $J_0$ is an even function. Applying the replacements $4 \cos^2{(\frac{\theta}{2})} = 2 \left[ 1+ cos (\theta) \right]$ and $u= \frac{\theta}{2}$ to this integral then yields
\begin{align}   \label{eq:theta-integral-2}
 \int_0^{\pi} du \, 4 \big[1+ \cos{(2 u)} \big] J_0 \Big( 2 kx_{\fd} \cos{(u)}\Big) .
\end{align}
Using the well-known identity \cite{Gradshteyn}
\begin{align} \label{eq:bessel-relation1}
\int_0^{\pi} du \, \cos{(2nu)}J_0 \Big(2 z \cos{(u)}\Big)  = (-1)^n \pi J_n^2 (z),
\end{align}
Eq. \eqref{eq:hessian-exact1} collapses to
\begin{align} \label{eq:hessian-exact2}
H(\bx,\bx') = \int_0^{K} d  k  \,  2 k  \,  \Big[    J_0^2(kx_{\fd}) - J_1^2(kx_{\fd}) \Big].
\end{align}

\subsubsection*{Preconditioned Hessian at single frequency}
\label{sec:preconditioned-Hessian-single-frequency}

In \cite{Javaherian2}, the objective function defined in Eq.~\eqref{eq:objective-function} was linearized and minimized sequentially over consecutive frequency intervals (two frequencies per linearization step) using a low-to-high frequency continuation scheme.

Below, we study the asymptotic behaviour of $H$, derived in Eq.~\eqref{eq:hessian-exact2}, at a single frequency $k$. 
\begin{enumerate}
\item In the limits $z \to 0 $, the asymptotic expansions of the Bessel functions yield \cite{Abramowitz}
\begin{align} \label{eq:bessel-series}
\begin{split}
J_0(z) = 1 - \frac{1}{4}z^2 + \frac{1}{64}z^4 + O\Big( z^6 \Big), \\
J_1(z) = \frac{1}{2}z - \frac{1}{16}z^3 + O\Big( z^5 \Big),
\end{split}
\end{align}
which yields the following approximation for $kx_{\fd} \ll 1$:
\begin{align} \label{eq:krll1}
2 k  \,  \Big[    J_0^2(kx_{\fd}) - J_1^2(kx_{\fd})  \Big] \approx  2k \Big[  1- \frac{3}{4} (kx_{\fd})^2 +\frac{5}{32} (kx_{\fd})^4 \Big]
\end{align}

\item In the limits $z  \to \infty $, the Bessel function of the first kind with fixed order $\nu$ can be approximated as \cite{Abramowitz}
\begin{align} \label{eq:bessel-z-inf}
J_{\nu}(z) \approx \big( \frac{2}{\pi z}\big)^{1/2}\cos{\big( z - \frac{\nu \pi}{2}-\frac{\pi}{4}\big) },
\end{align}
which yields the following approximation for $kx_{\fd} \gg 1$:
\begin{align}   \label{eq:krgg1}
2 k  \,  \Big[    J_0^2(kx_{\fd}) - J_1^2(kx_{\fd})  \Big] \approx  \frac{4}{\pi x_{\fd}} \sin{\big( 2kx_{\fd} \big) }
\end{align}
\end{enumerate}

Correspondingly, for the $kx_{\fd} \ll 1$ regime, the main lobe becomes progressively narrower as $k$ increases, while its peak amplitude grows linearly with $k$. In contrast, for the $kx_{\fd} \gg 1$ regime, Eq.~\eqref{eq:krgg1} exhibits an oscillatory behavior whose amplitude is independent of $k$ and decays as $1/x_{\fd}$, while the oscillation frequency increases linearly with $kx_{\fd}$. However, as shown below, when the objective function in Eq.~\eqref{eq:objective-function} is linearized and minimized over the entire frequency bandwidth of the ultrasound transducers in a single step, the corresponding preconditioned Hessian operator approaches a two-dimensional Dirac delta distribution and, consequently, the identity operator.

\subsubsection*{Preconditioned Hessian over a finite frequency band}
Below, we investigate the behaviour of the preconditioned Hessian for a finite frequency bandwidth, i.e., we assume that \(K\) in Eq.~\eqref{eq:hessian-exact2} is finite. Using the identity
\begin{align} \label{eq:bessel-derivative-identity}
\frac{d}{dz} \Big[ z J_{\nu}(z) J_{\nu+1}(z) \Big]
= z \Big[ J_{\nu}^2(z) - J_{\nu+1}^2(z) \Big],
\end{align}
and performing the change of variables \( k \to \tilde{k} \), with \( \tilde{k} = kx_{\fd} \), the integral over \( k \) in Eq.~\eqref{eq:hessian-exact2} yields
\begin{align}  \label{eq:hessian-exact3}
H(\bx,\bx')
= \frac{2}{x_{\fd}^2} \int_{0}^{Kx_{\fd}} d\tilde{k} \, \tilde{k} \, \Big[ J_0^2(\tilde{k}) - J_1^2(\tilde{k}) \Big]
= \frac{2K}{x_{\fd}} J_0(Kx_{\fd}) J_1(Kx_{\fd}).
\end{align}

Note that, in deriving the identity \eqref{eq:bessel-derivative-identity}, we have used the relations \cite{Abramowitz}
\begin{align} \label{eq:bessel-relation2}
\begin{split}
\frac{d}{dz} J_{\nu}(z)
&= J_{\nu-1}(z) - \frac{\nu}{z} J_{\nu}(z), \\
J_{\nu-1}(z)
&= \frac{2\nu}{z} J_{\nu}(z) - J_{\nu+1}(z).
\end{split}
\end{align}

Applying the same asymptotic analysis as above to the preconditioned Hessian operator derived in Eq.~\eqref{eq:hessian-exact3} yields
\begin{align}
\label{eq:hessian-exact4}
H(\bx,\bx')
\approx
\begin{cases}
K^2
\left[
1-\dfrac{3}{8}(Kx_{\fd})^2
\right],
& Kx_{\fd}\ll1, \\[2mm]
-\dfrac{2}{\pi x_{\fd}^2}
\cos\!\left(2Kx_{\fd}\right),
& Kx_{\fd}\gg1.
\end{cases}
\end{align}

The first asymptotic expression shows that the amplitude of the main
lobe increases quadratically with the maximum wavenumber \(K\), while
its width decreases proportionally to \(1/K\).
Additionally, the second asymptotic expression shows that the
oscillatory sidelobes decay as \(1/x_{\fd}^{2}\). Consequently, the off-diagonal elements of the preconditioned Hessian operator are strongly attenuated relative to its diagonal component, causing it to approach a two-dimensional Dirac delta distribution and, consequently, to act as the identity operator.

\subsubsection*{Preconditioned Hessian over an infinite frequency band}

The Hessian operator given by Eq.~\eqref{eq:hessian-exact3} was derived under the assumption that \(K\) is finite. We now show that, in the limit \(K \to +\infty\), the Hessian kernel converges to a two-dimensional Dirac delta distribution; that is,
\begin{align} \label{eq:hessian-limit1}
\lim_{K \to +\infty}
\left[
H(\bx,\bx')
:= \frac{2K}{x_{\fd}} J_0(Kx_{\fd}) J_1(Kx_{\fd})
\right]
=
2\pi \delta^{(2)}(\bx_{\fd}).
\end{align}

For Eq.~\eqref{eq:hessian-limit1} to hold, it suffices to show that, for any smooth test function \(f\),
\begin{align}   
\lim_{K \to +\infty}
\int_{\mathbb{R}^2} d^2\bx_{\fd} \,
 \frac{2K}{x_{\fd}} J_0(Kx_{\fd}) J_1(Kx_{\fd})   \, f(\bx_{\fd})
=
2\pi f(\mathbf{0}).
\end{align}

Using the measure \( d^2\bx_{\fd} = 2\pi x_{\fd} \,dx_{\fd} \), the integral becomes
\begin{align}
\lim_{K \to +\infty}
4\pi K
\int_{0}^{+\infty}
dx_{\fd} \,
J_0(Kx_{\fd}) J_1(Kx_{\fd})\, f(x_{\fd}).
\end{align}

Performing the change of variables \( x_{\fd} \to z \), with \( z = Kx_{\fd} \), yields
\begin{align} \label{eq:hessian-limit2}
\lim_{K \to +\infty}
4\pi
\int_{0}^{+\infty}
dz \,
J_0(z) J_1(z)\,
f\!\left(\frac{z}{K}\right).
\end{align}

Since \(f\) is smooth, \(f(z/K) \to f(0)\) as \(K \to +\infty\). Therefore,
\begin{align} \label{eq:hessian-limit3}
\lim_{K \to +\infty}
4\pi
\int_{0}^{+\infty}
dz \,
J_0(z) J_1(z)\,
f\!\left(\frac{z}{K}\right)
=
4\pi f(0)
\int_{0}^{+\infty}
dz \,
J_0(z) J_1(z).
\end{align}

Using the identity \cite{Abramowitz}
\begin{align}  \label{eq:bessel-derivative0}
J_1(z) =
-\frac{d}{dz} J_0(z)
\end{align}
the integral over $z$ becomes
\begin{align}
\int_{0}^{+\infty}
dz \, J_0(z) J_1(z)
=
-\frac{1}{2}J_0^2(z)
\Big|_{0}^{+\infty}
=
-\frac{1}{2}\bigl[0-1\bigr]
=
\frac{1}{2},
\end{align}
where we have used the asymptotic property \(J_0(z)\to 0\) as \(z\to+\infty\).

Substituting this result into Eq.~\eqref{eq:hessian-limit3} gives
\begin{align}
4\pi f(0)
\int_{0}^{+\infty}
dz  \, J_0(z)J_1(z)
=
2\pi f(0),
\end{align}
thereby proving Eq.~\eqref{eq:hessian-limit1} and, consequently, establishing the validity of the identity approximation of the Hessian operator given by Eq.~\eqref{eq:hessian-diagonalised}.
}

\end{document}